\documentclass[aps,twocolumn,prd,superscriptaddress]{revtex4-2}

\usepackage[utf8]{inputenc}

\usepackage{mathtools}
\usepackage{amsfonts}
\usepackage{mathrsfs}
\usepackage{bbm}
\usepackage{physics}
\usepackage{slashed}
\usepackage{tensor}

\usepackage{tikz}
\usetikzlibrary{shapes, arrows}

\usepackage{graphicx}
\usepackage{color, float}
\usepackage{array}
\usepackage[abs]{overpic}

\usepackage{placeins}

\usepackage{makecell}
\usepackage{subcaption}

\usepackage{xspace}
\usepackage{siunitx}
\usepackage{xfrac}
\usepackage{hyperref}
\usepackage[nameinlink]{cleveref}
\usepackage{appendix}

\usepackage{xifthen}
\usepackage{xcolor}
\hypersetup{
	colorlinks,
	linkcolor={red!75!black},
	citecolor={blue!75!black},
	urlcolor={blue!75!black}
}

\usepackage{booktabs}
\usepackage{multirow}

\newcolumntype{C}{>{$}c<{$}}
\AtBeginDocument{
	\heavyrulewidth=.08em
	\lightrulewidth=.05em
	\cmidrulewidth=.03em
	\belowrulesep=.65ex
	\belowbottomsep=0pt
	\aboverulesep=.4ex
	\abovetopsep=0pt
	\cmidrulesep=\doublerulesep
	\cmidrulekern=.5em
	\defaultaddspace=.5em
}

\graphicspath{{./figures/}}

\newcommand{\gettitle}{Physics-informed renormalisation group flows}

\newcommand{\getHeidelbergAffiliation}{\affiliation{Institut f{\"u}r Theoretische Physik, Universit{\"a}t Heidelberg, Philosophenweg 16, 69120 Heidelberg, Germany}}
\newcommand{\getZuerichAffiliation}{\affiliation{Institut f{\"u}r Theoretische Physik, ETH Z{\"u}rich, Wolfgang-Pauli-Str. 27, 8093 Z{\"u}rich, Switzerland}}
\newcommand{\getEMMIAffiliation}{\affiliation{ExtreMe Matter Institute EMMI, GSI, Planckstr. 1, 64291 Darmstadt, Germany}}

\hypersetup{
	pdftitle={\gettitle},
	pdfauthor={Ihssen},
	pdfkeywords={machine learning}
	{generalised RG flows} {renomalisation group} ,
	bookmarksopen=true,
	bookmarksopenlevel=2,
	bookmarksnumbered=true
}

\begin{document}
	
	\title{\gettitle}
	\author{Friederike Ihssen} 
	\getHeidelbergAffiliation
	\getZuerichAffiliation
	\author{Jan M. Pawlowski}\getHeidelbergAffiliation\getEMMIAffiliation

\begin{abstract} 

The physics of strongly correlated systems offers some of the most intriguing physics challenges such as competing orders or the emergence of dynamical composite degrees of freedom. Often, the resolution of these physics challenges is computationally hard, but can be simplified enormously by a formulation in terms of the dynamical degrees of freedom and within an expansion about the physical ground state. Importantly, such a formulation does not only reduce or minimise the computational challenges, it also facilitates the access to the physics mechanisms at play. The tasks of finding the dynamical degrees of freedom and the physical ground state can be systematically addressed within the functional renormalisation group approach with flowing fields which accommodates both, emergent composites as well as the physical ground state. 

In the present work we use this approach to set up \textit{physics-informed renormalisation group flows} (PIRG flows): Scale-dependent coordinate transformations in field space induce emergent composites, and the respective flows for the effective action generate a large set of \textit{target actions}, formulated in these emergent composite fields. This novel perspective on RG flows bears a great potential both for conceptual as well as computational applications: to begin with, PIRG flows allow for a systematic search of the dynamical degrees of freedom and the respective ground state that leads to the most rapid convergence of expansion schemes, thus minimising the computational effort. In combination this allows us to approach the lower simplicity bound for computations in a given theory. Secondly, the resolution of the remaining computational tasks within a given expansion scheme can be further reduced by optimising the physics content within a given approximation. Thirdly, the maximal variability of PIRG flows can be used to reduce the analytic and numerical effort of solving the flows within a given approximation. 

We demonstrate the great potential of physics-informed flows at the example of a scalar theory within the derivative expansion. Here, PIRG flows with fully fixed target actions allow us to convert the numerical task of solving coupled sets of partial differential equations of the convection-diffusion type into the qualitatively simpler task of solving a system of ordinary differential equations. We also show that PIRG flows can be set up such, that the higher order terms in the derivative expansion do not \textit{feed back} into the lower order ones, they only \textit{feed down} into higher orders of the expansion scheme. This feed-down structure leads to a further significant simplification of the typically fully coupled set of flow equations. While demonstrated within the derivative expansion, it constitutes an inherent, high simplification potential of any expansion scheme within PIRG flows. We explicitly solve the O(1) system with a classical target action in the local potential approximation, as well as the first order feed-down correction of the derivative expansion to a standard LPA computation in terms of the standard flow. These results compare well to those obtained with the standard flows, while being obtained with a qualitatively reduced numerical effort. Finally, we discuss the simple computation of specific correlation functions of the fundamental fields as well as general reconstruction schemes for all correlation functions and the effective action in terms of fundamental fields. 

\end{abstract}

\maketitle 
\ 
\newpage 
\ 

\newpage 
\
\onecolumngrid
\tableofcontents
\newpage 
\twocolumngrid

\section{Introduction}
\label{sec:intro}

Quantum field theories (QFTs) accommodate an infinite number of degrees of freedom in a quantum mechanical sense. This is mirrored in the fact that any computational approach for accessing generating functionals of QFTs or only observables is based on systematic expansions, in which this infinite number of degrees of freedom is approached in an asymptotic way. Best known and studied is perturbation theory ordered in powers of couplings or number of scatterings in the theory. Typically, perturbation theory is based on an expansion about the free Gau\ss ian theory, but there are also upgrades with expansions about (simple) interacting theories or expansions about a Gau\ss ian theory, whose covariance (kinetic operator) is given by that of the full theory, see e.g.~\cite{Ihssen:2022xjv, Ihssen:2023nqd}. Non-perturbative examples are lattice field theories, typically formulated on a space-time grid with a finite number of points, e.g.~$N_\tau \times N^3$ in four dimensions. Then, the QFT is defined in the continuum limit (a second order phase transition) with $N_\tau, N\to \infty$. Finally, the functional renormalisation group (fRG) offers a non-perturbative setup, where the limit is approached within systematic expansion schemes, commonly used ones being the derivative expansion and the vertex expansion, for a recent review see \cite{Dupuis:2020fhh}. For a recent detailed analysis including the discussion of systematic error estimates we refer to \cite{Ihssen:2024miv}. 

If applied to strongly correlated systems, the convergence properties can only be accessed in terms of \textit{apparent convergence}, based on the computational observation of convergence in the sense of an extrapolation. On the lattice this is signalled by apparent (critical) continuum and finite volume scaling. In the fRG approach, apparent convergence is encoded in the stability of an extrapolation in $1/N$ with small error bars, where $N$ is the order of the truncation. In the derivative expansion, $N$ is simply the order of momentum squared that is considered, while in the vertex expansion it is the number of fields in the vertex. In most of these cases, convergence is only obtained if the effective number of degrees of freedom is finite and is already captured by a finite number of lattice points or a finite order of the systematic approximation scheme in the fRG. Accordingly, both the convergence and, equally important, the access to the underlying physics mechanisms, can be qualitatively improved by formulating the theories in the relevant degrees of freedom that emerge dynamically. This goes hand in hand with an expansion about the physical ground state of the theory in terms of these degrees of freedom. In a correlation function approach such as the fRG or other functional approaches, the latter is captured in the full covariance or propagator of the theory. 

This asks for an approach, that accommodates general dynamical transformations of the fundamental degrees of freedoms as well as the emergence of additional ones. This is almost mandatory in strongly correlated systems, where the relevant degrees of freedom in different regimes differ qualitatively. Two chiefly important and very illustrative examples for such a situation are the theory of strong interactions (QCD) and ultracold atomic gases: The QCD dynamics at large momenta is well-described by scatterings of quarks and gluons within a perturbative expansion due to asymptotic freedom. In turn, the hadronic regime at asymptotically low momenta is well-described by chiral perturbation theory, an expansion in terms of multi-pion exchanges. While the asymptotic regimes allow for simple and rapidly convergent perturbative expansion schemes, the transition regime is strongly correlated and poses sizeable residual computational challenges. The second example are ultracold atomic gases, which are well-described by perturbative scatterings of atoms for sufficiently large temperatures. In turn, at temperatures below the critical one, ultracold atomic gases feature a superfluid phase with either tightly bound Bose-Einstein condensates (BEC phase) or Cooper pairs (BCS phase), connected by a Feshbach resonance. In particular in the BEC phase, a formulation in terms of tightly bound molecules is far more advantageous and allows for a simple physics interpretation. Like in QCD, the transition regime as well as the regime with the Feshbach resonance are strongly correlated and pose residual but hard computational challenges. 

The two examples above are but two of many strongly correlated physics systems with intriguing but challenging dynamics. As argued above, the quantitative or even qualitative access to the related physics phenomena asks for a theoretical approach that accommodates general reparametrisations of the theory as well as a systematic optimisation framework for such reparametrisations. 

In the present work we put forward and develop such an approach on the basis of generalised functional flows \cite{Wegner_1974, Pawlowski:2005xe}. It is specifically built on the generalised flow for the one-particle irreducible effective action $\Gamma[\phi]$  \cite{Pawlowski:2005xe}, that allows for general reparametrisations, captured by the composite field $\phi[\varphi]$ of given fundamental fields $\varphi$. 
Its most common use so far has been the description of emergent bilinear fermionic composites \cite{Gies:2001nw, Pawlowski:2005xe, Floerchinger:2009uf}, such as mesons in QCD \cite{Gies:2002hq, Braun:2009ewx, Mitter:2014wpa,Braun:2014ata, Rennecke:2015eba, Cyrol:2017ewj, Fu:2019hdw, Fukushima:2021ctq} or molecules and Cooper pairs or tightly bound molecules in ultracold atoms \cite{Floerchinger:2008qc,Floerchinger:2009pg,Scherer:2010sv}. Extensions of this setup to formulations in terms of general scale-dependent fields are the natural next step of improvement and have been considered previously in the context of optimising a given expansion scheme \cite{Lamprecht2007, Isaule:2018mxt, Isaule:2019pcm, Daviet:2021whj}, expansions about the covariance of the theory \cite{Salmhofer:2006pn, Ihssen:2022xjv} and the absorption of emerging couplings \cite{Baldazzi:2021orb, Baldazzi:2021ydj, Ihssen:2023nqd}. 

We advance this formalism qualitatively, based on a novel powerful interpretation of generalised functional flows as a flow equation for pairs $(\Gamma_T,\phi)$ of a given target action $\Gamma_T$ and the respective composite field $\phi$. The standard effective action $\Gamma_T=\Gamma$ is obtained for the trivial choice $\phi=\varphi$, where $\varphi$ denotes the unchanged field basis in terms of the fundamental field. The present approach allows for a rather general set of pairs $(\Gamma_T,\phi)$. Notably, this generality can be used to optimise the physics content of the target action in terms of best convergence of any approximation scheme. This implies in turn, that the description of the dynamics of the physics system at hand and the underlying mechanisms is simplified as much as possible. Furthermore, the variability of the approach allows us to minimise the computational demand for solving the set of flow equations. In particular, the approach allows us to arrange for a \textit{feed-down} structure of the set of flow equations in a given approximation reminiscent of the feed-down structure of perturbation theory: There, higher orders do not feed-back in lower ones. This comprehensive approach with its optimisation potential of maximising the physics content of a given approximation and minimising the remaining computational demand is called \textit{physics-informed} renormalisation group flows. In our opinion, 
unlocking the great potential of \textit{physics-informed} renormalisation group flows should allow us to significantly further our understanding and quantitative access to the physics of strongly correlated systems. 

We close this introduction with a bird's eye view: In \Cref{sec:flowingField}, we recall the definition of composite operators in the effective action and the path integral. The generalised flow equation as well as its previous applications in the literature are discussed. In particular, we make an explicit analysis of stability and existence of the formulation. 
This is followed by the discussion of the shift in perspective that is pivotal for the advances reported on in the present work: The concept of target action flows, i.e.~flows that implement any desired effective action $\Gamma_{T}$ are introduced in \Cref{sec:PIFlow}. Together with the definition of the composite field we deal with a two-dimensional functional space which we call the \textit{physics-informed} RG (PIRG) setup. 
Notably, in the PIRG setup a second functional direction is introduced and used for optimisation, the reduction of numerical/computational complexity of the RG-machinery. 
For example, this could be the decoupling of layers of a given expansion scheme, similar to perturbation theory whose higher order corrections  do not feed back into the lower orders. This type of optimisation should be seen in contrast to the standard optimisation of the physics content of a given approximation which also can been addressed in the context of the PIRG flows. The introduction of additional composite fields or even the substitution of the fundamental fields by composite ones may require the use of reconstruction schemes for correlation functions of the fundamental fields, respective observables and the effective action of the fundamental field. In the present work we discuss the map between fundamental and composite fields as well as the computation and general reconstruction of correlation functions of the fundamental fields in the PIRG setup. 
In \Cref{sec:FeedDown} we discuss feed-down PIRG flows and their applications: We initiated this discussion with general remarks on the feed-down structure for PIRG flows. These flows are set up to eliminate the feed-back of higher orders in a given expansion scheme. This is a considerable simplification of the computational effort since higher orders can be evaluated on top of existing lower order calculations, and we discuss this at the example of the derivative expansion for a simple scalar theory in 
\Cref{sec:PI-Feed-down}. Finally, in \Cref{sec:GenConv+Feed} we put PIRG flows to work within selected illustrative examples with numerical applications: First we show the high potential of the target action concept within the extreme choice of the target action, the classical one, where all quantum fluctuations are stored in the flowing composite. We also implement a feed-down flow for the first order derivative expansion, which demonstrates explicitly the practical use of these flows. We summarise our findings in \Cref{sec:Outlook} and provide an outlook on future developments.

\section{General flows and flowing fields}
\label{sec:flowingField}

The generalised flow equation \cite{Pawlowski:2005xe} for the effective action $\Gamma$ relies on a scale-dependent non-linear (in field space) transformation of the mean field $\phi$, the argument of $\Gamma[\phi]$. Such a change substitutes or augments the mean field $\varphi$ of the fundamental degrees of freedom with general composite flowing field $\phi_k[\varphi]$, where $k$ is the scale labelling the transformation, typically a momentum scale. Importantly, this transformation is fully encoded in the differential change of the field $\phi$ at a given scale $k$, 
\begin{align}
	\dot \phi_k[\phi]\,, 
	\label{eq:dotphikIntro}
\end{align}  
where the dot in $\dot{\phi}$ signals its infinitesimal nature. \Cref{eq:dotphikIntro} has no explicit reference to the fundamental field $\varphi$ and its local nature is key to the tractability of the transformation.  

In the present Section we provide a detailed discussion of our setup with flowing composite fields. This includes in particular an assessment of the global existence of the field transformations with the infinitesimal form \labelcref{eq:dotphikIntro} as well as limitations for $\dot\phi_k$ itself. 

In \Cref{sec:CompOp} we discuss the notion of composite fields in the effective action. This is followed by a brief recapitulation of the generalised flow equation in \Cref{sec:GenFlow}, including reviewing some of its previous applications. In 
\Cref{sec:FlowingFields} we provide, for the first time, a comprehensive analysis of the local and global existence and stability of the general field transformation \labelcref{eq:dotphikIntro}. This important aspect of the approach has so far only been discussed rather implicitly in previous works, see \cite{Pawlowski:2005xe, Ihssen:2023nqd}. Specifically, we show that the generalised flow equation is well-defined for general local flows of the composites. The notion of locality will be detailed below, but loosely speaking it refers to the property, that the flow of the composites is local in momentum space (\textit{momentum-local}, see \cite{Christiansen:2015rva, Dupuis:2020fhh}), as well as (uniformly) convergent in field space in the given systematic expansion scheme.

\subsection{Generating functionals with composite fields } 
\label{sec:CompOp}

For the following conceptual discussions we restrict ourselves to scalar O(N)-theories. The use of this simple set of example theories makes the discussion far more accessible. However, we emphasise that the structural results presented below do not depend on specific properties of the scalar theory they are derived in and trivially extend to general quantum field theories.  

We initiate this discussion with the remark, that the present approach is entirely based on the \textit{existence} of the generating functionals or rather their derivatives. Specifically, $n$-point correlation functions of a quantum field theory with the fundamental quantum field $\hat \varphi$, including their disconnected pieces are defined by $n$th order derivatives of the generating functional $Z[J_\varphi]$ with respect to the source $J_\varphi$ of the fundamental field. The one-particle irreducible (1PI) parts of these correlation functions are given by the $n$th order derivatives of the effective action $\Gamma[\varphi]$ with respect to the mean field $\varphi =\langle \hat\varphi\rangle$. We find structurally, 
\begin{align}
\langle \hat \varphi^n  \rangle = \frac{1}{Z[0]}	\frac{\delta^n Z[J_\varphi]}{\delta J_\varphi^n} \,,\qquad \langle \hat \varphi^n  \rangle^{\ }_\textrm{1PI}= \frac{\delta^n \Gamma[\varphi]}{\delta \varphi^n} \,,
	\label{eq:nPointvarphi}
\end{align}
for all $n$, see \cite{Pawlowski:2005xe}. The normalisation with $Z[0]$ in \labelcref{eq:nPointvarphi} leads to normalised correlation functions and specifically it leads to $\langle 1\rangle =1$. A further relevant set of correlation functions, the connected correlation functions, is derived from the Schwinger functional $\log Z[J_\varphi]$, and can be expressed in terms of either derivatives of $Z$ or of $\Gamma$. In \labelcref{eq:nPointvarphi} we suppressed the space-time arguments of the fields and the currents, $\hat \varphi^n = \hat\varphi(x_1)\cdots \hat\varphi(x_n)$ and similarly for the derivatives w.r.t.~the currents and the mean field.  

The derivatives \labelcref{eq:nPointvarphi} are nothing but the moments of the underlying probability distribution, 
\begin{align}
	d\mu[\hat\varphi] = [d\hat\varphi]_\textrm{ren}\, e^{-S[\hat\varphi]}\,,
	\label{eq:dmu}
\end{align}
where the hat indicates the microscopic fields or operators. With \labelcref{eq:dmu}, the existence of the generating functionals simply entails the convergence of the respective Taylor expansion at least as an asymptotic series. Note, that such a setup does not explicitly rely on a path integral representation with the measure \labelcref{eq:dmu}. While this is a mere tautology as proving the existence of a generating functional is tantamount to proving the existence of a path integral representation, it highlights the fact that solving the flow equation is equivalent to constructing the path integral. 

As discussed in \cite{Pawlowski:2005xe}, the functionals \labelcref{eq:nPointvarphi} with the fundamental fields $\hat \varphi$ or the respective currents $J_\varphi$ may not be an optimal representation of the moments of the statistical measure \labelcref{eq:dmu}. In extreme cases, the Taylor expansion with the moments \labelcref{eq:nPointvarphi} may not even converge though the theory exists. In both cases one may consider a different set of moments, namely that of a composite field $\hat\phi[\hat\varphi]$,
\begin{align}
	\frac{\delta^n Z_\phi[J_\phi]}{\delta J_\phi^n} \simeq 	\langle \hat\phi \cdots \hat\phi\rangle\,. 
\label{eq:CorrelationsComposites} 
\end{align}
Here, the subscript ${}_\phi$ indicates that the functional $Z_\phi[J_\phi]$ has a different functional form than $Z_\varphi[J_\varphi] = Z[J_\varphi]$ as it generates correlation functions of $\hat\phi$. \Cref{eq:CorrelationsComposites} may include correlations of the fundamental field or may only involve correlations of a given composite. A commonly used example is the inclusion of products of fields, e.g.
\begin{align} 
	\hat \phi= \Bigl(\hat\varphi(x)\,,\, \hat\varphi(x)\hat\varphi(y)\Bigr)\,,\quad \textrm{or}\quad 	\hat \phi= \Bigl(\hat\varphi(x)\,,\, \hat\varphi^2(x)\Bigr)\,,
	\label{eq:hatphi2PI2PPI}
\end{align}
with the currents $J_\phi(x, y)= (J_\varphi(x), J_{\varphi\varphi}(x,y))$ and $J_\phi= (J_\varphi(x), J_{\varphi^2}(x))$ respectively. The first example in \labelcref{eq:hatphi2PI2PPI} is related to the two-particle irreducible (2PI) effective action, while the second one is related to the two-point particle irreducible (2PPI) effective action and density functional theory (DFT). In DFT one drops the current of the fundamental field $\hat\varphi$. \Cref{eq:hatphi2PI2PPI} also highlights the fact that the flowing field approach does not only change the coordinate system (or frame) in field space, it can also be used to include additional fields. Note that in the examples provided in \labelcref{eq:hatphi2PI2PPI}, the composite mean field also includes the disconnected parts of the respective operators. In the 2PI and 2PPI formulations these parts are usually subtracted. 

In the specific examples in \labelcref{eq:hatphi2PI2PPI}, the generating functionals reduce to the original generating functionals for vanishing currents of the composite fields $J_{\varphi\varphi}$ or $J_{\varphi^2}$. Accordingly, we maintain full access to the physics of the underlying theory. In summary, this provides a powerful versatile functional approach to general quantum field theories. 

The correlation functions of the composites are readily obtained from a generalisation of the path integral representation of the generating functional $Z_\phi$: We introduce a source term, where the current $J_\phi$ couples to general composite fields $\hat \phi[\hat \varphi]$,  
\begin{align} 
	Z_\phi[J_\phi ]  =\frac{1}{{\cal N}_\phi} \int [d\hat\varphi]_\textrm{ren}\, e^{-S[\hat\varphi]+\int_{\boldsymbol{x}} J_\phi \hat\phi[\hat \varphi]}  \,, 
	\label{eq:ZGenComposites}
\end{align}
with a normalisation ${\cal N}_\phi$ at our disposal and $\int_x$ indicates the integration over Euclidean space-time. Then, $J_\phi$ derivatives of 	\labelcref{eq:ZGenComposites} lead to \labelcref{eq:CorrelationsComposites}. We note that the normalisation ${\cal N}$ leading to normalised correlation functions including $\langle 1\rangle =1$ is $\phi$-independent as it is simply given by the path integral without the source term. 

The source term consists of currents for the composite fields $\hat \phi^a$ with $J_\phi=(J_\phi^1,J_\phi^2,...,J_\phi^n) $ and reads  
\begin{align}
	\int_{\boldsymbol{x}} J_\phi \hat\phi =\sum_a  \int_{\boldsymbol{x}_a} J^a_\phi(\boldsymbol{x}_a)\,\hat\phi^a(\boldsymbol{x}_a) 	, \quad 	\boldsymbol{x}_a= (x_1,...,x_{n_a})\,,
	\label{eq:DefSuperindex}
\end{align}
with the possible dependence of $\hat \phi$ on multiple space-time points collected in  $\boldsymbol{x}_a$. The classical action is given in terms of the fundamental field $\hat \varphi$ or the respective mean field $\varphi$, 
\begin{align}
	S[\varphi] = \int_x \left\{ \frac{1}{2} \left(\partial_\mu \varphi\right)^2 +\frac{\mu_\varphi}{2} \varphi^2  + \frac{\lambda_\varphi }{8} \,  \varphi^4\right\} \,.
	\label{eq:ClassicalAction1-ON}
\end{align}
The composite field may be different from the fundamental field $\hat \varphi$, in terms of which the classical action is defined.
If one choses $\hat \phi[\hat \varphi] = \hat \varphi$, we simply obtain the standard definition of the path integral with the current $J_\varphi$.
The effective action $\Gamma_\phi[\phi]$ of the composite field $\phi$ is obtained by a Legendre transform with respect to the current $J_\phi$,
\begin{align}
	\Gamma_\phi[\phi] = \sup_{J_\phi}\left(\int_{\boldsymbol{x}} J^a_\phi \,\phi^a - \ln Z_\phi[J_\phi]\right)\,, \quad J_{\phi}^a = \frac{\delta \Gamma_\phi}{\delta \phi_a}\,. 
	\label{eq:GenG}
\end{align}
As for the generating functional $Z_\phi$, the subscript $_\phi$ used in the effective action $\Gamma_\phi$ indicates  that the Legendre transform was taken with respect to the current $J_\phi$. Hence, its argument is the mean field $\phi$. 
This leaves us with an effective action that depends on a collection of composite (mean) fields $\phi= \langle \hat \phi \rangle $, 
\begin{align}
	\phi(\boldsymbol{x})=\bigl(\phi_1(\boldsymbol{x}_1),...,\phi_n(\boldsymbol{x}_n)\bigr)\,.  \label{eq:GenComposite}
\end{align}
The advantages of such a formulation are multifold. To name two important ones, subject to the chosen composite $\phi$, the effective action $\Gamma_\phi[\phi]$ in \labelcref{eq:GenG} may have a more rapid convergence in terms of a Taylor expansion in $\phi$, or its computation may be significantly simpler.

\subsection{Generalised functional flows} 
\label{sec:GenFlow}

In the presence of composite fields with a possible scale dependence $\hat \phi_k [\hat \varphi]$, the generalised flow equation for the effective action takes the form \cite{Pawlowski:2005xe}, 
\begin{align}\nonumber 
	\Biggl( \partial_t + \int_{\boldsymbol{x}}  \dot{\phi} \frac{\delta}{\delta \phi}\Biggr)\left(  \Gamma_{\phi}[\phi]  -c_\phi \int \phi
	\right)  \\[1ex]
	&\hspace{-4.5cm}=\frac{1}{2} \Tr\Bigg[ G[\phi]\left(\partial_t +2 \frac{\delta \dot{\phi}}{\delta \phi} \right) \, R_\phi\Bigg] + \partial_t \ln{\cal N}_{\phi}\,,
	\label{eq:GenFlow} 
\end{align}
where the $\dot \phi$-insertions govern the infinitesimal change of the mean field basis $\phi$ with the cutoff scale $k$,
\begin{align}
	\dot \phi_k[\phi] = \bigl\langle \partial_t \hat\phi_k[\hat\phi]\bigr\rangle \,,\qquad  t=\log \frac{k}{k_\textrm{ref}}\,, 
	\label{eq:dotphi} 
\end{align}
where $t$ is the (negative) RG-time and $k_\textrm{ref}$ is some reference scale, commonly the initial scale 
$ k_\textrm{ref}=\Lambda$ is chosen. In \labelcref{eq:GenFlow} we have dropped all explicit reference to the cutoff scale $k$ for the sake of readability. We have also introduced the regulator function $R_{\phi,k}$ that encodes the details of the cutoff procedure. The subscript ${}_\phi$ indicates that the regulator couples to the composite fields, as discussed later around \labelcref{eq:Reghatphi}. In the present work we concentrate on infrared momentum cutoffs, for more details see \Cref{app:thrs+reg}. However, the approach is far more general as we may substitute the momentum scale $k$ by a general flow parameter $s$. This flow parameter may be a length scale or a time, or simply comprise a general reparametrisation. In this more general case, $R_{\phi,s}$ implements the respective change. The general flow uses the propagator of the composite fields, 
\begin{align}
	G_{\phi_i \phi_j}[\phi](\boldsymbol{p},\boldsymbol{q}) = \left[ \frac{1}{\Gamma_{\phi}^{(2)}+R_{\phi}}\right]_{\phi_i \phi_j}(\boldsymbol{p},\boldsymbol{q}) \,, 
	\label{eq:Prop}
\end{align}
with 
\begin{align} 
	\Gamma_\phi^{(n)}[\phi](\boldsymbol{p}_1,...,\boldsymbol{p}_n) =  \frac{\delta}{\delta \phi_1(\boldsymbol{p_1} )}\cdots   \frac{\delta}{\delta \phi_n(\boldsymbol{p_n} )} \Gamma_\phi[\phi]\,.
	\label{eq:Gammaderphi}
\end{align}
The notation used in \labelcref{eq:Gammaderphi} is extended to $n$th derivatives of general functionals and specifically also to $\dot \phi[\phi]$.

The constant flow $\partial_t \ln{\cal N}_{\phi}$ on the right-hand side of \labelcref{eq:GenFlow} originates in a $k$-dependent normalisation of the path integral. In standard Wetterich flows, this constant is typically either dropped or chosen such that the value of the effective action or the effective potential at the minimum is vanishing. However, in \labelcref{eq:GenFlow} it is not a mere constant, instead it also induces a field-dependent change of $\dot \phi$, and hence influences the field dependence of the effective action. This is discussed later in an explicit $\phi^4$-example in \Cref{app:c1fixing}.

The shift $c_\phi  \int \phi$ in \labelcref{eq:GenFlow} occurs in the presence of linear (in $\phi$) explicit symmetry breaking terms in the classical action. It only affects the zero mode or collective excitation of the field $\phi$. Such a breaking can be absorbed in a shift of the current $J_\phi$, and the combination $\Gamma_\phi -c_\phi \int \phi$ is symmetric: This accommodates the fact that linear breaking terms do not enter the quantum dynamics of the effective action and hence the flow equation is that of the symmetric theory. The occurrence and necessity of such a term has been first discussed in detail in \cite{Fu:2019hdw} for explicit chiral symmetry breaking in QCD. There, the respective linear terms in the fermionic bilinear composite $\phi \propto \bar q q$ due to the non-vanishing current quark mass term proportional to $\int \bar q q$ in the fundamental QCD action. More details can be found in Section II in \cite{Fu:2019hdw}. In the present work we incorporated this term in \labelcref{eq:GenFlow} only for the sake of completeness. All examples considered here are fully O(N)-symmetric scalar theories, i.e.~$c_\sigma = 0$, and we shall consider the general case with $c_\sigma \neq 0$ in future work.

Finally, if the composite field equals the fundamental field, $\phi=\varphi$, the terms proportional to $\dot \phi$ in \labelcref{eq:GenFlow} vanish and the flow equation reduces to the standard Wetterich flow \cite{Wetterich:1992yh}, see also \cite{Ellwanger:1993mw, Morris:1993qb}. In comparison to the Wetterich equation, the addition on the left-hand side of the equation considers the immediate effect of the change of field-basis on the effective action. It only has an implicit regulator dependence and could also be used to change the field-basis without lowering the cutoff scale $k$, hence inducing a regulator-independent flow of the field. Finally, on the right hand side of the equation we obtain a correction proportional to the regulator $R_{\phi}$, which corresponds to a correction of the RG flow.

\subsubsection{Key properties and examples} 
\label{eq:GenFlowExplanation}

A central property of the generalised flow \labelcref{eq:GenFlow} is the regularisation of the propagator of the composite field. In the path integral representation this regularisation is introduced with an additional quadratic term of the composite field $\phi$,  
\begin{align}
	\Delta S_k[\hat\phi_k] = \frac12 \int \hat \phi_k\, R_{\phi,k}\, \hat \phi_k\,,  
	\label{eq:Reghatphi}
\end{align}
see \cite{Pawlowski:2005xe}. In \labelcref{eq:Reghatphi} we concentrate on the quadratic nature in $\hat\phi$, a more detailed form is provided later in \labelcref{eq:GenRphi}. The $\hat\phi^2$-dependence is reflected by the occurrence of the $\phi$-propagator in \labelcref{eq:GenFlow} without any transformation matrices. Importantly this direct regularisation of the composite field suppresses potential soft or even massless composite modes. If such composite soft modes are not regularised, the flow is not momentum local, for a comprehensive discussion see \cite{Ihssen:2024miv}. This leads to a worse convergence of any expansion scheme and such a setup even fails in low orders of the given expansion scheme, for a respective analysis see e.g.~\cite{Fu:2022uow, Fu:2024ysj}. 

The generalised flow equation \labelcref{eq:GenFlow} allows us to enforce field transformations. The best-studied case in the literature is the introduction of dynamical composite degrees of freedom. This is well-tested and quantitatively successful in theories with bilinear fermionic composites such as QCD with the composite pions and $\sigma$-mode, as well as in ultracold atoms with composite atomic dimers and trimer fields. Then, the physics of the respective four-Fermi scattering vertex $\lambda_{\psi}$ is stored in the dynamics of the emergent composite. Practically this is implemented with the constraint,  
\begin{align}
	\dot\phi_k[\phi]:\quad  \partial_t \lambda_{\psi}\stackrel{!}{=}0\,. 
	\label{eq:lambda4f0}
\end{align}
This fixes the flow of the transformation $ \dot\phi_k[\phi]$, \cite{Gies:2001nw, Gies:2002hq, Braun:2009ewx, Mitter:2014wpa,Braun:2014ata, Rennecke:2015eba, Cyrol:2017ewj, Fu:2019hdw, Fukushima:2021ctq}. 

A first application with non-linear transformations can be found in \cite{Lamprecht2007} within an O(N)-theory. There, the composite field $\phi_k$ interpolates between the Cartesian field basis $(\varphi_1,...,\varphi_n)$ in the symmetric phase and the polar basis in the broken phase (\textit{Goldstonisation}). The specific approach and results from \cite{Lamprecht2007} have also been taken over in \cite{Isaule:2018mxt}. More recently, the generalised flow has been used to eliminate the wave function $Z_\phi[\phi](p)$, enforcing $Z_\phi\equiv 1$. In terms of the flow this entails, \cite{Baldazzi:2021ydj}, 
\begin{align}
	\partial_t Z_\phi[\phi](p)\stackrel{!}{=}0\, \quad \longrightarrow \quad \dot\phi_k[\phi]\,. 
	\label{eq:Zphi1}
\end{align}
This has been implemented within the first order of the derivative expansion with $Z_\phi=Z_\phi(\phi)$ \cite{Baldazzi:2021ydj, Ihssen:2023nqd, Bonanno:2025mon}, for applications in gravity see \cite{Baldazzi:2023pep, Baldazzi:2021fye, Knorr:2022ilz, Ohta:2025xxo}.

Both non-linear examples, the Goldstonisation in  \cite{Lamprecht2007, Isaule:2018mxt} and the removal of $Z_\phi[\phi]$ in \labelcref{eq:Zphi1} can be embedded in a \textit{ground state expansion}, see \cite{Ihssen:2023nqd}. In such an expansion, the effective action $\Gamma_\phi$ is expanded about the (on-shell) ground state of the theory: this entails 'classical' dispersions as well as a parametrisation in the dynamical degrees of freedom such as the Goldstone modes in the broken phase. We emphasise that this expansion scheme also incorporates the emergent bilinear composites discussed around \labelcref{eq:lambda4f0} as dynamical infrared degrees of freedom. 

The above examples are schematically depicted in \Cref{fig:schems}, including a comparison to the Wetterich flow. In \Cref{fig:TheorySpace} we sketch different flows in theory space. In \Cref{fig:schem2} we sketch the (functional) two-parameter space that visualises the additional degrees of freedom at our disposal.

\subsubsection{Generalised effective action} 
\label{eq:GenFlowGenAction}

In view of future applications to non-abelian gauge theories such as QCD, as well as quantum gravity we also emphasise again, that the effective action $\Gamma_\phi[\phi]$ in \labelcref{eq:GenFlow} is not simply a reparametrisation of $\Gamma_\varphi[\varphi]$ but a different generating functional. This will be discussed in detail in \Cref{sec:Reconstruction}. This change of the generating functional allows for setting up gauge-invariant or diffeomorphism-invariant flows by simply implementing gauge-invariance constraints via the flow of the composite. This is not possible, if a reparametrisation of $\Gamma_\varphi[\varphi]$ is considered which is by construction gauge-fixed.   

We close this discussion of the generalised effective action with a remark on \textit{microscopic} and \textit{macroscopic} transformations as introduced in \cite{Wetterich:2024uub}. There, macroscopic transformations are done on the level of the effective action and hence re-parametrise the 
effective action of the fundamental field, leading to $\Gamma_\varphi\left[\varphi[\phi]\right]$. In turn, microscopic transformations are done on the level of the fields $\hat \phi$ which are also coupled to currents. The latter source terms change the effective action, leading to $\Gamma_\phi[\phi]$. Only on the equations of motions the two actions are identical as the currents vanish. In \cite{Wetterich:2024uub} the reconstruction problem for the effective action $\Gamma_\varphi[\varphi]$ as well as for observables from the effective action  $\Gamma_\phi[\phi]$ was raised. We will discuss comprehensively the computation of  observables from $\Gamma_\phi[\phi]$ as well as the general reconstruction in \Cref{sec:Reconstruction}. In our opinion, the change or rather generalisation of the effective action, that is inherent to the generalised flow equations, is not a deficiency but rather is key to its potential, which is unlocked in \Cref{sec:PIFlow}.

\subsection{Existence of flowing field transformations}
\label{sec:FlowingFields}

The initial effective action of the scalar theory \labelcref{eq:ClassicalAction1-ON} in $d<4$ space-time dimensions is given by 
\begin{align}
	\Gamma_{\phi,\Lambda}[\varphi] = S[\varphi]\,, \qquad \varphi=\langle \hat\varphi \rangle \,, 
	\label{eq:InitialG}
\end{align}
with the \textit{flat} fundamental field $\hat\varphi$. While it is not necessary, we have assumed in \labelcref{eq:InitialG} that 
\begin{align}
	\phi_\Lambda[\varphi]= \varphi \,,
\end{align}
for the sake of simplicity. \Cref{eq:InitialG} is approached in the presence of the regulator $R_{\Lambda}$ that suppresses all quantum fluctuations in $\phi$ and $\varphi$, and the couplings and wave functions in the classical action possibly depend on $\Lambda$. For example, in three dimensions the RG flow of the mass squared scales linearly with $\Lambda$, while the self-coupling and the wave function have a finite UV limit. Note that \labelcref{eq:InitialG} implies that the generating functional of the theory admits the path integral representation \labelcref{eq:ZGenComposites}. 
This explains the notion \textit{flat} field: It describes the fact that the path integral measure is flat (translation invariant) and the suppression of all quantum fluctuations leads to the initial effective action \labelcref{eq:InitialG}. If, on top of this, the initial effective action has canonical dispersions, the quantum field $\hat\varphi$ obeys canonical commutation relations in the absence of ultraviolet anomalous dimensions.

In the functional renormalisation group approach, the composite fields can be included step-by-step in the flow of the theory with the infrared cutoff scale $k$ via the generalised flow for the effective action $\Gamma_k[\phi]$ \labelcref{eq:GenFlow}. 
Interestingly, this equation only depends on $\dot \phi_k[\phi]$ and not on the global transformation $\phi_k[\varphi]$. Of course the latter can be obtained by integrating the former, which we discuss in \Cref{sec:Reconstruction}. Therefore, the \textit{global existence} of such a transformation is proven within a two-step procedure:  
\begin{itemize} 
	\item[(i)] \textit{Local} existence of $\dot \phi_k[\phi]$ in \labelcref{eq:dotphi}: One has to show that there is an operator $\partial_t \hat\phi_k[\hat\phi]$ leading to $\dot \phi_k[\phi]$, which is an inverse problem. 
	\item[(ii)] \textit{Global} existence of $\phi_k[\varphi]$: Its flow is integrated from $k=\Lambda$ to $k=0$. The result has to be free of divergences.   
\end{itemize} 
In summary, the crucial step is (i), which is discussed in the following, while (ii) always can be arranged for. For illustrative examples discussing singularities in the flow and their removal see \cite{Ihssen:2023nqd}.

\subsubsection{Local existence of $\dot \phi_k[\phi]$}
\label{sec:local}

Let us now discuss potential obstructions for $\dot \phi_k[\phi]$ that come with its implicit definition \labelcref{eq:dotphi}. To that end we consider the expansion of $ \partial_t \hat\phi_k[\hat\phi]$ in terms of operators $\hat{\cal O}_n[\hat\phi](x_1,...,x_{m_n})$, 
\begin{align}
	 \partial_t \hat\phi_k[\hat\phi](\boldsymbol{x}) = \sum_n \int\limits_{\boldsymbol{x}} \,c_n\left( \boldsymbol{x}, \boldsymbol{x}^{\ }_{{\cal O}_n}\right) \hat{\cal O}_n[\hat\phi]\left(\boldsymbol{x}^{\ }_{{\cal O}_n}\right)\,, 
 \label{eq:hatphiExpand}
\end{align}
where $\boldsymbol{x}$ is the vector of space-time arguments of $\phi$, see  \labelcref{eq:GenComposite}, and $	\boldsymbol{x}^{\ }_{{\cal O}_n}$ is the vector of space-time arguments of ${\cal O}$,
\begin{align} 
	\boldsymbol{x}^{\ }_{{\cal O}_n} = (x_1,...,x_{m_n})\,.
\end{align}
An obvious choice is $\hat{\cal O}_n[\hat\phi](x_1, \dots ,x_n)=\hat\phi(x_1)\cdots \hat \phi(x_n)$ with $m_n=n$. A further choice is e.g.~a basis of rational functions of the composite fields $\hat\phi$ that admit convergent Taylor expansions. The latter constraint of the existence of a Taylor expansion simplifies the considerations below but it is not a necessary requirement. 
 
\Cref{eq:hatphiExpand} readily leads to the flow of the mean field basis 
\begin{align}
	\dot\phi_k[\phi] =\sum_n \int\limits_{\boldsymbol{x}} \,c_n\left( \boldsymbol{x}, \boldsymbol{x}^{\ }_{{\cal O}_n}\right){\cal O}_n[\phi]\left(\boldsymbol{x}^{\ }_{{\cal O}_n}\right)\,.
	\label{eq:phiExpand}
\end{align}
In \labelcref{eq:phiExpand}, the functionals ${\cal O}_n$ are the expectation values of the respective operators $\hat{\cal O}_n$. In terms of the composite fields $\phi$ and their correlation functions it reads  
\begin{align}
{\cal O}[\phi] = \hat{\cal O}\left[\hat\phi(\boldsymbol{x}) = \int_{\boldsymbol{y}} G[\phi](\boldsymbol{x},\boldsymbol{y})\frac{\delta}{\delta \phi(\boldsymbol{y})} +\phi(\boldsymbol{x})\right] \,, 
\label{eq:CalO}
\end{align}
where $\boldsymbol{x}, \boldsymbol{y}$ carry the space-time dependence of $\phi$, and 
$G[\phi]$ is the full field-dependent propagator of the composite field, 
\begin{align} 
G[\phi](\boldsymbol{x},\boldsymbol{y}) = \langle \phi(\boldsymbol{x}) \phi(\boldsymbol{y})\rangle[\phi] - \phi(\boldsymbol{x})  \phi(\boldsymbol{y})\,.
\label{eq:Gphi}
\end{align}
For operators $\hat{\cal O}[\hat \phi]$ that admit a Taylor expansion in $\hat\phi$, \labelcref{eq:CalO} is well-defined, and so is \labelcref{eq:phiExpand}. Note that the task is not to surmise an explicit definition of the right-hand side of \labelcref{eq:phiExpand} for a given left-hand side, as we only need the left-hand side explicitly. The only task is to construct the set of allowed transformation flows $\{\dot\phi_k[\phi]\}$. Importantly, this implies that we do not have to explicitly solve the inverse problem underlying (i).  

In order to appreciate the generality of the implicit definition \labelcref{eq:CalO} within the functional renormalisation group approach, we use the Fourier representation of the summands in \labelcref{eq:phiExpand}, 
\begin{align}
  \int\limits_{\boldsymbol{p}} \,c_n(\boldsymbol{p}){\cal O}_n[\phi](\boldsymbol{p})\,.
 \label{eq:CalOp}
\end{align}
Note that the definition \labelcref{eq:CalO} implies a loop representation of ${\cal O}$. These loops do not indicate perturbation theory, and depend on full propagators and vertices and vertices as the flow equation (or a Dyson-Schwinger equation) for the effective action itself. In contradistinction to the flow equation the loop representation of a general ${\cal O}$ includes higher loop orders. 

Now we restrict ourselves to transformations, which are \textit{momentum-local}: 
\begin{itemize}
	\item[(a)] \textit{$\dot \phi[\phi](\boldsymbol{p})$ is momentum-local}: It decays for large momenta and large fields and is stable for small momenta and small fields. 
	 \item[(b)] \textit{\Cref{eq:CalOp} is momentum-local}: All its loops only involve loop momenta $l_i^2\lesssim k^2$. In particular, this constraint renders all loops finite.  
\end{itemize}
Both constraints (a), (b) imply and require an infrared regularisation of the composite fields $\hat\phi$. In the functional renormalisation group approach this is achieved with cutoff terms 
\begin{align} 
 \Delta S_k[\hat\phi_k] =\frac12 \sum_{ab} \int_{\boldsymbol{p},\boldsymbol{q}}  \hat \phi_{a,k}^\dagger(\boldsymbol{q})  \, R^{ab}_{\phi,k}(\boldsymbol{q},\boldsymbol{p}  ) \, \hat \phi_{b,k}(\boldsymbol{p})\,. 
\label{eq:GenRphi}
\end{align}
The indices ${a,b}$ in \labelcref{eq:GenRphi} sum over different composite field species, and the regulator function $R_{\phi}(\boldsymbol{q},\boldsymbol{p})$ is typically diagonal in the momenta and suppresses the propagation of modes with momenta $p_i^2\lesssim k^2$. 
We emphasise that \labelcref{eq:GenRphi} is also stabilising expansion schemes as it guarantees that the effective action is formulated in degrees of freedom, whose flows 'live' on the same momentum shell. Amongst other stabilising properties, this optimises momentum expansions. In turn, the restriction of the regulator terms to that of the fundamental fields leads to emergent non-localities during the flow. This is most obvious in the case where the composite fields are massless. Notably this happens if the composite fields describe dynamical Goldstone modes such as in QCD with dynamical chiral symmetry breaking and the composite pion fields. 

In summary, we have achieved two goals in one stroke within this setup: First of all, it restricts the flow of the field transformations to those, that have a similar cutoff scale as the flow of the effective action. Secondly, it allows for momentum and field expansions of the right-hand side of \labelcref{eq:phiExpand} that mirror the given systematic expansion scheme in the effective action itself. 
Importantly, we readily conclude that in a given order of any systematic expansion scheme, the series on the right-hand side of \labelcref{eq:phiExpand} is dense in the set of momentum-local $\dot \phi[\phi]$, and the latter can be chosen freely. This finalises our discussion of the constraint (i). 

As discussed before, the constraint (ii), i.e.~global existence of the map $\phi_k[\varphi]$, is monitored during the flow. This leaves us with the optimal combination of a free choice of the \textit{local} field transformation flows in (i) and a guaranteed \textit{global} existence by simply integrating $\dot \phi$ over the RG-time $t$. We add that the \textit{locality} of the flow of $\phi$ is guaranteed in the most common use of the flowing fields in the literature initiated in \cite{Gies:2001nw}, where bilinear flowing fields have been considered. There it is used to trivialise the flow of a four-point function or coupling. For its most common application in QCD, see \cite{Gies:2002hq, Braun:2009ewx, Mitter:2014wpa,Braun:2014ata, Rennecke:2015eba, Cyrol:2017ewj, Fu:2019hdw, Fukushima:2021ctq}. However, beyond the bilinear application, general flowing fields \cite{Pawlowski:2005xe} are required. This has been used for trivialising the flow of the wave function $Z_\phi(\phi)$ in a scalar theory, see below \labelcref{eq:Zphi1} and \cite{Lamprecht2007, Isaule:2018mxt, Baldazzi:2021ydj, Ihssen:2023nqd}, leading to a classical dispersion for the flowing field $\phi$. Within these applications, $\dot \phi$ is either related to (minus) the flow of the coupling or that of the wave function $Z(\phi)$. By definition these flows satisfy momentum-locality (a) and (b), as the respective flow equations of the coupling or the wave function field in the effective action are momentum-local for the class of regulators with \labelcref{eq:GenRphi}.

\section{Physics-informed RG flows} 
\label{sec:PIFlow} 

In this Section we introduce the novel perspective on functional renormalisation group flows, which underlies the conceptual and computational advances in this work: Instead of viewing the generalised functional RG \labelcref{eq:GenFlow} as a means to compute the effective action or rather correlation functions of a given set of fields, including composite ones, we should view the generalised flow as a defining equation for the effective action $\Gamma_\phi[\phi]$ \textit{and} the composite $\phi[\varphi]$ on an equal footing. Then, the physics information of a given theory is not stored in the effective action but in the pair 
\begin{align} 
	\left( \Gamma_\phi[\phi] \,,\, \phi[\varphi] \right) \,. 
	\label{eq:PI-Pair}
\end{align}
In general, both entries, the effective action $ \Gamma_\phi[\phi]$ and the composite field $\phi[\varphi]$ carry part of the physics of the system under consideration. This novel perspective allows us to unlock the full power of the generalised flow equation. 

\begin{figure*}
	\centering
	\begin{subfigure}{.48\linewidth}
		\centering
		\includegraphics[height=0.5\linewidth]{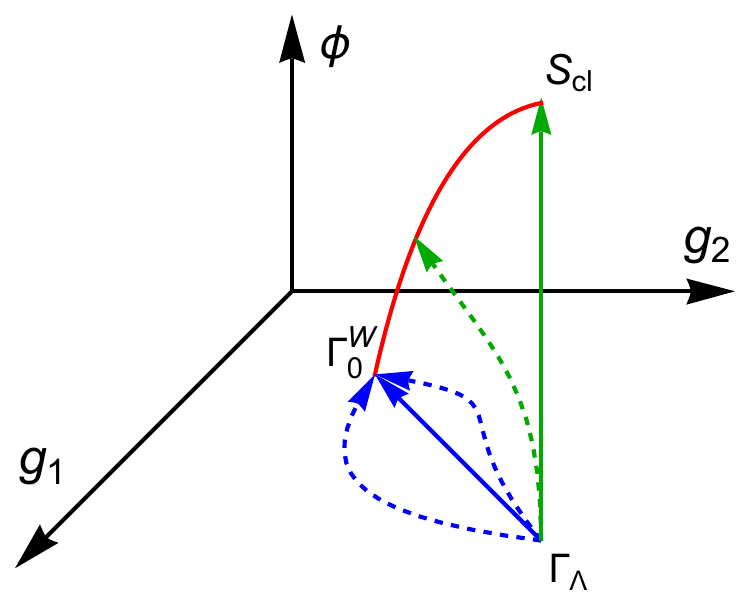}
		\subcaption{Representation in theory-space: The $\phi$-axis comprises axes for all expansion coefficients $f_i$ of the functional $\phi[\varphi]$. The effective action is given in terms of the expansion coefficients $g_i$. \hspace*{\fill}}
		\label{fig:TheorySpace} 
	\end{subfigure}%
	\hspace{0.02\linewidth}%
	\begin{subfigure}{.48\linewidth}
		\centering
		\includegraphics[height=0.5\linewidth]{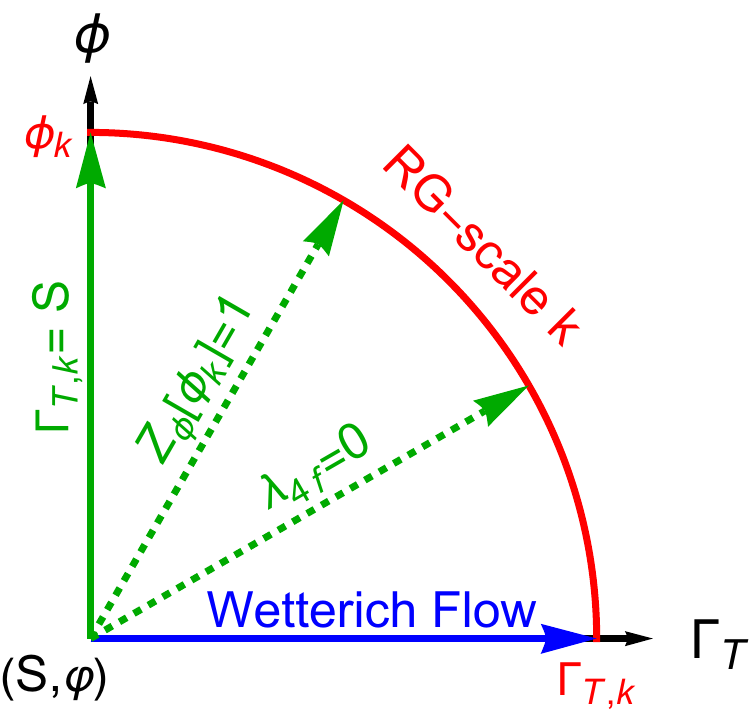}
		\subcaption{Representation in terms of pairs $(\Gamma_{T},\phi)$: The horizontal axis is that with $\phi=\varphi$, the vertical one is that with $\Gamma_\phi=S$.\vspace{4mm} \hspace*{\fill}}
		\label{fig:schem2} 
	\end{subfigure}%
	\caption{Schematic depiction of physics-informed RG flows: The flow is either captured in the (composite) fields $\phi$ or in the effective (target) action $\Gamma_{T}$. Starting from the initial conditions given by the classical action and fundamental field $(S=\Gamma_\Lambda, \varphi=\phi_\Lambda)$, the flow of $(\Gamma_{T,k}, \phi_k)$ is fixed at every RG-scale $k$ \textit{(red)} by the generalised flow equation \labelcref{eq:GenFlow} and an implicit condition for $\dot{\phi}$. Whilst all depicted flows are contained in \labelcref{eq:GenFlow}, we depict the Wetterich flow with $\dot{\phi} = 0$ in \textit{(blue)} and the non-trivial PIRG flows in \textit{(green)}. We point out the classical target action flow on the y-axis \textit{(green, no dashing)}, which can be viewed as 'orthogonal' to the Wetterich flow with $\dot \Gamma_{T} = 0$. \hspace*{\fill}} 
	\label{fig:schems}
\end{figure*}

Note also, that the practical use of \labelcref{eq:GenFlow} is facilitated by the occurrence of the differential local form of the transformation $\dot\phi[\phi]$, instead of the global transformation $\phi[\varphi]$. Hence, instead of the field $\phi[\varphi]$ in the pair \labelcref{eq:PI-Pair} we may also consider its differential form $\dot \phi[\phi]$ together with the RG-time derivative of the effective action, 
\begin{align} 
	\left(   \partial_t \Gamma_\phi[\phi] \,,\, \dot \phi[\phi]\right) \,. 
	\label{eq:PI-dPair}
\end{align}
Substituting also the effective action in the pair \labelcref{eq:PI-Pair} by its $t$-derivative in	\labelcref{eq:PI-dPair} is a matter of choice, we could also have kept $\Gamma_\phi[\phi]$.  

With this radical shift in perspective both entries in \labelcref{eq:PI-Pair,eq:PI-dPair} are on the same level. For example, in the same spirit as we have fixed $\dot\phi$ by some condition such as \labelcref{eq:lambda4f0,eq:Zphi1}, we can also fix the entire effective action $\Gamma_\phi$. Then, the generalised flow equation is rewritten to implement a given target action $\Gamma_\phi=\Gamma_{T}$. 
Instead of using the additional freedom of the flowing field to maximise the physics content of an expansion, the target action simply dictates an expression for $\Gamma_k[\phi]$ at all scales. Then physical information is no longer contained in $\Gamma_{\phi,k}[\phi]$, but in the pair $(\phi, \Gamma_T)$, which we call physics-informed (PI). 
The physics-informed RG (PIRG) setup opens the door to all kinds of simplification and optimisation procedures of the computational task at hand, and some aspects are illustrated in \Cref{fig:schems}. Finally we would like to highlight one important simplification, the so-called \textit{feed-down} flows. This concept is introduced and discussed at a later point in \Cref{sec:FeedDown}.

Importantly, the PIRG setup stores a significant amount of information in the definition of the field. This raises the issue of reconstructing the information from the infinitesimal transformation $\dot \phi$ which appears in the flow equation.
In \Cref{sec:Reconstruction} we specifically discuss the reconstruction of correlation functions of the fundamental field $\hat\varphi$.

\subsection{Target action flows} 
\label{sec:TargetAction}

As is clear from the previous examples, the generalised flow \labelcref{eq:GenFlow} can be used to eliminate single terms in the effective action, see \labelcref{eq:lambda4f0} and \labelcref{eq:Zphi1}. This procedure can be further generalised for fixing the full effective action: The transformation can be cast into the form 
\begin{align}
	\partial_t \Gamma_\phi[\phi] -\partial_t \ln {\cal N}_{\phi}\stackrel{!}{=}\partial_t \Gamma_T[\phi]\,,
	\label{eq:TargetG}
\end{align} 
where $\Gamma_T[\phi]$ is a cutoff-dependent target action that is at our disposal. We have absorbed the flow of the normalisation ${\cal N}_\phi$ for the sake of simplicity. However, we emphasise that the choice of ${\cal N}_\phi$ and hence the choice of the constant part of the target action will play a pivotal rôle in the computation of correlation functions and observables, see \Cref{sec:NoJ-Fund} for a detailed discussion. 

Using \labelcref{eq:TargetG} in \labelcref{eq:GenFlow}, we arrive at the target action representation of the generalised flow
\begin{align}\nonumber 
	\int_{\boldsymbol{x}} \dot{\phi}[\phi]\, \Gamma_T^{(1)}[\phi] - \Tr\Bigg[ G_T[\phi]\,  \dot{\phi}^{(1)}[\phi]\, R_{\phi}\Bigg]  & \\[2ex]
	&\hspace{-3.5cm} =\frac{1}{2} \Tr\Bigg[ G_T[\phi]\,\partial_t R_{\phi}\Bigg] - \partial_t\Gamma_T [\phi] \,, 
	\label{eq:GenFlowT} 
\end{align}
with the full propagator of the target action $G_T[\phi]$. In comparison to \labelcref{eq:GenFlow}, the target action flow \labelcref{eq:GenFlowT} has been rearranged such that it can be read as a differential equation for the flowing fields $\dot \phi[\phi]$. This facilitates the discussion of general properties and constraints of the flowing fields. Indeed, \labelcref{eq:GenG,eq:GenFlowT} already lead to a first restriction on, or rather a specification of the pair \labelcref{eq:PI-Pair} of, target space actions and corresponding fields $\phi$. The latter comes with a set ${\cal S}_\phi$ of allowed values of $\phi$ and one should consider the pair $\{\phi, {\cal S}_\phi\}$:  $\Gamma_T+\Delta S_k$ is a Legendre transform, and hence is convex. This entails that $\Gamma^{(2)}_T[\phi] + \Delta S^{(2)}_k[\phi]\geq 0$ for $\phi\in {\cal S}_\phi$. As in the Wetterich equation this convexity constraint is carried by the propagator $G[\phi]$ in \labelcref{eq:GenFlowT}, which developed a singularity if the convexity of the full two-point function including the regulator is lost. This can be formulated as 
\begin{align}\nonumber 
(\{\phi, {\cal S}_\phi\},\Gamma_T) &\\[1ex] 
& \hspace{-1.7cm}\in \Bigl\{ (\phi,\Gamma_\phi[\phi]) \,\Big|\, 	\Gamma^{(2)}_{\phi,k}[\phi]+R^{\ }_{\phi,k} \geq 0\quad  \forall \,\phi \in {\cal S}_\phi\Bigr\}\,.  
\label{eq:SetGT}
\end{align}
For example, for a real scalar composite field, the set ${\cal S}_\phi$ is a subset of the real numbers, $\phi\in\mathbbm{R}$. Then, for target actions with a convex effective action or rather $	\Gamma^{(2)}_{\phi,k}[\phi]+R^{\ }_{\phi,k} \geq 0$ for all values of the field, the set ${\cal S}_\phi$ is simply given by the real numbers, ${\cal S}_\phi=\mathbbm{R}$. In turn, for $	\Gamma^{(2)}_{\phi,k}[\phi]+R^{\ }_{\phi,k}< 0$ for $\phi\in {\cal D}_\phi$, at least this subset of the real numbers has to be excluded from ${\cal S}_\phi\subset \mathbbm{R} \backslash{\cal D}_\phi$ . 

Evidently, the set ${\cal S}_\phi$ is not necessarily the same as that for the fundamental fields, ${\cal S}_\varphi$. 
A well-known example for such a case is the effective scalar field used in the large N limit of scalar O(N) theories. The large N limit can be solved by the method of characteristics, which introduces composite fields via the backdoor. In this case, the full potential, or rather its first derivative is given by the UV potential, the argument being the composite field. The values of the latter are restricted to values larger than the solution of the EoM. This analogy is explored further in \Cref{app:LargeN}.

For the two examples with \labelcref{eq:lambda4f0,eq:Zphi1}, the target action $\Gamma_T$ is the full RG-scale-dependent action with the exception of one term, which is trivialised. Either it is put to zero, \labelcref{eq:lambda4f0}, or it is kept classical or rather on its UV value, \labelcref{eq:Zphi1}. Due to the convexity restoring property of the flow, see e.g.~\cite{Litim:2006nn}, \labelcref{eq:SetGT} is satisfied. We emphasise that this dynamical convexity restoration requires a dynamical target action. 

Opposed to that, let us now concentrate on the interesting subset of target actions which are fixed completely. Then, instead of dynamical convexity restoration of the effective action, the set \labelcref{eq:SetGT} is evolving dynamically. This subset includes extreme cases, whose evaluation will help us gain a comprehensive understanding of the power of the setup. For these cases, $\dot\phi$ and its $\phi$-derivative $\dot \phi^{(1)}$ are the only functionals that are determined in the flow and they are not fed back. Then, \labelcref{eq:GenFlowT} constitutes an ordinary functional differential equation for the functional $\dot\phi[\phi]$ at each $k$. 

In summary, the task of solving the flow factorises into one of solving the ODE for $\dot \phi[\phi]$ at each $k$, and then integrating the flow. Accordingly, we have stored all physics of the theory in the pair 
\begin{align}
\bigl(\Gamma_T \,,\, \dot\phi[\phi]\bigr)\,, \qquad \textrm{with}\qquad \dot\phi[\phi](\Gamma_T)\,, 
\label{eq:GT+dotphi} 
\end{align}
and the flow of the transformation is solely determined by $\Gamma_T$. An example for \labelcref{eq:GT+dotphi} is provided by the vertical line in \Cref{fig:schem2} with the classical target action. We call the combined setup \labelcref{eq:GenFlowT,eq:GT+dotphi} the physics-informed RG (PIRG). We emphasise that also the previous setups with \labelcref{eq:lambda4f0,eq:Zphi1} are partially PI. 

The full transformation from the ultraviolet field $\phi_\Lambda$ to the composite field $\phi[\phi_\Lambda]$ is obtained by integrating $\dot \phi$ from the initial scale to the final scale, 
\begin{align}
	\phi[\phi_\Lambda] = \phi_\Lambda- \int^\Lambda_0 \frac{d k}{k} \, \dot\phi[\phi]\,,
\label{eq:reconstruction}
\end{align} 
which we discuss further in \Cref{sec:Reconstruction}.
We emphasise that within this fully factorised form the task of solving the original flow equation, which is the task of solving a functional PDE, is reduced to simply integrating the solution of a functional ODE. Evidently, this is a qualitative reduction of the numerical complexity. 
We can again pick up the example of the large N limit in O(N) theories, which is solved (in LPA) with the method of characteristics: Roughly speaking, the PIRG allows us to transfer the method of characteristics to general theories. This comes at the price of some additional subtleties that will be discussed in \Cref{sec:Reconstruction}. The method of characteristics translates into picking the classical action as target action, which we consider in \Cref{sec:ClassicalTarget-ON}.

\subsection{Functional optimisation of physics-informed renormalisation group flows} 
\label{sec:FunOptPI}

The remaining problem is that of \textit{finding} a suitable target action that allows the integration of the coordinate transformation. For example, PIRG flows can be applied to a given solution of the original flow equation and the non-trivial field transformation can be used to upgrade the solution. Moreover, we can also do this in parallel for a set of target actions and try to optimise the physics content in a given approximation. This extends the setup of functional optimisation in \cite{Pawlowski:2005xe} to a combined optimisation of the regulator function and the flowing fields. 

In this Section we only lay the foundation of generalised functional optimisation, further considerations and applications of functional optimisation in PIRG flows will be considered elsewhere. In particular, we have completed the PIRG approach with general operators flows in \cite{Ihssen:2025hyl}. Here we just briefly recapitulate functional optimisation based on operator flows as introduced in \cite{Pawlowski:2005xe} and extended in \cite{Pawlowski:2015mlf}, and discuss the required generalisation. Functional optimisation is based on the minimisation of RG-trajectories. While this is a global criterion, in most practical applications local variants are used, partly because of the difficulty to define proper distances in the space of theories. The latter can be derived from the Fisher information metric, but its practical implementation is intricate. Still, minimising the RG-trajectory entails that at a given \textit{physical} cutoff scale the remaining flow is minimal. In absence of flowing fields the flow operator has the representation 
\begin{align}
	\partial_t = -\frac12 \Tr \,\left[  G_k[\phi] \partial_t R_{\phi} \,  G_k[\phi] \frac{\delta^2}{\delta \phi^2}\right] \,, 
	\label{eq:dtFlow}
\end{align}
on operators ${\cal O}_k$ in the set of full correlation functions as well as functions of the source term,  
\begin{align} 
	{\cal O}[\phi,R_{\phi}] \in \left\{\hat{\cal O}\left[J,\frac{\delta}{\delta J}\right] Z_k[J]\right\}\,, 
	\label{eq:setofO}
\end{align}
where $\hat{\cal O}[x,y]$ is a general function of two variables, for more details see \cite{Pawlowski:2005xe}. For a specific class of flowing fields, \labelcref{eq:dtFlow} admits a similar structure, see \cite{Ihssen:2025hyl}. Therefore, the arguments below are readily extended to PIRGs.
This representation can be readily obtained by a $\phi$-derivative of the flow of the effective action: The current itself is part of the set of operators $\{\cal O\}$, while the effective action is not. 

This criterion specifically entails that the remaining flow of the propagator is minimal. Moreover, optimised flows minimise the violation of the derivative property of the right-hand side of \labelcref{eq:dtFlow}
\begin{align}
	\left|\oint \partial_t  {\cal O}[\phi,R^\textrm{opt}_k] \right| = \textrm{minimal} \,, 
\end{align}
see \cite{Pawlowski:2005xe, Pawlowski:2015mlf}. This entails that for identical \textit{physical} cutoffs the propagator with the optimised regulator is already as close as possible to the full propagator $G[\phi]=G[\phi,0]$, 
\begin{align}
	\bigl\|G[\phi,0]- G[\phi,R^\textrm{opt}_k] \bigr \| \leq \bigl\|G[\phi]- G[\phi,R_{\phi}]\bigr\|\,, 
\end{align}
for all hyper surfaces $R_{\phi}\in \{ R_\bot\}$ with the same physics. The latter condition is an intricate one and is related to the definition of a metric on the space of theories. A practical workaround is the definition of this hyper surface via the spectrum of the theory with a given cutoff. In short, we define $R_{\phi}\in \{ R_\bot\}$ with 
\begin{subequations}
	\label{eq:Rbot}
	\begin{align} 
		\{ R_\bot\} = \left\{R_{\phi} \, \Bigl| \, k^2_\textrm{phys} =\min_{\textrm{Re}\left(k_\textrm{pol}\right)}\{\left(\textrm{Re}\left(k_\textrm{pol}\right)\right)^2\}  \right\}\,,
		\label{eq:HypersurfacePhys}
	\end{align}
	with $k_\textrm{pole}$ are the positions of the zeros / cuts in the complex plane, 
	\begin{align} 
		\{k_\textrm{pol}\}  \ =  \left\{k\, \Bigl|\, \Gamma^{(2)}[\phi,R_{\phi}](-k^2_\textrm{pole})  = 0  \right\}\,.
		\label{eq:0-CP}
	\end{align}
\end{subequations} 
In the presence of flowing fields the $t$-derivative \labelcref{eq:dtFlow} is only a partial one, the total $t$-derivative also involves terms proportional to $\dot\phi$ as well as its derivatives. These terms can be readily derived from \labelcref{eq:GenFlow} by one further $\phi$-derivative even though this is not the most concise representation. In any case, we have to deal with a different total $t$-derivative that acts on the combined space $(\Gamma_T,\phi)$. Trajectories have to be minimised in this space and neither $\dot \phi=0$ nor $\partial_t\Gamma_T$ will be minimal. While this offers very exciting possibilities, a full discussion goes far beyond the scope of the present work, and more details will be presented elsewhere.

\subsection{Flowing composites and observables} 
\label{sec:Reconstruction} 

Finally, we address the computation of correlation functions of the fundamental field and observables within the PIRG setup. This issue is essential in the target action formulation, but is also relevant for the physically motivated examples with one or several fixed vertices from \Cref{sec:GenFlow}:

Since the composite field $\hat \phi$ is introduced via the flow of the transformation \labelcref{eq:dotphi}, it carries a genuine regulator dependence
\begin{align}
	\hat\phi=\hat\phi[\hat \varphi, R]\,,\qquad \textrm{and}\qquad 	\phi=\phi(\varphi, R)\,,
\end{align} 
where the argument $R$ expresses the fact that there is an integrated dependence on the regulator, see also \cite{Pawlowski:2005xe}. 
Consequently, the generating functional $Z[J_\phi]$ and the corresponding effective action $\Gamma[\phi]$ defined in \labelcref{eq:ZGenComposites} and \labelcref{eq:GenG} respectively, depend on the chosen regulator path $R$. 
One readily concludes, that the effective action $\Gamma[\phi]$ is not the standard effective action, while it still comprises the full physics of the given theory. We note in passing that this statement is even true for composite operators that do not depend on $R$. In any case this begs the question of how we can compute observables built from correlation functions such as scattering amplitudes, as well as characteristic scales such as critical temperatures or resonance spectra within the present framework. 

We need to distinguish flowing field setups that allow a direct access to correlation functions of the fundamental fields in terms of field derivatives of the effective action and those that do not. In the latter setups the fundamental field is effectively integrated out. 
For example, both \labelcref{eq:lambda4f0} and \labelcref{eq:Zphi1} do not require it, since observables are directly related to the composite field $\phi$. The definition of the flowing field is such that $\phi$ is a better representation of the physical degree of freedom than the initial fundamental (mean) field of the derivative expansion $\varphi = \langle \hat \varphi \rangle$. 

In the following, the perspective is shifted away from \textit{physically motivated} definitions of the composite. In these cases reconstruction schemes of correlation functions and observables become vital. We emphasise that the reconstruction schemes we discuss in this work are direct computational schemes, while in a further work we shall also discuss machine-learning based reconstruction schemes that face conceptual intricacies such as spectral reconstructions. In the present work we discuss computational reconstruction schemes for two qualitatively different situations: In the first one, discussed in \Cref{sec:J-Fund}, we still keep the fundamental field, coupled to its own current. In \Cref{sec:NoJ-Fund} we discuss the setup, where the fundamental field evolves into a composite one.

\subsubsection{Effective action with fundamental and composite fields}
\label{sec:J-Fund} 

We consider the general formulation with a composite field $\phi_a$, \labelcref{eq:GenComposite}, where the index $a$ runs over different (composite) field species, which also include the fundamental field with 
\begin{align}
	\phi_1(x)= \varphi(x)\,, \qquad 
 \label{eq:multiplefields}
 \end{align}
implying $	\boldsymbol{x}_1= x$ in \labelcref{eq:DefSuperindex}. For example, this is the case in QCD, where one keeps the quarks but also introduces composite fields for mesonic interaction channels, for recent discussions and progress see \cite{Fu:2019hdw, Fukushima:2021ctq} and the reviews \cite{Dupuis:2020fhh, Fu:2022gou}. 
This allows us to readily reconstruct the effective action of the fundamental field, 
\begin{align} 
	\Gamma_\varphi[\varphi] = \Gamma_\phi\bigl[\varphi, \Delta\phi_\textrm{\tiny{EoM}}[\varphi]\bigr]\,, \quad \Delta \phi =(\phi_2,..., \phi_n)\,,  
\end{align}
and the $\phi^a_\textrm{\tiny{EoM}}$ are the solutions of the equation of motion (EoM) for $\phi^a$,
\begin{align} 
	\phi^a_\textrm{\tiny{EoM}}: \quad  \left. \frac{\delta \Gamma_\phi}{\delta \phi_a}\right|_{\phi^a =\phi^a_\textrm{\tiny{EoM}}}=0\,.
	\label{eq:EoMphia}
\end{align}
Note, that we have taken the Legendre transform with respect to all fields $\phi = (\varphi,\Delta \phi)$. 
\Cref{eq:EoMphia,eq:GenG} entails $J^a_\phi=0$ and the path integral \labelcref{eq:ZGenComposites} reduces to that with a source to the fundamental field, as does the Legendre transform \labelcref{eq:GenG}, 
\begin{align}
	\Gamma_\phi\bigl[\varphi,  \Delta\phi_\textrm{\tiny{EoM}}[\varphi]\bigr]= \sup_{J_\varphi}\bigl(\int_x J_\varphi \,\varphi - \ln Z_\varphi[J_\varphi]\bigr)\,, 
	\label{eq:Recon1} 
\end{align}
with $Z_\varphi[J_\varphi]=Z_\phi[J_\varphi,J_{\Delta \phi}=0]$.
\Cref{eq:Recon1} also provides a relation between the correlation functions of the fundamental field and that of the composite field as well as $\phi_a^\textrm{EoM}[\varphi]$. We elucidate this at the example of the inverse propagator of the fundamental field, 
\begin{align}\nonumber 
	\Gamma_\varphi^{(2)} [\varphi](x,y) =&\, \left. \frac{\delta^2 \Gamma_\phi[\varphi,\Delta\phi]}{ \delta \varphi(x) \delta\varphi(y)} \right|_{\Delta\phi= \Delta\phi_\textrm{\tiny{EoM}}}\\[1ex] 
	&\hspace{-2cm} 	- \int\limits_{\boldsymbol{x}_a,\boldsymbol{x}_b} \Gamma_{\phi,ab}[\varphi, \Delta\phi_\textrm{\tiny{EoM}}] \frac{\delta \Delta \phi_\textrm{\tiny{EoM}}^a(\boldsymbol{x}_a)}{\delta \varphi(x)} \frac{\delta\Delta\phi_\textrm{\tiny{EoM}}^b(\boldsymbol{x}_b)}{\delta \varphi(y)} \,, 
	\label{eq:Gvarphi-Gphi}
\end{align}
where 
\begin{align} 
	\Gamma_{\phi,a_1\cdots a_n}[\phi] = \frac{\delta^n \Gamma_\phi[\phi]}{\delta\phi^{a_1} \cdots \delta \phi^{a_n}}\,.
	\label{eq:Gan}
\end{align}
In the following we denote an evaluation on the equations of motion with $\vert_{\textrm{\tiny{EoM}}}$ or omit it for the sake of readability. Note that all terms with factors 
\begin{align} 
	\frac{\delta^n } {\delta \varphi^n }\left[\left. \frac{\delta\Gamma_\phi}{\delta\Delta \phi_a}\bigl[\varphi, \Delta\phi[\varphi]\bigr] \right|_{\textrm{\tiny{EoM}}} f[\varphi] \right] \,,
	\label{eq:EoM0}
\end{align}
vanish identically due to \labelcref{eq:EoMphia}: The expression in the square bracket in \labelcref{eq:EoM0} vanishes identically. In \labelcref{eq:Gvarphi-Gphi} this removes the terms 
\begin{align} 
	\frac{\delta}{\delta\varphi} \left[ \frac{\delta\Gamma_\phi}{\delta\Delta \phi_a}\right]\,, 
	\label{eq:RemoveEoM} 
\end{align} 
but this structure holds true for general correlation functions. More explicitly we obtain 
\begin{align} \nonumber 
	\frac{ \delta^2 \Gamma_\phi\left[\varphi,  \Delta\phi_\textrm{\tiny{EoM}}[\varphi]\right]}{\delta\varphi^2} = &\, \frac{\delta}{\delta\varphi } \left[ \frac{ \delta \Gamma_\phi[\varphi,\Delta\phi]}{\delta\varphi}+ \frac{ \delta \Gamma_\phi}{\delta\Delta \phi^a}  \frac{\delta\Delta\phi^a }{\delta\varphi} \right] \\[1ex] 
	& \hspace{-3.5cm} = \frac{\delta}{\delta\varphi } \frac{ \delta \Gamma_\phi[\varphi,\Delta\phi]}{\delta\varphi} =  \frac{ \delta^2 \Gamma_\phi[\varphi,\Delta\phi]}{\delta\varphi^2}+ \frac{\delta\Delta\phi^a }{\delta\varphi}  \frac{ \delta^2 \Gamma_\phi}{\delta\Delta \phi^a\delta\varphi}  \,.
\end{align}
For the last term we use 
\begin{align}\nonumber 
	\left. \frac{ \delta^2 \Gamma_\phi}{\delta\Delta \phi^a\delta\varphi} \right|_{\Delta\phi= \Delta\phi_\textrm{\tiny{EoM}}} & \\[1ex] \nonumber 
	&\hspace{-2.5cm} =\frac{\delta}{\delta \varphi}  \left[ \frac{ \delta \Gamma_\phi}{\delta\Delta \phi^a}[\varphi,\Delta\phi_\textrm{\tiny{EoM}}]\right]-  \frac{\delta\Delta\phi^b}{\delta\varphi} \frac{ \delta^2 \Gamma_\phi[\varphi,\Delta\phi]}{\delta\Delta \phi^b \delta\Delta \phi^a}\\[1ex] 
	&\hspace{-2.5cm} = -  \frac{\delta\Delta\phi^b}{\delta\varphi} \frac{ \delta^2 \Gamma_\phi[\varphi,\Delta\phi]}{\delta\Delta \phi^b \delta\Delta \phi^a} \,. 
\end{align} 
This concludes our discussion of the direct reconstruction of the full effective action. 

This approach with emergent composites is reminiscent of nPI approaches and density functional theory and indeed includes both as specific examples. So far, the composite field approach has been mostly used in the spirit of density functional theory, in particular in QCD, \cite{Gies:2002hq, Braun:2009ewx, Mitter:2014wpa, Braun:2014ata, Cyrol:2017ewj, Fu:2019hdw, Fukushima:2021ctq}. Roughly speaking, a current is coupled to the scalar bilinear of the quarks, $J_\sigma(x) \bar q(x) q(x)$, and further bilinear invariants, a prominent example being the scalar (and pseudoscalar) current as well as the quark density, $\bar q(x) \gamma_0 q(x)$. 
Evidently, this does not incorporate the full fermionic two-point function $\bar q(x) q(y)$ but only its diagonal (local) part. This is sometimes also called \textit{two-point particle irreducible} (2PPI). Since also the quark is retained as a dynamical field, coupled to its own source, all correlation functions of the fundamental fields in QCD can be reconstructed, for a detailed discussion see \cite{Fu:2019hdw}.

\subsubsection{Effective action without fundamental fields and cumulants-preserving flows}
\label{sec:NoJ-Fund}

A relevant example for an effective action approach without the fundamental fields is density functions theory. For example, we can consider a formulation in QCD where a source is coupled to the quark density, $J_n \bar q(x)\gamma^0 q(x)$. This introduces the density field $n(x)$ in the Legendre transform with 
\begin{align}
	\phi=(q,\bar q, n)\,.	
\label{eq:phiDensity}
\end{align} 
In \labelcref{eq:phiDensity} we suppressed the glue sector for the sake of simplicity. Keeping the quark fields leaves us with an effective action $\Gamma_\phi[\phi]$. On the equations of motion of the density field, this action reduces to the effective action of the fundamental theory, $\Gamma_\textrm{fund}[q,\bar q] = \Gamma_\varphi[q,\bar q]$ with $\varphi=(q,\bar q)$, 
\begin{align}
	\Gamma_\textrm{fund}[q,\bar q] = 	\Gamma_\phi\bigl[q,\bar q, n^{\ }_\textrm{EoM}[q,\bar q]\bigr]\,,
	\label{eq:GFullTheory}
\end{align}
and the correlation functions of the quark in the fundamental theory are obtained by taking total $q,\bar q$-derivatives, also including that of $n_\textrm{EoM}[q,\bar q]$. In turn, integrating out the quarks leads to an effective action of the density $n$, that is $\Gamma_n[n]$. This is done by switching off the source term for the quarks in the path integral, that is solving the equations of motion for the quark 
\begin{align}
	\Gamma_n[n] = 	\Gamma_\phi\bigl[q^{\ }_\textrm{EoM}[n],\bar q^{\ }_\textrm{EoM}[n],n\bigr]=\Gamma_\phi\bigl[0,0,n]\,.
	\label{eq:GFullTheoryDensity}
\end{align}
\begin{figure*}
	\centering
	\begin{subfigure}{.48\linewidth}
		\centering
		\includegraphics[width=\textwidth]{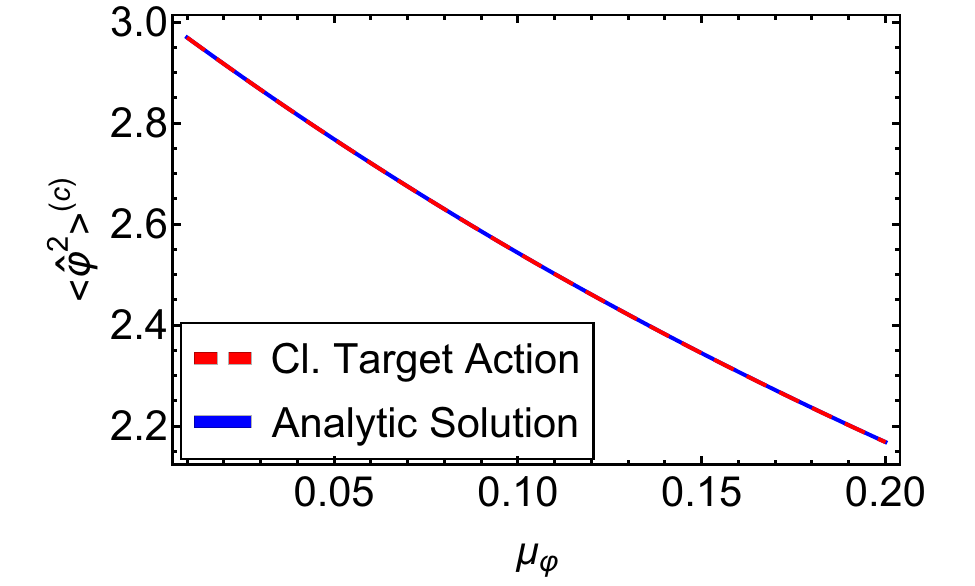}
		\subcaption{Two-point function.\hspace*{\fill}}
		\label{fig:d02pt}
	\end{subfigure}%
	\hspace{0.02\linewidth}%
	\begin{subfigure}{.48\linewidth}
		\centering
		\includegraphics[width=\textwidth]{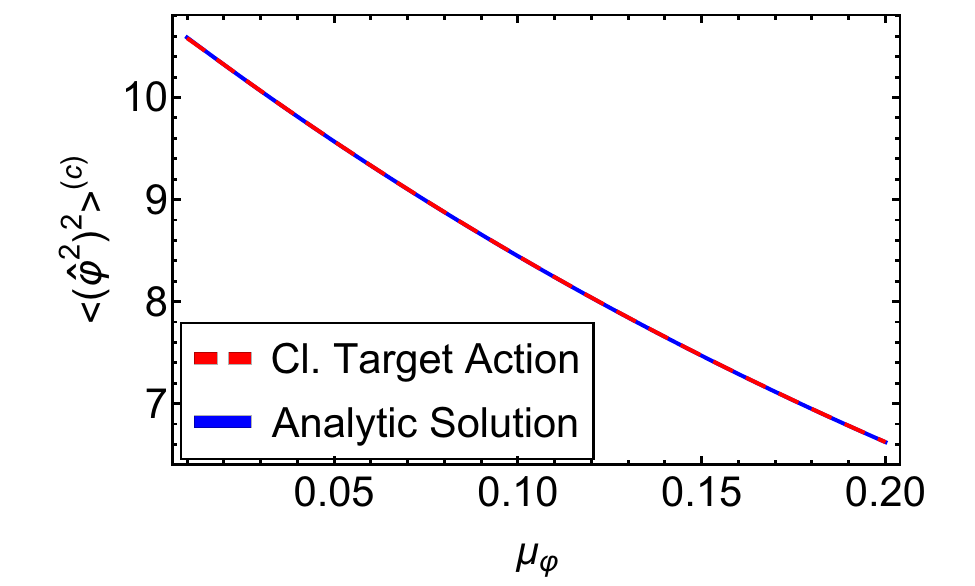}
		\subcaption{Four-point function.\hspace*{\fill}}
		\label{fig:d04pt}
	\end{subfigure}%
	\caption{Reconstruction \textit{cumulants} using \labelcref{eq:varphi2Correlations} in a zero-dimensional test-case, see \Cref{app:ClassicalTargetd0}. The coordinate map $\dot \phi$ implements a classical target action flow, which is discussed in \Cref{sec:ClassicalTarget-ON}. The solution can be computed by either taking $\mu_\varphi$-derivatives of the cumulants-preserving generating functional $Z^{(c)}(0)$ or by $J_\varphi$-derivatives of $Z_\varphi(J_\varphi)$, for details see \Cref{app:ClassicalTargetd0}. \hspace*{\fill}} 
	\label{fig:d0correlations}
\end{figure*}
This is a variant of density functional theory and in contradistinction to the approach discussed above, the correlation functions of the fundamental quark field cannot be reconstructed. Still, the respective effective action is one in the fundamental theory and not that of a model, even though it only allows us to compute the correlation functions of the density. In QCD, these correlations, or more generally, fluctuations of conserved charges are measured in heavy ion experiments and are chiefly important for the resolution of the QCD phase structure at finite density; for recent work within the fRG approach with flowing fields using \labelcref{eq:GFullTheory}, see \cite{Fu:2023lcm}. 

\Cref{eq:GFullTheoryDensity} is but one example of using the composite field as the only degree of freedom. It can be readily used, since the composite field has a direct physical interpretation. In the following we discuss the composite field approach, that retains the full information of all correlation functions, but does not introduce additional degrees of freedom. For that purpose we define a one component field
\begin{align} 
	\phi=\phi_1\,,\qquad \textrm{with}\qquad \phi_\Lambda=\varphi\,.
\end{align}
This leaves us with an effective action of the composite field $\phi$, which carries the full information about the underlying fundamental field $\varphi$ via the initial condition. We are left with the pair 
\begin{align} 
	\Gamma_\phi[\phi]\,,\qquad \phi= \phi_\Lambda - \int^\Lambda_k \frac{d k'}{k'}  \dot \phi[\phi_{k'}]\,. 
	\label{eq:G+CompositeField} 
\end{align} 
and since the field $\phi$ at the initial scale $\Lambda$ is the fundamental field $\varphi$, \labelcref{eq:Recon1} implicitly defines the map from $\varphi \to \phi$ with 
\begin{align} 
	\phi[\varphi]= \varphi - \int^\Lambda_0 \frac{d k'}{k'}  \dot \phi[\phi_{k'}]\,. 
	\label{eq:phi-varphi}
\end{align} 
This allows us to define an effective action of the fundamental field with 
\begin{align}
	\tilde \Gamma_{\phi}[\varphi] = \Gamma_\phi\bigl[\phi[\varphi]\bigr]\,.
	\label{eq:Gammaphi}
\end{align}
Note that \labelcref{eq:phi-varphi} provides a map from the fundamental field to the composite one, but this knowledge does not suffice to compute correlation functions of the fundamental fields 
\begin{align}
	\langle \hat\varphi(x_1)\cdots \hat \varphi(x_n)\rangle_\textrm{1PI} \,.
	\label{eq:Fund1PICorrelations}
\end{align}
However, the latter can be obtained from the flow equation for general operators, also put forward in \cite{Pawlowski:2005xe}. Moreover, many relevant observables such as the density or its correlations can be obtained by taking derivatives of the effective action $\Gamma_\phi$ or the generating functional $Z_\phi$ with respect to the classical coupling parameters $\boldsymbol{g}$ of the theory at hand with 
\begin{align}
\boldsymbol{g} = (g_1,...,g_m)\,. 
\label{eq:CouplingParameters}
\end{align}
For our example scalar theory with the classical action \labelcref{eq:ClassicalAction1-ON} this reads  $\boldsymbol{g}=(\mu_\varphi,\lambda_\varphi)$. 

We exemplify this with the present scalar theory and derivatives with respect to the mass of the fundamental field. To begin with, we can use the logarithm of the generating functional for generating cumulants of the operators in the classical action \labelcref{eq:ClassicalAction1-ON}. For example, the cumulants of $\varphi(x)^2$ are obtained by 
\begin{align}
	\Bigl\langle \prod_{i=1}^n\, \int\limits_{x_i}\hat \varphi^2(x_i)\Bigr\rangle^{(c)} =&\, (-2)^n \frac{d^n \log Z_\phi^{(c)}[0]}{d (\mu_\varphi)^n}  \,,  
	\label{eq:carphi2Cumulants}
\end{align}
where the superscript in the expectation value, $\langle \cdot \rangle^{(c)}$ stands for cumulants of $\int_x \varphi^2(x)$, and the same superscript in the generating functional, $Z^{(c)}_\phi$ stands for the cumulants-preserving subclass of generating functionals: For \labelcref{eq:carphi2Cumulants} to hold, the normalisation ${\cal N}^{(c)}$ of $Z_\phi^{(c)}$ in \labelcref{eq:Zphi} cannot depend on any of the parameters of the classical action of the theory at hand. For our example of the $\phi^4$-theory with the action 	\labelcref{eq:ClassicalAction1-ON} this entails, 
\begin{align} 
	 \frac{d^n {\cal N}^{(c)}}{d (\mu_\varphi)^n} =0= \frac{d^n {\cal N}^{(c)}}{d (\lambda_\varphi)^n}\,,\qquad \forall n\,. 
\label{eq:Nk-muIndep}	
\end{align}
We emphasise that \labelcref{eq:Nk-muIndep} excludes the common choice of the normalisation as the path integral without the current term. This choice leads to $Z_\phi[0]=1$ and normalised expectation values of $Z_\phi$, but evidently it is not cumulants-preserving. This entails in turn, that cumulants-preserving generating functionals are \textit{not} normalised, $Z^{(c)}_\phi[0]\neq 1$. Note also in this context that 
\begin{align} 
	Z_\phi[0]= Z_\varphi[0]\,,
	\label{eq:ZEoM-ID}
\end{align}
for the same normalisation, and a given cumulants-preserving normalisation can be used in all generating functionals $Z_\phi$. We elaborate on this in \Cref{app:c1fixing}. 

The first two cumulants of $\hat \varphi^2(x)$ are related to the variance $\sigma$ and kurtosis $\kappa$ (non-Gau\ss ianity) of the field $\hat \varphi$. The variance 
is obtained via 
\begin{align}
	\sigma = &\,	\Bigl\langle  \int\limits_{x}\hat \varphi^2(x)\Bigr\rangle-\int_x \varphi^2(x)\stackrel{\varphi(x)=0}{=} -2 \frac{d \log Z^{(c)}[0]}{d \mu_\varphi}\,,
	\label{eq:sigma}
\end{align}
where we have assumed $\varphi(x)=0$ in the second step. For $\varphi=0$, the kurtosis is given as 
\begin{align}
	\kappa  = &\,	\frac{1}{\sigma^2}\Bigl\langle \prod_{i=1}^2\, \int\limits_{x_i}\hat \varphi^2(x_i)\Bigr\rangle - 1 = \frac{4}{\sigma^2} \frac{ d^2 \log Z^{(c)}[0]}{d (\mu_\varphi)^2}\,. 
	\label{eq:kappa}
\end{align}
Both relations can be extended to $\varphi(x)\neq 0$, but we would like to keep these examples simple.  

The logarithm of the generating function in the absence of currents, $J_\phi=0$, is related to the effective action, evaluated on a solution of the equation of motion, 
\begin{align} 
	\log Z[0] = -  \Gamma_\phi[\phi_\textrm{EoM}]\,.  
\label{eq:preZ-Gammacumulants}
\end{align}
\Cref{eq:preZ-Gammacumulants} implies that the choice of the constant part of $\Gamma_\phi$ is essential for deriving cumulants from the effective action, evaluated on a solution of the EoM. For this task we have to restrict ourselves to cumulants-preserving effective actions $\Gamma_\phi^{(c)}$ with 
\begin{align} 
	\log Z^{(c)}[0] = -  \Gamma^{(c)}_\phi[\phi_\textrm{EoM},\dot {\cal C}]\,.  
	\label{eq:Z-Gammacumulants}
\end{align}
\Cref{eq:Z-Gammacumulants} maps the cumulants-preserving property \labelcref{eq:Nk-muIndep} for the normalisation of the generating functional $Z^{(c)}$ into conditions for the constant part of the effective action. In the PIRG flow \labelcref{eq:GenFlowT}, this constant is part of the target action $\Gamma_T$, see \labelcref{eq:TargetG} and we write 
\begin{align}
	\Gamma_T=\Gamma_T[\phi,\dot {\cal C}]\,, \quad \textrm{with} \quad  \partial_t \Gamma_T[\phi,\dot {\cal C}]= \partial_t \Gamma_T[\phi,0]+\dot {\cal C}\,.
	\label{eq:GTC}
\end{align} 
The term $\dot {\cal C}$ accommodates the flow of a non-trivial normalisation as well as other more physical parts. While it is introduced as a constant part of the flow, it also induces additional $\phi$-dependent terms in the transformation $\dot\phi[\phi]$. Its pivotal rôle for both, adjusting cumulants-preserving flows and simplifying field transformations, is discussed in \Cref{app:c1fixing,app:ClassicalTargetd0}.

Then, the integrated $n$th order or cumulants of the mass operator $\hat \varphi^2(x)$ and the $\phi^4$-interaction operator $\hat \varphi^4(x)$ in \labelcref{eq:ClassicalAction1-ON}, including mixed ones, can be obtained by $n$th order $-\mu_\varphi$ (or $-\lambda_\varphi$) derivatives of $- \Gamma[\phi^{\ }_\textrm{EoM}]$. For the present example of cumulants of $\hat \varphi^2(x)$ we obtain 
\begin{align}
	\Bigl\langle \prod_{i=1}^n\, \int\limits_{x_i}\hat \varphi^2(x_i)\Bigr\rangle^{(c)} =&\, -(-2)^n \frac{d^n  \Gamma^{(c)} _\phi[\phi_\textrm{EoM}]}{d (\mu_\varphi)^n} \,. 
	\label{eq:varphi2Correlations}
\end{align}
This computational approach for relevant observables, and specifically \labelcref{eq:varphi2Correlations}, is illustrated and tested comprehensively with the extreme choice of a classical target action in zero dimensions in \Cref{app:ClassicalTargetd0}. In zero dimensions, the flows provide exact results which can be validated by analytically computing the generating function: The main result of \Cref{app:ClassicalTargetd0} is showcased in \Cref{fig:d0correlations}, which shows the successfully reconstructed first and second cumulants calculated using \labelcref{eq:varphi2Correlations}. Higher-dimensional applications with a classical target action are also discussed in \Cref{sec:ClassicalTarget-ON}. 

In summary, the reconstruction of correlation functions in terms of $\varphi$ from the effective action of the composite fields $\phi$ requires full knowledge of $\phi[\varphi]$. Moreover, we emphasise that \labelcref{eq:varphi2Correlations} does not imply that the effective action of the composite field and that of the fundamental field coincide in general or even for cumulants-preserving effective actions, 
\begin{align}
	\tilde \Gamma_\phi[\varphi] \neq \Gamma_\varphi[\varphi] \,,
\label{eq:gammaNotGamma}
\end{align}
with $\tilde \Gamma_\phi$ defined in \labelcref{eq:Gammaphi}. Still, we can readily reconstruct specific correlation functions as well as physics properties such as the phase structure and critical scaling.

\section{Feed-down flows}
\label{sec:FeedDown}

A further intriguing option for PIRG flows is the possibility to break the feed-back loop in the flow equation: Roughly speaking, generic systematic expansion schemes centre around an expansion of the effective action, accompanied by potential further truncations on the right-hand side of the flow equation. All of these schemes have in common, that the low order results  \textit{feed down} into the additional relations that govern the higher orders.  Most of the schemes have in common, that higher orders terms in the given expansion \textit{feed back} into the flows of lower order terms. Notable exceptions are perturbation theory and exact $n$-body hierarchies in non-relativistic systems such as \cite{Diehl:2007xz, Floerchinger:2013bty, Tanizaki:2013uma} (Skorniakov, Ter-Martirosian, (STM)), see also \cite{Dupuis:2020fhh}. Such feed-back loops potentially destabilise the lower orders and slow down the (apparent) convergence of systematic expansion schemes. An illustrative example with computational evidence has been presented in \cite{Pawlowski:2014zaa}, where different expansion points in the Taylor expansion of the effective potential enhanced or minimised the feed-back loop of the higher order couplings to the flow of the lower order ones: Fastest convergence was obtained with minimal feedback. Indeed, for some cases, only the minimal feedback formulation showed convergence at all. 

Hence, the question arises whether one can use the flexibility of PIRG flows to break the feed-back loop inherent in the standard flow equation by means of flowing fields or rather an appropriate choice of the target action. We shall see below that this is indeed possible and we call the respective flows \textit{feed-down flows}. 
For a full discussion of the feed-back structure of expansion schemes we have to discuss pairs of target actions $\Gamma_T[\phi]$ and flowing fields $\dot\phi[\phi]$, and the expansion scheme has to be applied to both, leading to 
\begin{align} 
	( \Gamma_T,\dot \phi)_n \,,
	\label{eq:PairGT+dotphi}
\end{align} 
where $n$ labels the order of the expansion scheme. Schematically we write 
\begin{align}
	\Gamma_{T,n}= \sum_{i=0,j_i}^n \omega^{\ }_{\Gamma_T, j_i} {\cal O}_{j_i}\,,\quad  \dot \phi_n  =  
	\sum_{i=0,m_i}^n \dot \omega_{\dot\phi,m_i} {\cal I}_{m_i}\,.
	\label{eq:PairExpansion}
\end{align}
where the operator sets $ \{ {\cal O}_{m_n}[\phi]\},\,  \{ {\cal I}_{m_n}[\phi]\}$ are bases of the operators in the $n$th order $( \Gamma_T,\dot \phi)_n$ of the given expansion scheme. With \labelcref{eq:PairExpansion} we define the feed-down condition for flows
\begin{align} 
	\frac{\delta  \dot \omega_{j_n}  } {\delta \omega_{j_m}}=0=	\frac{\delta  \dot \omega_{j_n}  } {\delta \dot \omega_{j_m}}\,,\qquad \forall m > n \,,
	\label{eq:FeedDownCondition}
\end{align}
where $\dot \omega$ stands for the flows of $\omega^{\ }_{\Gamma_T}$ and for the functional equations that determine the expansion coefficients $ \omega_{\dot\phi}$. 

We visualise the \textit{feed-down} structure of the flow 
in \Cref{fig:FeedDown}. \Cref{eq:FeedDownCondition} breaks the feed-back loop as no higher order operator is present in the functional equations for the lower order ones \textit{(dashed red arrow)}. 

\begin{figure}[t!]
	\begin{tikzpicture}
		\node at (0,0) (g0) {\color{black}$( \Gamma_T,\dot \phi)_n \hspace{2mm}$};
		\node at (0,-1) (p0) {\color{black}$( \Gamma_T,\dot \phi)_{n+1}$};
			\draw [thick, black, ->] (g0) -- (p0);
			\draw [thick, dashed,red,  -|>] 
			(p0) edge[out=0, in=0] (g0);
		\end{tikzpicture}
		\caption{Feed-down structure of the flow with the feed-down condition \labelcref{eq:FeedDownCondition}. The straight back arrow indicates the feed-down of the $n$th-order pair into the higher order flows, specifically the $n+1$st order. The dashed red arrow indicates the potential feed-back of the higher order terms into the lower order ones. The feed-down condition \labelcref{eq:FeedDownCondition} breaks up this feed-back.\hspace*{\fill}}
		\label{fig:FeedDown}
	\end{figure} 
For example, the feed-back structure of the vertex is a very simple one: The flow of $n$-point functions $\Gamma_T^{(n)}$ depends on $m$-point functions $\Gamma_T^{(m)}$ with $m\leq n+2$ and on $\dot\phi^{(j)}$ with $j\leq n+1$. In terms of the expansion coefficients $\omega_{n}$ we deduce from \labelcref{eq:GenFlowT}, 
\begin{subequations} 
	\label{eq:eq:FeedDownVertexI}
	\begin{align}
		\dot \omega_{j_n} =\textrm{PI}_{j_n}\left[ \{ \omega^{\ }_{\Gamma_T,j_{m_1}}, \dot\omega_{\dot\phi , j_{m_2}}  \}\right]\,,
		\label{eq:FeedDownVertexI1}
	\end{align}
	with 
	\begin{align}
		m_1\leq n+2\,\quad \textrm{and} \quad  m_2\leq n+1\,. 
		\label{eq:FeedDownVertexI2}
	\end{align}
\end{subequations}
In \labelcref{eq:FeedDownVertexI1}, PI stands for the flows of $\omega_{\Gamma_T}^{\ }$ and the functional equations for $\omega_{\dot\phi}$ derived from the physics-informed flow \labelcref{eq:GenFlowT}. We conclude that feed-down flows in the vertex expansion satisfy 
\begin{align}
	\frac{\delta \textrm{PI}_{j_n}}{\delta \omega_{j_m} }= 0 =  \frac{ \delta \textrm{PI}_{j_n} }{\delta \dot \omega_{j_m}} \,,\quad \textrm{for}\quad m>n\,.
	\label{eq:FeedDownVertexII}
\end{align}
In general, \labelcref{eq:FeedDownVertexII} constitutes a hierarchy of constraints for the expansion coefficients $\{\omega_{\Gamma_T}\,,\,\dot \omega_{\dot\phi}\}$. The most straightforward application is its partial implementation as a hierarchy of constraints for $\{\dot \omega_{\dot\phi}\}$ with 
\begin{align}
	\frac{\delta \textrm{PI}_{\Gamma_T,j_n}}{\delta \omega_{\Gamma_T,j_m} }= 0\,,\quad m>n\,.  
	\label{eq:FeedDownVertexPractical}
\end{align}
This concludes our conceptual analysis and first example of feed-down flows.
Explicit results for the vertex-expansion will be presented elsewhere.

In the following we give a derivation of feed-down flows for the derivative expansion in \Cref{sec:FeedDownDerivative}.
Furthermore, we discuss optimisation and regulator dependence of feed-down flows in \Cref{sec:GenConv+Feed}. 
A first application of feed-down flows is considered in \Cref{sec:CompleteFeedDown1stOrder}.

\subsection{Feed-down flows in the derivative expansion}
\label{sec:FeedDownDerivative}

In this Section we discuss the question, whether we can choose or rather restrict our flowing fields such, that the coupled set of PDEs for the effective action reduce to a feed-down system of ODEs, the \textit{feed-down flows}. 
We illustrate this at the example of the derivative expansion. Here, the label indicates the order of derivatives applied to the fields in the effective action. Schematically we write 
\begin{subequations} 
\label{eq:DerExpand}
\begin{align}
	\Gamma[\phi]= &\, \sum_{n} \sum_{m,\boldsymbol{r}_m} \int_ x \frac{1}{n!}\gamma^{\boldsymbol{a}_m} _{\boldsymbol{\mu}_{2n}}(\phi,\boldsymbol{r}_m) \left[\prod_{i=1}^{m} \partial^{r_i} \phi^{a_i}\right]_{\boldsymbol{\mu}_{2n}} \hspace{-.3cm}\,, 
	\label{eq:DerExpandGa}
\end{align}
with 
\begin{align}
\boldsymbol{r}_m=(r_1,...,r_m)\,,\quad \boldsymbol{a}_m=a_1\cdots a_m\,,\quad \boldsymbol{\mu}_{2n}=\mu_1\cdots \mu_{2n}\,,
\label{eq:vectors} 
\end{align} 
and 
\begin{align}
 r_1\leq r_2\cdots \leq r_m\,,\quad \sum_{i=1}^m r_i=2n\,,\quad 1\leq m\leq 2 n\,.
\label{eq:sums} 	
\end{align} 
\end{subequations} 
The contraction over field indices and Lorentz indices comprises all possible Lorentz invariant combinations. If we restrict ourselves to O(N)-symmetric composite fields $\phi$, the expansion in \labelcref{eq:DerExpand} also is O(N)-symmetric. 

Note also that the unconstrained sum in \labelcref{eq:DerExpand} has redundancies, and different terms of the same order can be mapped into each other by partial derivatives. For example, in the first order of the derivative expansion in an O(N)-theory with O(N)-symmetric composites we have the effective potential and a sum of three first order terms, 
\begin{align}\nonumber 
\Gamma[\phi]= &\,\int_x\Biggl\{V(\rho)\,+  Z(\rho) \,(\partial_\mu \phi^a)^2 +   Y(\rho)\, (\phi^a \partial_\mu \phi^a )^2\\[1ex] 
&\hspace{.8cm}-W(\rho) \phi^a \partial_\mu^2  \phi^a\Biggr\}\,,
\label{eq:3terms1storder}
\end{align}
with the invariant $\rho = \phi^2/2$.
However, a partial integration of the $W$-term leads to 
\begin{align}\nonumber 
	\Gamma[\phi]= &\,\int_x\Biggl\{ V(\rho)+	\left[ Z(\rho)+W(\rho) \right]\, (\partial_\mu \phi^a)^2\\[1ex] 
	&\hspace{.8cm}+  \left[Y(\rho)+W'(\rho)\right]\, (\phi^a \partial_\mu \phi^a )^2 \Biggr\}\,,
	\label{eq:2terms1storder}
\end{align}
and with $Z\to Z - W$ and $Y\to Y-W'$ we are left with the two derivative terms in the first line in \labelcref{eq:3terms1storder}. These two terms constitute the standard basis of the first order terms in the derivative expansion. As discussed before, the feed-down label indicates the order of the derivative expansion and the effective potential carries the label 0 and the first order terms, or rather $Z(\rho)$ and $Y(\rho)$ carry the label 1. 

Now we show, that the flow of the target potential has no contributions from $Z,Y$ for an appropriate choice of $\dot\phi$. With this choice it only involves the potential $V$ and its derivatives $V^{(n)}$. In turn, the flow of the target $Z,Y$ depends on $V,Z,Y$ and the field transformation. In the above case with trivial targets $Z,Y=1$ this flow reduces to a constraint equation for the field transformation. We also discuss, how this feed-down structure extends to higher orders in the derivative expansion leading to a complete feed-down flow for the full derivative expansion. 

We emphasise again that the following arguments are valid for general transformations, but we restrict ourselves to O(N)-symmetric transformations for the sake of simplicity. We complement the derivative expansion of the effective action, or more precisely the target action,  with a derivative expansion of the flow of the field transformation, 
\begin{subequations}
	\label{eq:dotPhiDerExp}
\begin{align} 
	\dot\phi^a[\phi] =\sum_{n=0}^\infty  \dot \phi^a_n[\phi]\,, 
		\label{eq:dotPhiDerExpExplicit}
\end{align}
where the $\dot \phi^a_n[\phi]$ are the $n$th order terms in the derivative expansion with $2n$ derivatives. They are expanded in a complete basis of derivative operators, 
\begin{align} 
	 \dot \phi^a_n[\phi]  =  \sum_{m,\boldsymbol{r}_m} f^{a \boldsymbol{b}_m} _{\boldsymbol{\mu}_{2n}}(\phi,\boldsymbol{r})\left[\prod_{i=1}^{m} \partial^{r_i} \phi^{b_i}\right]_{\boldsymbol{\mu}_{2n}}  \,,
	\label{eq:fn}
\end{align} 
\end{subequations} 
with $1\leq m\leq 2n$ and $\sum r_i=2n$ for all $m$. In \labelcref{eq:fn}, only O(N) and Lorentz invariant combinations are considered. \Cref{eq:fn} mirrors the expansion of the effective action in \labelcref{eq:DerExpand}. 

We illustrate the general expressions with the first two terms, $\dot \phi^a_0$ and $\dot \phi^a_1$. The respective expansion coefficients can be parametrised with 
\begin{align} \nonumber 
	f^{ab}_{\mu_1\mu_2}(\phi) =&\,\delta_{\mu_1\mu_2} \left(  \delta^{ab} f_1 +\phi^a  \phi^b \bar f_1 \right)\,, \\[2ex]\nonumber   
	f^{ab_1b_2}_{\mu_1\mu_2}(\phi) =&\,\delta_{\mu_1\mu_2} \phi^a \left(  \delta^{b_1 b_2} f_2 +\phi^{b_1}\phi^{b_2} \bar f_2\right) \\[1ex]
	& + \delta^{a b_1} \phi^{b_2} f_3\,,
\label{eq:dotPhiDerExpGenfs}
\end{align} 
with $\rho$-dependent coefficients $f_i(\rho)$ and $\bar f_i(\rho)$. \Cref{eq:dotPhiDerExpGenfs} covers all tensor structures, allowed by Lorentz and O(N)-symmetry. After contraction we arrive at 
\begin{align}\nonumber 
	 \dot \phi^a_0[\phi] =&\, \phi^a f_0(\rho)\,, \\[1.5ex] \nonumber 
	\dot \phi^a_1[\phi] =&\,  \Bigl[\delta^{ab}\, f_1(\rho)\,  + \phi^a \phi^b \,  \bar f_1(\rho) \Bigr]\left( \partial_\mu^2 \phi^b \right)  \\[1ex] \nonumber 
	& + \phi^a \Bigl[  \delta^{bc} \,   f_2(\rho)+  \phi^b\phi^c \,\bar f_2(\rho)\Bigr]\, \left( \partial_\mu \phi^b \partial_\mu \phi^c \right) \\[1ex]
	& + \delta^{ab}\phi^c \, f_3(\rho) \left( \partial_\mu \phi^b \partial_\mu \phi^c \right) \,.
	\label{eq:dotPhiDerExpCoeffs}
\end{align}
The zeroth order flowing field, $\dot \phi^a_0$, has already been introduced and used in \cite{Ihssen:2023nqd} with the notation $-\frac12 \eta(\rho)$ for $f_0(\rho)$. If contracting $\dot \phi^a_1$ with $\phi^a$, the five $f_i$ and $\bar f_i$ terms reduce to three terms with the factors $\phi\partial^2\phi, (\partial \phi)^2$ and $(\phi\partial \phi)^2$, the coefficients being $f_1+\phi^2 \bar f_1, f_2$ and $f_3 +\phi^2 \bar f_2$ respectively. However, if contracting $f^a_1$ with derivative terms, e.g.~with $\partial^2 \phi^a$, the terms stay apart. Accordingly, the five terms feed down differently to higher orders in the derivative expansion. Note also that in the first order of the derivative expansion of the effective action the five $O(\partial^2)$-invariants used in \labelcref{eq:dotPhiDerExpCoeffs} are related by partial integration, and typically one only considers the second and third invariant. 

Finally we remark, that while \labelcref{eq:dotPhiDerExpCoeffs} holds for all $N$, the higher orders $\dot\phi_{n>0}$ simplify considerably for $N=1$ as we have less independent operators for $N=1$. In particular, the first order field $\dot \phi_1$ reads 
\begin{align} 
	\dot \phi_1[\phi] =&\, \left( \partial_\mu^2 \phi \right) \, \tilde  f_1(\rho)+ \phi \,(\partial_\mu \phi)^2\,  \tilde f_2(\rho)  \,, 
	\label{eq:dotPhiDerExpCoeffsO1}
\end{align}
with $\tilde f_1= f_1 + \phi^2 \bar f_2$ and $\tilde f_2 = (f_2+ \phi^2 \bar f_2) +f_3$. 

Inserting this expansion \labelcref{eq:dotPhiDerExp} into the target action flow \labelcref{eq:GenFlowT} for the effective potential, one readily sees that this flow depends on $f_0$ and $f_1,\bar f_1$, but not on $f_{i>1}$ and $\bar f_{i>1}$: The derivative terms potentially enter the flow via the $\dot\phi'$ term, while they do not enter directly via $\dot\phi$, evaluated on constant fields. However, the additional field derivative in $\dot\phi'$ can only remove one field carrying derivatives. Accordingly, all terms with more than one derivative factor drop out. The $f_1$-term contributes as $\dot\phi'$, evaluated on constant fields $\phi$ and is given by 
\begin{align}
	\frac{\delta \dot \phi_1^a}{\delta \phi^b(q)}=  q^2 z^{ab}(\phi)\,,
	\label{eq:No-feeddown}
\end{align}
with the loop momentum $q$ and 
\begin{align}
z^{ab}(\phi) = - f_1(\rho)\delta^{ab}-  \phi^a \phi^b \bar f_1(\rho)  \,.
	\label{eq:No-feeddownz}
\end{align}
\Cref{eq:No-feeddown} is readily extended to higher order derivatives, 
\begin{align} 
\dot\phi\propto \left[ \left(\partial_\mu^2\right)^n \phi \right]\,f(\phi) \longrightarrow \frac{\delta \dot\phi}{\delta \phi(q)}=  \left(-q^2\right)^n f(\phi)\,,
	\label{eq:feeddownn}
\end{align}
while all other derivative terms of $n$th order drop out: The derivatives of the field in \labelcref{eq:feeddownn} lead to terms proportional to $(q^2)^n$ in the loop proportional to $\dot\phi'$, where $q$ is the loop momentum. all other higher derivative terms feature at least derivatives of two fields, and hence the respective contributions to the flow of the target effective action are proportional to at least one $\partial_\mu\phi$ and vanish on constant fields. Accordingly, there are only two terms in $\dot \phi$, that contribute to the flow of the effective potential. We are led to 
\begin{align}
	\dot \phi^a=\phi^a \, f_0(\phi) -z^{ab}\left(\phi,-\partial^2\right)\,\partial^2 \phi^b  + \Delta \dot \phi  \,,
	\label{eq:phiFullV}
\end{align}
where all derivatives in $z^{ab}(\phi,-\partial^2)$ act on $\phi^b$ to the right, and $\Delta \dot \phi $ only comprises terms with more that one field with derivatives. The final expression for $\dot\phi'$ is 
\begin{align}\nonumber 
\dot \phi'{}(q^2)=&\, 	\frac{\delta \dot\phi^a}{\delta \phi^b(q)}\\[1ex]
=&\,  f_0(\rho)\delta^{ab} +  \phi^a \phi^b\frac{\partial f_0(\phi)}{\partial \rho} +q^2 z^{ab}(\phi,q^2) \phi^b   \,.
	\label{eq:phiprimeFullV}
\end{align}
\Cref{eq:phiprimeFullV} is the exact expression for $\dot\phi'$, evaluated on constant fields. This allows the following important conclusion: Flowing fields with the choice 
\begin{align} 
z^{ab}\left(\phi,q^2\right)\equiv 0\,,
\label{eq:zab0}	
\end{align}
lead to \textit{feed-down} flows for the effective potential.

In the first order of the derivative expansion, $z^{ab}(\phi,q^2)$ reduces to \labelcref{eq:No-feeddownz}. While $z^{ab}(\phi)=0$ is a sufficient condition for feed-down flows in the 1st order, it is not a necessary one. We also remark, that for $N=1$, \labelcref{eq:zab0} simply implies $f_1+\phi^2 \bar f_1=0$ in the first order of the derivative expansion, see \labelcref{eq:No-feeddownz}, and hence no $\partial^2 \phi$-part in $\dot  \phi_1$ in \labelcref{eq:dotPhiDerExpCoeffsO1}. 

We now discuss the minimal constraint which gives the maximal flexibility also in higher orders of the expansion scheme. To that end we note that in the flow equation for the target effective potential, $z^{ab}(\phi,q^2)$ only occurs in the $\dot \phi'{}^a$-loop in \labelcref{eq:GenFlowT}, which can be written as 
\begin{align}\nonumber 
\Tr\Bigg[ G[\phi]\,\dot \phi'{}^a_1  \, R_{\phi}\Bigg]=&z^{ab}(\phi) \!\int \frac{d^d q}{(2\pi)^d }  G^{ca}[\phi,q^2]\, q^2   R^{b c}_k(q^2)\\[1ex]
= &z^{ab}(\phi) \Bigl[ \delta^{ca}  L(\rho) + \phi^c \phi^a \bar L(\rho)\Bigr]\,, 
\label{eq:1storderphi'V} 
\end{align}
where $\delta^{ca} L(\rho) + \phi^c \phi^a \bar L(\rho)$ constitutes the results of the loop integration on the right-hand side of the first line. Hence, we are left with 
\begin{align} 
\left( f_1 + 2 \rho \bar f_1\right)  \,\left(   L  + 2 \rho \bar L\right)  + (N-1) f_1 \,L\, =0\,.  
	\label{eq:Contractzab0}
\end{align}
where we also have used \labelcref{eq:No-feeddownz}, and we have dropped the $\rho$ dependence of $f_1,\bar f_1,L,\bar L$ for the sake of readability. In conclusion, feed-down flows in the first order of the derivative expansion are obtained for 
\begin{align}
2\rho\,	 \bar f_1 = - f_1  \left( 1   +( N -1)\,\frac{ L }{ L  + 2 \rho \bar L}\right) \,, 
	 \label{eq:f1-barf1}
\end{align}
leaving us with four free parameters in the first order terms. These free parameters reflect the redundancy of the general $\dot\phi_1$, if contracted with functions of the field and not its derivatives, as discussed before. Moreover, we can choose $f_1,\bar f_1$ such that \labelcref{eq:f1-barf1} is guaranteed while still generating a non-trivial contribution to the first order term $\phi\partial^2\phi$. 

For $N=1$, the constraint \labelcref{eq:f1-barf1} simply enforces 
\begin{align}
\tilde f_1 =0\,, 
\label{eq:f1-barf1O1}
\end{align}
in \labelcref{eq:dotPhiDerExpCoeffsO1} and no $\partial^2 \phi$-term is present. Then, we are left with a single free coefficient function $\tilde f_2$, and we will study the O(1)-theory explicitly in the first order of the derivative expansion for different target actions in \Cref{sec:CompleteFeedDown1stOrder}. 

We proceed with extending the above discussion to higher orders of the derivative expansion: Assume we have already constructed a feed-down flow for the $n$th order of the derivative expansion. The $n+1$st order term $\phi'{}^a_{n+1}$ in the flowing field is given in \labelcref{eq:fn}. Its potential feed-back contribution to lower order terms solely originates in \labelcref{eq:1storderphi'V} with $\phi'{}^a_1 \to \phi'{}^a_{n+1}$. The latter has the form 
\begin{subequations} 
\label{eq:feeddownn+1}
\begin{align} \nonumber 
\dot \phi'_{n+1}(q)  = &\, \sum_{m,\boldsymbol{r}_m, r_{m+1}}  {f'}^{a \boldsymbol{b}}_{\boldsymbol{\mu}_{2n+2}}(\phi,\boldsymbol{r}_{m}) \\[1ex]
&\hspace{1cm}\times  \sum_{j=1}^m \left[q^{r_{m+1}} \prod_{i=1}^{m} \partial^{r_i} \phi_{b_i}\right]_{\boldsymbol{\mu}_{2n+2}} \hspace{-.3cm} \,,
\label{eq:feeddownn+1dotphi'}
\end{align} 
with 
\begin{align} 
	m\leq 2 n\,,\qquad r_1+\cdots + r_{m+1} = 2n+2\,,
\end{align}
\end{subequations} 
derived from \labelcref{eq:fn}. Note also, that $f'$ in \labelcref{eq:feeddownn+1} are not derivatives of $f$ in \labelcref{eq:feeddownn+1} but rather weighted sums of the $f$-coefficients. 

A general $\dot \phi'_{n+1}(q)$ leads to a feed-back to all lower orders: The powers of the derivatives $0\leq r_1+\cdots r_m\leq 2 n+2$ determines the lowest order of the feed-back for each term in \labelcref{eq:feeddownn+1dotphi'}. The lowest order is the $(q^2)^{n+1}$-term in $z^{ab}(\phi,q^2)$ in \labelcref{eq:phiFullV} and it contributes to the feed-down condition \labelcref{eq:zab0} for the effective potential. The next order contributes to the first order equations and leads to three constraints similar to \labelcref{eq:Contractzab0}. Evidently, it can be satisfied without restricting the generality of the flowing fields. 

In summary we have shown that we can successively construct flowing fields that lead to feed-down flows for the target effective action. 
This optimises the derivative expansion as the potentially destabilising feed-back terms from the higher orders are missing. 

\Cref{eq:feeddownn+1} expresses the fact that the contributions of the $n+1$st order terms in $\dot\phi$ do not contribute to the orders $i<n+1$, the sufficient condition for feed-down flows. Note, that seemingly this may not allow us to satisfy all constraints for the flowing fields required to generate a given target action effective action. However, as in the first order, the derivative terms in a general expansion of $\dot \phi$, that have to be dropped due to \labelcref{eq:feeddownn+1}, are linked to other terms still allowed by \labelcref{eq:feeddownn+1}, if inserted in the effective action. For formulating this property we remark that a general $\dot\phi$ simply is a general functional in the given order of the derivative expansion, and hence all functionals can be described by it. In particular we have 
\begin{align} 
\Gamma_T= \Gamma_T[\phi_T]\,,\quad \textrm{with} \quad \dot\phi \in \{ \dot \phi\}^{\ }_\textrm{feed-down}\,. 
\label{eq:Completedotphifd}
\end{align} 
This concludes our structural discussion of feed-down flows in the derivative expansion of the $\phi^4$-theory.

\subsection{Regulator dependences and optimisation in feed-down flows}
\label{sec:GenConv+Feed} 

On the basis of the explicit example in the derivative expansion in the last Section we provide further insight in feed-down effective actions and more generally target actions $\Gamma_T\left[\phi[\varphi]\right]$. On the basis of the explicit example in the derivative expansion in the last Section we provide further insight in feed-down effective actions and more generally target actions $\Gamma_T\left[\phi[\varphi]\right]$. We first remark that the full effective action $\Gamma_\varphi[\varphi]$ at $k=0$ is regulator-independent,
\begin{align}
	\frac{\delta \Gamma_\varphi[\varphi]}{\delta R_{\phi}}=0\,.
	\label{eq:GammvarphiRegindep}
\end{align}
\Cref{eq:GammvarphiRegindep} is also the cornerstone of RG-consistency, see \cite{Pawlowski:2005xe, Pawlowski:2015mlf, Braun:2018svj}. 
Importantly, the property \labelcref{eq:GammvarphiRegindep} is absent for generalised flows: The infinitesimal transformation $\dot\phi[\phi,R]$ carries an explicit $R_{\phi}$ dependence and so does $\phi_{k=0}[\varphi,R]$. 
Therefore, the generating functionals $Z_{k=0}[J_\phi, R]$ and $\Gamma_T[\phi[\varphi,R],R]$ depend on the trajectory $R$. 
We infer from this, that the regulator dependence of expansion coefficients of $\Gamma_T[\phi[\varphi]]$ is not a good convergence test and systematic error estimate. 
However, observables are regulator-independent and we can extract convergence properties and error estimates. 
This is a direct consequence of the reformulation of the task to compute the effective action $\Gamma_\varphi$ to that of computing the pair $(\Gamma_T,\phi)$. In \Cref{sec:FunOptPI} we have already discussed the consequences for functional optimisation. 

This general statement is apparent in the derivative expansion with feed-down: The zeroth order, LPA is not changed by higher orders and is given by the LPA results in the Wetterich equation that carry a large regulator dependence. If computing observables such as the higher order cumulants \labelcref{eq:varphi2Correlations}, the regulator dependences have to cancel out.

\section{Applications of PIRG flows}
\label{sec:PI-Feed-down}

PIRG flows can be employed to significantly reduce the computational complexity. This has already been demonstrated at the example of the zero-dimensional case in \Cref{sec:NoJ-Fund}, see the discussion there and \Cref{fig:d0correlations}. In the present Section we extend this analysis with the example of the O(1) model in $d>2$. The physics of this model is well studied and the effective potential including the fixed point regime and critical exponents at the phase transition have been computed in the symmetric and broken phases with high accuracy, see the recent review \cite{Dupuis:2020fhh} and literature therein. Therefore it is a good testing ground for our novel approach.

In \Cref{sec:ClassicalTarget-ON} we test the concept of a general target action, combined with a flowing field, which underlies the PIRG approach, within an extreme choice: The target action is chosen to be the classical action or potential, $V_{T,k}=V_\textrm{cl}$. Then, all the fluctuation physics is contained in the map $\phi[\varphi]$. We compute the effective potential $V_T(\phi(\varphi))$ in the local potential approximation (LPA) and contrast it with the LPA result for the standard Wetterich flow. This illustrates both the great potential of PIRG flows and its intricacies: The task of solving an non-algebraic PDE is recast to the task of solving a simple set of decoupled, linear ODEs. As expected, the effective potential $V_\varphi(\varphi)$ derived from the Wetterich flow does not agree with the effective action $V_{T,k}=V_\textrm{cl}$: the full target potential is non-trivial in the sense of \labelcref{eq:gammaNotGamma} and the computation of $V_\varphi(\varphi)$ requires reconstruction: The reconstruction of correlation functions from \labelcref{eq:varphi2Correlations} in classical target action flows is studied in \Cref{app:ClassicalTargetd0}. We consider the zero-dimensional case as a comprehensive and decisive benchmark.

In \Cref{sec:CompleteFeedDown1stOrder} we use feed-down flows for the first order of the derivative expansion. 
This first application showcases the power of PIRG flows and the derived concept of feed-down flows: We can utilise the advances and results obtained so far in a given approximation scheme, and go beyond the respective order without any modification (back-feeding) of the former results. 
The effective potential is computed with the standard Wetterich flow, resulting in the well-known results in LPA. The field-dependent  wave function is absorbed in the order $\partial^2$ part of the flowing field such that the anomalous dimension does not feed back into the effective potential. 
We provide solutions at the critical point and in the broken phase and compare to the ground state expansion with \labelcref{eq:Zphi1} or \cite{Ihssen:2023nqd}.

\subsection{Classical target action flows}
\label{sec:ClassicalTarget-ON}

The physics-informed flows provide the freedom of choosing a particular target action $\Gamma_{T,k}$, which defines how the quantum information in the PIRG flow is distributed between the flowing field and effective action $(\Gamma_{T,k}, \, \phi_k)$. We have seen, that this freedom can be used to achieve a specific task, i.e.~store specific couplings in the flow or implement feed-down flows. This raises the question if we can go all the way: We devise a lowest order flowing field transformation (using the notation in \labelcref{eq:dotPhiDerExp})
\begin{align}\label{eq:phidot0}
	\dot \phi[\phi] = \dot \phi^{\ }_0[\phi]\,,
\end{align}
which stores the entirely of the quantum fluctuations of the RG flow in the field transformation $ \dot \phi$ with the target action 
\begin{align}
	\Gamma_T[\phi]  = \frac12 \int_x\left(\partial_\mu\phi\right)^2  +\int_x \, V_\textrm{cl}(\phi)  + \mathcal{C}\,, 
	\label{eq:GTClassical}
\end{align}
and the classical action 
\begin{align}
	V_\textrm{cl}(\rho) = \mu_\varphi \rho +\frac{\lambda_\varphi}{2}  \rho^2 \,, 
	\label{eq:Vclassical}
\end{align}
where $\mu_\varphi \in \mathbbm{R}$ can be positive or negative and $\mathcal{C}$ is a scale-dependent constant, recall \labelcref{eq:GTC}.
The solution of the classical EoM is given by 
\begin{align} 
	\rho_\textrm{min} = -\frac{\mu_\varphi}{\lambda_\varphi}\theta(-\mu_\varphi)  \,. 
	\label{eq:rhomin}
\end{align}
If the classical action is chosen as the target action, then, trivially, \labelcref{eq:rhomin} also defines the solution to the EoM for the full theory. 
Defining a classical target action \labelcref{eq:GTClassical} implements the minimal flow of the effective action
\begin{align} 
	\frac{1}{{\mathcal{V}}_d} \partial_t \Gamma_{T,k}  = \dot{\mathcal{C}}_k  \,.
	\label{eq:FlowGTClassical}
\end{align}
This can be viewed as the orthogonal approach to the Wetterich flow, which uses $\dot{\phi} \equiv 0$. The constant $\mathcal{C}_k$ contains the normalisation $ \mathcal{N}_\phi$ and is absent in the Wetterich flow. However, in the PIRG approach, $\mathcal{C}_k$ is essential for cumulants-preserving flows and a valuable tool for optimisation, see \Cref{app:c1fixing}. 

We begin our analysis with a discussion of the minimal truncation: The \textit{classical target action} LPA and compare the results to the \textit{Wetterich} LPA in \Cref{sec:LPA-LPAT}. This is followed by a derivation of the \textit{classical target action} flow in \Cref{sec:CTAderiv}. The computation and discussion of the corresponding field transformation is given in \Cref{sec:Ini+Bd}.

\begin{figure*}
	\centering
	\begin{subfigure}{.48\linewidth}
		\centering
		\includegraphics[width=\linewidth]{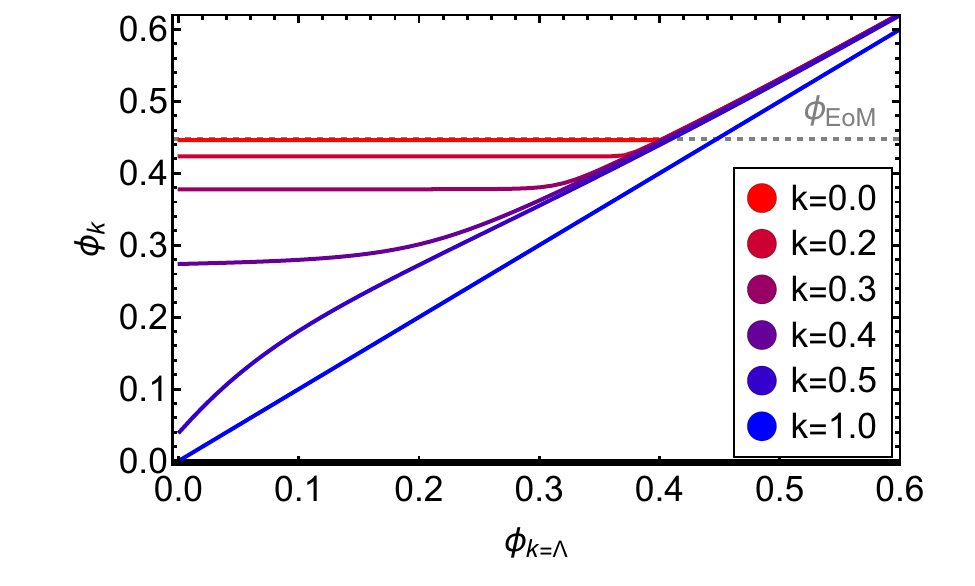}
		\subcaption{Scale-dependent, flowing field $\phi_k$ as a function of the initial field $\varphi=\phi_\Lambda$. The location of the minimum $\phi_{\textrm{EoM}}$ is indicated by the grey dashed line.\hspace*{\fill}}
		\label{fig:reconstructPhi}
	\end{subfigure}%
	\hspace{0.02\linewidth}%
	\begin{subfigure}{.48\linewidth}
		\centering
		\includegraphics[width=\linewidth]{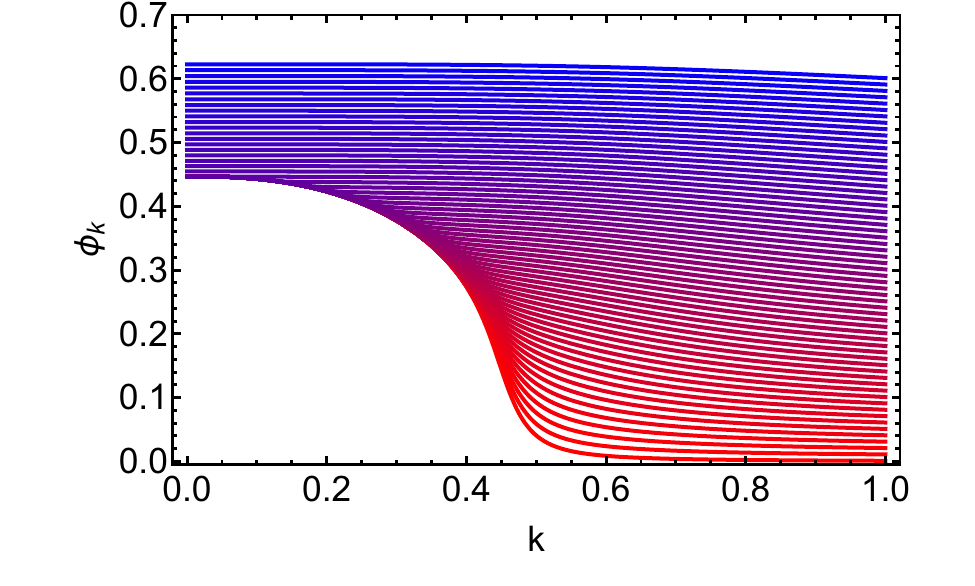}
		\subcaption{Flowing field $\phi_k$ as a function of scale for different values of the initial field $\phi_\Lambda = \varphi$. Importantly, every single field value can be calculated independently of the others. \hspace*{\fill}}
		\label{fig:flowPhi}
	\end{subfigure}%
	\caption{Flowing field transformation of the \textit{classical target action} flow in LPA, in the broken phase of an O(1) theory, compare \labelcref{eq:GTClassical} in $d=3$ with $m_\varphi^2 = -0.1$ and $\lambda =1$ at an initial scale $\Lambda = 1$. One can discern the flattening of the unphysical regime $\rho < \rho_{\textrm{EoM}} $.\hspace*{\fill}} 
	\label{fig:exampleflows}
\end{figure*}
%

\subsubsection{Wetterich LPA versus classical target action LPA}
\label{sec:LPA-LPAT}

In the zeroth order derivative expansion of the \textit{Wetterich} flow as well as the zeroth order of PIRG flows all derivative terms in the effective action and the in the flowing field are dropped. Note that this does not imply that the two expansion schemes agree even in this lowest order. This makes it even more interesting to compare the respective results. 

We start our investigation with the (\textit{Wetterich}) LPA-effective action
\begin{subequations}
	\begin{align}
		\Gamma_{k}[\phi] = \frac12 \int_x\left(\partial_\mu\phi\right)^2  +\int_x V_{k}(\phi)\,, 
		\label{eq:Gphi-LPA}
	\end{align}
	and the coordinate transformation 
	\begin{align}
		\phi_k[\varphi]\,,
		\label{eq:phi-varphiLPA}
	\end{align}
\end{subequations}
which is implicitly defined via its flow $\dot\phi[\phi]$ and $\varphi=\phi_\Lambda$, see \labelcref{eq:reconstruction}. In LPA this transformation carries no derivative terms.
Inserting \labelcref{eq:phi-varphiLPA} into \labelcref{eq:Gphi-LPA}, leads to 
\begin{align}
	\tilde \Gamma_{\phi,k}[\varphi] = \frac12 \int_x\phi'(\varphi)^2 \left(\partial_\mu\varphi \right)^2  + \int_x V_{k}\bigl(\rho[\varphi]\bigr)\,, 
	\label{eq:Gvarphi-GphiLPA}
\end{align}
with $\phi'(\varphi)=\partial_\varphi \phi(\varphi)$. Hence, \labelcref{eq:Gvarphi-GphiLPA} seemingly contains a field-dependent  wave function 
\begin{align}
	Z_\varphi(\varphi) =\phi'(\varphi)^2\,,  
	\label{eq:SeemingZ}
\end{align}
and goes beyond LPA in $\varphi$. However, if the approximation is expanded beyond LPA, while keeping the feed-down structure intact, i.e.~$f_1=0$, the wave function $Z_\varphi$ also gets contributions from $f_2,f_3$ (\labelcref{eq:dotPhiDerExpCoeffs} and further higher order terms) that do not feed back into the flow of the effective potential. Accordingly, they can be taken into account without changing the flow equation for $V_k$. We obtain the same potential, but the wave function is modified to
\begin{align}
	Z_\varphi(\varphi) =\phi'(\varphi)^2 + f_{2,3}-\textrm{contributions}\,.  
\end{align}
Consequently, the wave function in	\labelcref{eq:Gvarphi-GphiLPA} has to be interpreted with caution: 
Indeed, we arrive at the standard LPA expression for the wave function of $\varphi$ with 
\begin{align}
	=\phi'(\varphi)^2 + f_{2,3}-\textrm{contributions} = 1\,,
	\label{eq:EnforceLPA}
\end{align}
which highlights that a first stabilising approximation is given by LPA' or the full first order of the derivative expansion. This is well-known from the discussion of pole masses in the fRG, \cite{Helmboldt:2014iya}. There it has been shown, that the correct pole masses in O(N)-theories and  Yukawa models require at least an LPA' approximation.

The correction between LPA and the first order derivative expansion for \textit{Wetterich} flows is relatively small in O(N) models in $d>2$, since the generated wave functions or anomalous dimensions are also small. The \textit{classical target action} flow changes this: Due to \labelcref{eq:Gvarphi-GphiLPA} the wave function is as big as it can get within LPA, since the entirety of the flow is stored in the map $\phi[\varphi]$ and accordingly $\phi'(\varphi)$. From this point of view, this flow should be viewed as a sub-optimal expansion scheme in the sense of rapid convergence. Nevertheless, the concept merits investigation due to its tremendous potential for computational simplification which we discuss in the following.

\subsubsection{Classical target action flow in O(1)}
\label{sec:CTAderiv}

We proceed with a discussion of the target action setup for the O(1) theory in LPA. We restrict ourselves to the classical target potential \labelcref{eq:GTClassical} later. The generalised RG flow of the potential is derived by evaluating \labelcref{eq:GenFlow} at constant background $\phi(x) = \phi$. The full target action flow reads
\begin{align}
	& 	\dot{\phi}\, V_T^{(1)} -\frac{2}{d+2} \dot{\phi}' \frac{A_d k^{d+2}}{k^2 + V_T^{(2)}} 
	= \frac{A_d k^{d+2}}{k^2 + V_T^{(2)}} 
	- \partial_t V_T \,,
	\label{eq:LPAtarget} 
\end{align}
where $A_d= \frac{2 \pi^{d/2}}{d \ \Gamma(d/2) (2 \pi)^d}$. We use the definition from \labelcref{eq:phidot0} where we have restricted ourselves to transformations $\dot\phi$ without derivative terms. 

Now we use the classical target action \labelcref{eq:GTClassical} with $V_T=V_\textrm{cl}$ and arrive at 
\begin{align}\nonumber 
	\dot{\phi}\,\phi \, \left(\mu_\varphi + \lambda_\varphi \frac{\phi^2}{2}\right)=&\\[1ex]
	&\hspace{-1.7cm} \left[1+ \frac{2}{d+2} \dot{\phi}'\right] \frac{A_d k^{d+2}}{k^2 + \mu_\varphi + \frac{3}{2} \lambda_\varphi \phi^2
	} - \dot{\cal C}_k \,,
	\label{eq:LPAO1} 
\end{align}
which is an ordinary differential equation (ODE) for $\dot \phi$ at every RG-scale $k$. In LPA we have replaced a partial differential equation in $V_k$ by an ODE in $\dot \phi$.
Note, that the current choice of classical target action may be replaced by any other analytic expression $V_T=V_k$, also including a scale dependence, and still yield the same technical simplification. In particular, one might use a perturbative potential $V_T=V_{\mathrm{1-loop}}$ and include higher order terms via the flowing field.

Now we solve the \textit{classical target action} flow: Firstly, the constant part of the flow is rewritten to match the general structure of the flow
\begin{align}
	\dot{\cal C}_k = c_1 A_d k^{d}\,.
	\label{eq:ChoiceCk}
\end{align}
\Cref{eq:LPAO1} is solved using the method of variation of constants, which yields an integral 
\begin{subequations}\label{eq:numSol}
	\begin{align}\label{eq:phidotInt}
		\dot \phi = &\int_{\phi}^{\infty} dx \, \frac{d+2}{2} \left(1- c_1 
		\frac{k^2 + \mu_\varphi +3 \lambda_\varphi \frac{ x^2}{2}}{k^2} \right) \notag \\[1ex]
		& \hspace{8mm}\times \exp \left( f\left(\frac{\phi^2}{2}\right)-f\left(\frac{x^2}{2}\right)\right) \,,
	\end{align}
	with
	\begin{align}
		f(y) = \frac{(d+2) \, y }{2 A_d k^{d+2}}\left[ (\mu_\varphi+ \lambda_\varphi y)^2 + k^2\left( \mu_\varphi + \frac{\lambda_\varphi}{2} y\right)\right]  \,.
	\end{align}
\end{subequations}
The upper integration boundary of \labelcref{eq:phidotInt} corresponds to the boundary condition of the ODE and is set to $\infty$ to implement locality as discussed in \Cref{sec:local} \textit{(a)}.
Then, the transformation $\dot \phi$ is always well defined and finite: Due to the integration boundary, the integration variable $y$ is always bigger than $\phi$, which turns the exponential function in the integrand $<1$ for $\phi \geq 0$.
Furthermore, we fix the constant $c_1$ to implement a $\phi \to - \phi$ symmetry in $\dot \phi$. For any other choice $\dot \phi$ diverges as $\phi \to - \infty$, this is discussed further in \Cref{app:c1fixing}.

The scale-dependent coordinate map $\phi_k(\varphi)$ is evaluated by inserting $\dot \phi$ in \labelcref{eq:G+CompositeField} and performing the integration, see \Cref{fig:reconstructPhi} for an explicit example with initial conditions specified in \Cref{sec:Ini+Bd}.
The map is monotonous, and even strictly monotonous above the equations of motion \labelcref{eq:EoM0}. From this we have shown that the transformation exists globally according to \Cref{sec:local} (b).

The evolution with the RG-scale is depicted in \Cref{fig:flowPhi} for different values of the initial field $\varphi$. \Cref{fig:flowPhi} manifests how different initial values of $\varphi$ evolve independently from each other; each line in the plot is computed separately and may be parallelised in a trivial manner. 

By comparison, the Wetterich flow, i.e.~\labelcref{eq:LPAtarget} with $\dot \phi =0$ and \labelcref{eq:numSol} is a second order partial differential equation, whose solution entails many numerical challenges that have been studied for many years now \cite{Grossi:2019urj, Koenigstein:2021rxj, Koenigstein:2021syz, Steil:2021cbu, Ihssen:2022xkr}.

\subsubsection{Comparison to Wetterich LPA}
\label{sec:Ini+Bd}

In this Section we present results for the LPA potential using the \textit{classical target action} flow and compare to results obtained in the standard \textit{Wetterich} formulation.
In particular, we solve \labelcref{eq:numSol} in $d=3$ for a set of initial conditions deep in the broken phase, i.e.~$\mu_\varphi = -0.1$ and $\lambda = 1$, measured in terms of the UV-cutoff scale $\Lambda = 1$.

The coordinate map in \Cref{fig:exampleflows} shows that all values $\varphi$ are either mapped onto the classical minimum or some value $\phi_k \geq \phi_{\textrm{EoM}} = \sqrt{ 2 \,  \, 0.1} \approx 0.447$. This corresponds to the flattening of the potential in the Wetterich flow. While the latter typically requires an ever increasing precision of the flow,  the 
$\phi_k$ are simply mapped onto $\phi_{\textrm{EoM}}$. This is achieved similar to the solution using the method of characteristics in the large N limit, which is recapitulated in \Cref{app:LargeN}. The \textit{classical target action} flows have no dynamical mechanism for convexity restoration as is known from the Wetterich flow, where the flat regulator implements a pole in the flow which drives the potential to convexity. Instead, the structure of the exponential function in \labelcref{eq:numSol} is such, that field values corresponding to the unphysical part of the potential are always increased and thus pushed towards the minimum $\phi_{\mathrm{EoM}}$. Accordingly, this goes hand in hand with a dynamical deformation of the set of allowed $\phi$ in \labelcref{eq:SetGT}. 

\begin{figure}[t]
	\centering
	\includegraphics[width=\linewidth]{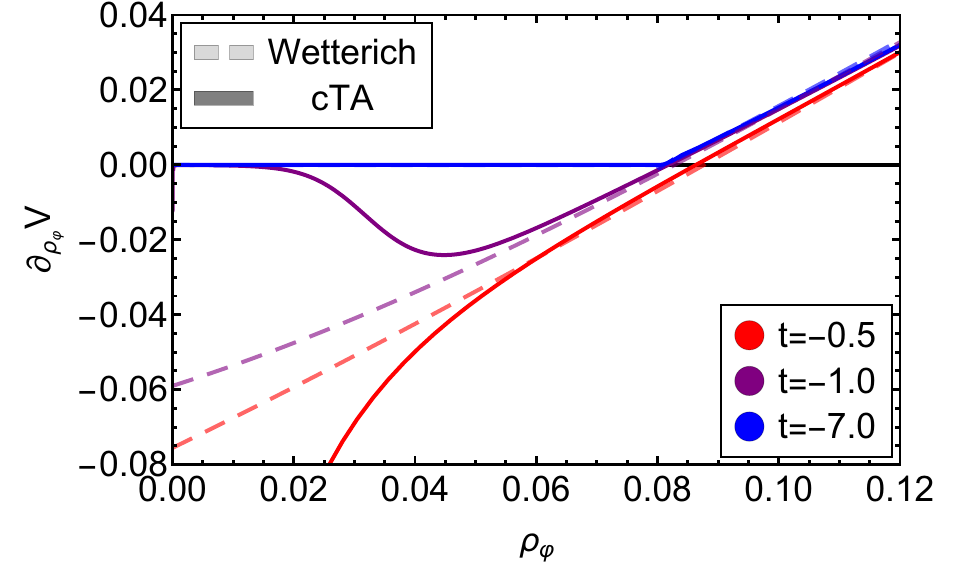}
	\caption{Solution of the \textit{Wetterich} and \textit{classical target action} (cTA) flows at different RG-times. We find that the location of the EoMs coincides well at $k \to 0$, see \labelcref{eq:solsEom}. However, we find qualitative differences in the RG-time evolution as well as deviations away from the EoM. We have depicted the first derivative of the potential $\partial_{\rho_\varphi} V$ to highlight the equations of motion.\hspace*{\fill}}
	\label{fig:GalerkinComparison} 
\end{figure}

For a direct comparison to the Wetterich result, we solve \labelcref{eq:LPAtarget} for $\dot \phi =0$ using the same initial conditions. In the numerical computation we use a continuous Galerkin scheme with an implicit solver for the time stepping, which was tested in \cite{Ihssen:2023qaq, Ihssen:2023xlp}. A graphical comparison is given in \Cref{fig:GalerkinComparison}. The figure shows the effective potential in terms of the original field $\varphi$. More explicitly it shows
\begin{align}
 \partial_{\rho_\varphi} V_{\mathrm{cl}}(\rho(\rho_\varphi)) = (\mu_\varphi + \rho \lambda) \partial_{\rho_\varphi} \rho  \,,
\end{align}
in comparison to the (first derivative of the) Wetterich potential $\partial_\rho V_{k}(\rho)$.

The solution to the equations of motion in terms of $\varphi$ coincides well between both schemes, with the respective values
\begin{align}
	\varphi_{\mathrm{EoM, cTA}} = 0.4020(7) \,, \quad \varphi_{\mathrm{EoM, W}} = 0.403(3) \,.
\label{eq:solsEom}
\end{align}
for reference, the classical equation of motion is located at $\varphi_{\mathrm{EoM, cl}} = 0.447$. 
The error on the Wetterich result is given by the numerical error derived from the size of the bulge created from resolving a kink using a finite element method. The error on the PIRG flows is related to the resolution in the field, which could technically be decreased even more.

The genuinely different regulation of momentum modes, i.e.~the regulator acts on the composite field $\phi$ in \textit{classical target action} flows, whereas it acts on $\varphi$ in the Wetterich flow, shows in the deviation of results at intermediate $k$ in \Cref{fig:GalerkinComparison}. Furthermore, we can see a vastly different behaviour in the flows in the unphysical regime, where momentum dependences play a big role. This is also due to the genuinely different truncation of \textit{classical target action} LPA and Wetterich LPA.

We conclude that the relatively straightforward identification of the map $\phi[\varphi]$ deep in the broken phase leads to an impressive agreement of the solution of the EoM, see \labelcref{eq:solsEom}. Still, there remain open questions whose solution will allow us to extend this straightforward computation to the full phase structure: 
\begin{itemize}
	
\item[(1)] If approaching the phase transition, i.e.~$\mu_\varphi \to 0$, the good agreement between $V_\phi(\phi(\varphi))$ and $V_\varphi(\varphi)$ does not hold any more, recall the discussion around \labelcref{eq:gammaNotGamma}. This simply entails that higher order correlations are getting successively more important in the vicinity of the phase transition and the differences between the two effective actions and hence also the potentials show. This fluctuation dominance is already indicated by the large deviations at small field values in \Cref{fig:GalerkinComparison}. 
	
\item[(2)] The constant $c_1$ plays an essential role in the optimisation of the reparametrisation. Amongst other deformations, it also encodes a field-dependent  shift between $\phi$ and $\varphi$ which is difficult to resolve. In combination with the first point (i) this complicates the extraction of observables such as $\varphi_\textrm{EoM}$ and alike. The choice used here is detailed in \Cref{app:c1fixing}, see \labelcref{eq:NormaliseMap}. This procedure cannot be straightforwardly extended to target actions in the symmetry broken phase, which is also addressed in \Cref{app:c1fixing}.
\end{itemize}
Both matters will be investigated further in the future, but are beyond the scope of the present work. In principle, information on the correlation functions can be obtained from \labelcref{eq:varphi2Correlations} and we envisage an investigation of critical behaviour using the PIRG flows.

\subsection{First order derivative expansion in the O(1)-model with the $V_W$ target action}
\label{sec:CompleteFeedDown1stOrder}

We can also use the PIRG flows to construct feed-down flows based on existing results and known approximations:
We study the first non-trivial order of complete feed-down flows within the first order of the derivative expansion. In order to optimise/minimise the computational effort and the use of existing results we base our computation on the LPA effective action obtained from the standard Wetterich flow as the target action, 
\begin{align}
	\Gamma_T[\phi] = \int_x \left\{  \frac{1}{2} \left(\partial_\mu \phi^a\right)^2 +V_{W,k}(\phi)\right\}\,,
	\label{eq:VWk-ON}
\end{align}
where $V_W$ is the fully $k$-dependent LPA potential. It is computed with a simple finite difference implementation, which was outlined in \cite{Ihssen:2023qaq}. Alternatively, we can also use an implementation in terms of Continuous or Discontinuous Galerkin methods, which allow us to capture the intricate dynamics in the potential which is linked to first order phase transitions \cite{Grossi:2019urj, Grossi:2021ksl, Ihssen:2022xkr}. 
Our present developments utilise simple, ready-to-use codes which are tailor-made for the LPA task and beyond. Codes are openly accessible in the GitHub repositories linked to \cite{Ihssen:2022xkr, Ihssen:2023qaq}.

\subsubsection{Flowing fields for complete feed-down flows: O(N)}
\label{eq:FlowingFieldsON}

The fluctuation physics, which is not contained in the LPA potential $V_W$, is stored in the flowing field, i.e.~we consider a complete feed-down flow which leaves the LPA part of the system $(\Gamma_T,\dot \phi)$ untouched. To that end we expand the field transformation in derivatives, see \labelcref{eq:dotPhiDerExp}, 
\begin{align} 
	\dot \phi= \dot \phi_0 + \sum_{n=1}^\infty \dot \phi_n\,. 
	\label{eq:OnfeeddowndotphiPre}
\end{align} 
Each term $\dot \phi_n$ carries $\partial^{2n}$ derivatives and the complete feed-down condition requires that a higher order term $\dot\phi_n$ does not feed-back to the lower order terms in both the target action and the transformation, $(\Gamma_{T,i},\dot\phi_i)$ for $i<n$. A slight extension that does not break the feed-down property in terms of the target action would be the cancellation of $\dot\phi_n$ and $\dot \phi_{i<n}$ terms that contribute to $\Gamma_{T,i}$. Such a cancellation of different order terms harbours the potential danger of implicitly generating $1/\partial^2$ terms in the transformation. Hence we discard this option in the following. 

In the present case the Wetterich LPA target action requires to set the lowest order term to zero, 
\begin{align} 
	\dot \phi_0\equiv 0\,,
	\label{eq:dophhi00} 
\end{align}
as it contributes to the flow of $\Gamma_{T,0}$. Inserting \labelcref{eq:dophhi00} in \labelcref{eq:OnfeeddowndotphiPre}, leaves the simple transformation 
\begin{align} 
	\dot \phi= \dot \phi_1\,,
		\label{eq:Onfeeddowndotphi}
\end{align} 
\begin{figure*}
	\centering
	\includegraphics[width=\textwidth]{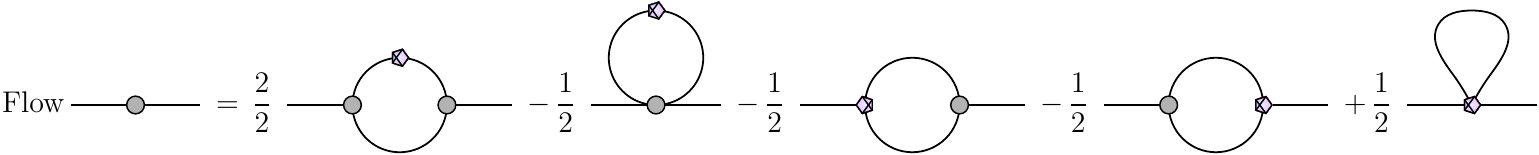}\vspace{2mm}
	\parbox{0.55\textwidth}{\caption{Diagrammatic part of the flow of the self-energy for $N=1$, which is used to derive the feed-down flows \labelcref{eq:dtZfeeddown}. The notation is that introduced in \cite{Ihssen:2024miv}. It is adapted to \textit{QMeS} \cite{Pawlowski:2021tkk} as well as accommodating the additional diagrammatic contributions in the generalised flow \labelcref{eq:GenFlow,eq:GenFlowT}. The first two diagrams are the standard contribution to the scalar self energy. Additionally, we have new diagrams with external legs connecting to the flowing field insertion. \hspace*{\fill}} 
		\label{fig:feeddown}
	}%
	\parbox{0.45\textwidth}{\includegraphics[width=0.8\linewidth]{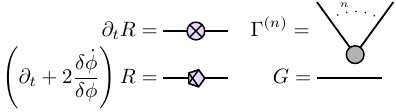}\vspace{4mm}}
\end{figure*}
which can also be augmented by higher order feed-down terms $\phi_{n\geq2}$ in the derivative expansion of the composite field. Here we concentrate on the first order and drop the higher order terms. Then, the map from the fundamental field to the composite one takes the form 
\begin{align}
	\phi_k[\varphi|=\varphi + \phi_{1,k}[\varphi]\,, \qquad \phi_{1,k} =\int_\Lambda^k \frac{d k'}{k'} \dot\phi_{1,k'}[\varphi]\,. 
	\label{eq:CompFieldVWk}
\end{align}
With $\dot\phi_1$ as defined in \labelcref{eq:dotPhiDerExpCoeffs}, the second term on the right-hand side of \labelcref{eq:CompFieldVWk} consists out of second derivative terms in $\phi$. Accordingly, we can also order the right-hand side in terms of $\varphi$ and its derivatives. For the sake of completeness we first provide the general expression
\begin{align} \nonumber 
	\phi^a_1[\varphi] =&\,  \Bigl[\delta^{ab}\, F_1(\rho_\varphi)\,  + \varphi^a \varphi^b \,  \bar F_1(\rho_\varphi) \Bigr]\left( \partial_\mu^2 \varphi^b \right)  \\[1ex] \nonumber 
	& + \phi^a \Bigl[  \delta^{bc} \,   F_2(\rho_\varphi)+  \varphi^b\varphi^c \,\bar F_2(\rho_\varphi)\Bigr]\, \left( \partial_\mu \varphi^b \partial_\mu \varphi^c \right) \\[1ex]
	& + \delta^{ab}\varphi^c \, F_3(\rho_\varphi) \left( \partial_\mu \varphi^b \partial_\mu \varphi^c \right) +O(\partial^4)\,, 
\label{eq:Feeddownphi1Gen}
\end{align}
where $F_i(\rho_\varphi)$ are the $k$-integrals of the coefficient functions $f_i$ in \labelcref{eq:dotPhiDerExpCoeffs} at fixed $\rho_\varphi$ and $O(\partial^4)$ collects higher derivative terms in $\varphi$, not in $\phi$. \Cref{eq:CompFieldVWk} entails that the derivative expansion in the composite $\phi$ is expanded order by order in the derivatives of $\varphi$ and the lower order terms only feed down into the higher order ones. This feed-down property for the derivative expansion of the fields is present for all target action flows, independent of the feed-down property of the flow itself. In the present case this allows us to insert \labelcref{eq:Feeddownphi1Gen} into \labelcref{eq:VWk-ON} and expand it in terms of the derivative expansion of $\varphi$. 

In contrast to \Cref{sec:ClassicalTarget-ON}, keeping the LPA potential $V_W$ removes the issue of fixing the constant $c_1$ (compare \labelcref{eq:numSol}). Additionally, \labelcref{eq:dophhi00} allows the interpretation of $\phi$ as the LPA mean field, which is upgraded with momentum corrections. Thus keeping the original spirit of the derivative expansion at lowest order and in the definition of field space.

\subsubsection{Results for the O(1)-model}
\label{eq:FeeddownO11storder}

In this Section we concentrate on O(1), which is the simplest showcase. For the sake of completeness we also provide the respective relations in \Cref{app:Gen1stOrder}, which elucidates the structural similarity to the O(1) case. A complete solution of the general O(N) case goes beyond the scope of the present work. 

The flowing field is parametrised by \labelcref{eq:dotPhiDerExpCoeffsO1}, leading to 
\begin{align}
	\phi_1[\varphi] =&\, (\partial_\mu^2\varphi) \,  \tilde F_1(\rho_\varphi) +  \varphi \left( \partial_\mu \varphi \right)^2 \,  \tilde F_2(\rho_\varphi)  \,, 
	\label{eq:Feeddownphi1O1}
\end{align}
where $\tilde F_1$ and $ \tilde F_2$ are obtained by integrating $\tilde f_1,\tilde f_2$ in \labelcref{eq:dotPhiDerExpCoeffsO1}. We use \labelcref{eq:Feeddownphi1O1}, only keeping terms with up to two derivatives. Inserting the field redefinition \labelcref{eq:CompFieldVWk} into the Wetterich LPA target action \labelcref{eq:VWk-ON} and only keeping the first order terms in $\varphi$, we are led to 
\begin{align} \nonumber 
	\frac12 \int_x (\partial \phi)^2 \to& \frac12 \int_x (\partial \varphi)^2 \,,\\[2ex]\nonumber 
	\int_x \,V_{W,k}(\rho) \to &	\int_x \,V_{W,k}(\rho_\varphi +\varphi \phi_1) \\[1ex]
	\to&  	\int_x\, 	V_{W,k}(\rho_\varphi ) + 	\int_x\, (\varphi \phi_1)\,V'_{W,k}(\rho_\varphi) \,,
\end{align}
i.e.~the momentum-dependent contribution $\dot \phi_1$ to the LPA field translates to a correction of the wave function. In the following we use $'$ to denote $\rho$ derivatives and $^{(1)}$ to denote $\phi$ derivatives.

We rewrite the action in terms of $\varphi$ as
\begin{align}
	\Gamma_\phi[\varphi] = \int_x \left\{ \frac12 \tilde Z_\phi(\rho_\varphi) \,\left(\partial_\mu \varphi\right)^2 +V_{W,k}(\rho_\varphi)\right\}\,,
	\label{eq:Vst-ONvarphi}
\end{align}
where the wave function $\tilde Z_\phi(\rho_\varphi)$ is neither the wave function of the standard effective action $\Gamma_\varphi$ nor is it the wave function of $\phi$. It is defined as 
\begin{align}\label{eq:DelatZ}
\tilde Z_\phi(\rho_\varphi) = 1+\Delta\tilde Z_\phi(\rho_\varphi)\,,
\end{align}
with 
\begin{align}\nonumber 
\Delta \tilde Z_\phi(\rho) = &\,V'_{W}(\rho) \,\Bigl[ 2 \rho  \,\tilde F_2(\rho) -\left( \tilde F_1(\rho)+ 2 \rho  \,\tilde F'_1(\rho)  \right) \Bigr] \\[1ex]
&\,- 2 \rho \,V_{W}''(\rho)\,  \tilde  F_1(\rho)\,.
\label{eq:Zphi}
\end{align}
We now proceed with the computation of $\tilde Z_\phi$. One of the feed-down constraints is satisfied by setting $\tilde f_1=0$. Secondly, the triviality of the wave function $Z_\phi(\phi)$ of $\phi$ implicitly defines $\phi \tilde f_2 = f$,
\begin{align} \label{eq:Z1withphi1}
	Z_\phi(\rho) \stackrel{!}{=} 1 \quad \longrightarrow \quad Z_{\phi,\Lambda}=1 \quad  \textrm{and} \quad  \partial_t Z_{\phi,k}(\rho) =0\,, 
\end{align}
with 
\begin{align}
	\partial_t Z_{\phi,k}(\rho) := 	\left[ \frac{\partial }{\partial p^2} \frac{\delta^2 \partial_t \Gamma_T }{\delta \phi(p) \delta \phi(-p)}\right]_{p^2=0} \,.
\label{eq:dtZfeeddown}
\end{align}
The diagrammatic part of the flow of the two point function is depicted in \Cref{fig:feeddown}. The field-dependent regulator insertion creates a new type of diagram in which external legs connect to the flowing field insertion. With the momentum-dependent field transformation \labelcref{eq:Feeddownphi1O1} this generates new terms in the flow of the wave function, but leaves the flow of the potential unchanged.

Inserting the Wetterich LPA target action \labelcref{eq:VWk-ON} into the flow, leads to the constraint 
\begin{widetext}
\begin{align}\nonumber 
	2  f \ V_W^{(1)} =&  \int \frac{d^d q}{(2\pi)^d}    R_{\phi}(q) G_W(q) \times  \left[2 f^{(1)} - 2 f V_W^{(3)} \frac{\partial}{\partial p^2} \Big\{  \left.
	G_W(q+p) \, (p + q) \cdot p + G_W(q-p) \,  (p - q) \cdot p 
	\Big\}\right|_{p=0}  \right]\notag \\[1ex]
&+
	\int \frac{d^d q}{(2\pi)^d} \Biggl\{   \partial_t R_{\phi}(q) G^2_W(q)\left(V_W^{(3)}\right)^2 \frac{\partial}{\partial p^2} \Big\{  \left.
	G_W(p+q) 
	\Big\}\right|_{p=0} \Biggr\}  \,,
	\label{eq:f2flow}
\end{align}
where the Wetterich LPA potential $V_W$ and the propagator $G_W$ are purely input and do not feed back.
Thus \labelcref{eq:f2flow} is a simple linear ordinary differential equation in $\phi \tilde f_2 = f$ which can be evaluated on top of a standard Wetterich LPA computation.

We can evaluate the momentum derivative and trace analytically for the flat regulator, \Cref{app:thrs+reg},
\begin{align}
		2 f  V_W^{(1)} = - A_d \frac{k^{d+2}\left(V_W^{(3)}\right)^2}{\left(k^2 + V_W^{(2)}\right)^4}
		+ 4\frac{A_d}{d+2} f^{(1)}  \frac{ k^{d+2}}{k^2 + V_W^{(2)}}- 8 \frac{A_d}{d+2} f \frac{ k^{d+2}}{(k^2 + V_W^{(2)})^2}V_W^{(3)}
		\,.
		\label{eq:f2tilde}
\end{align}
\end{widetext} 
\begin{figure*}
	\centering
	\begin{subfigure}{.48\linewidth}
		\centering
		\includegraphics[width=\textwidth]{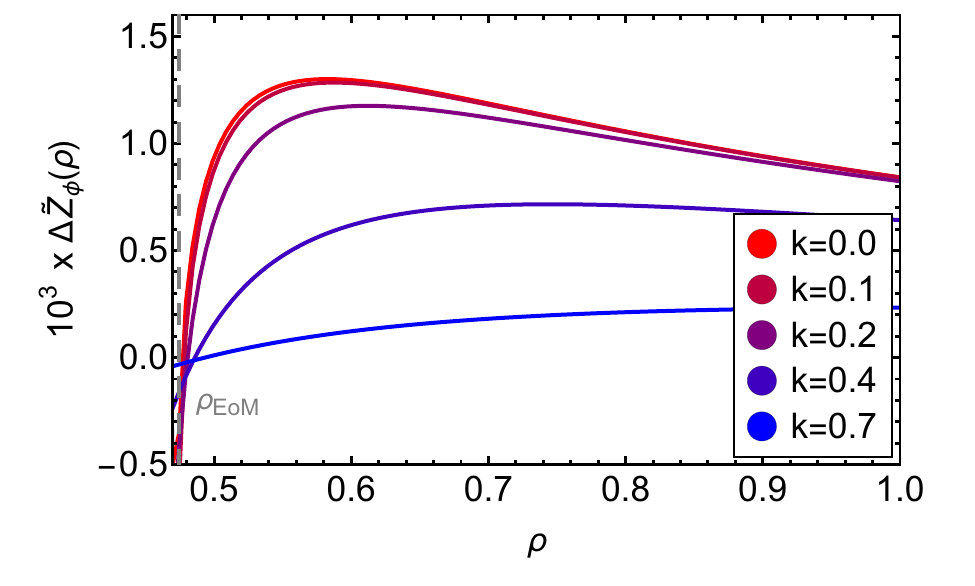}
		\subcaption{Deep in the broken phase: $\rho_{\mathrm{EoM}} =0.475 $. \hspace*{\fill}}
		\label{fig:brokenZ1}
	\end{subfigure}%
	\hspace{0.02\linewidth}%
	\begin{subfigure}{.48\linewidth}
		\centering
		\includegraphics[width=\textwidth]{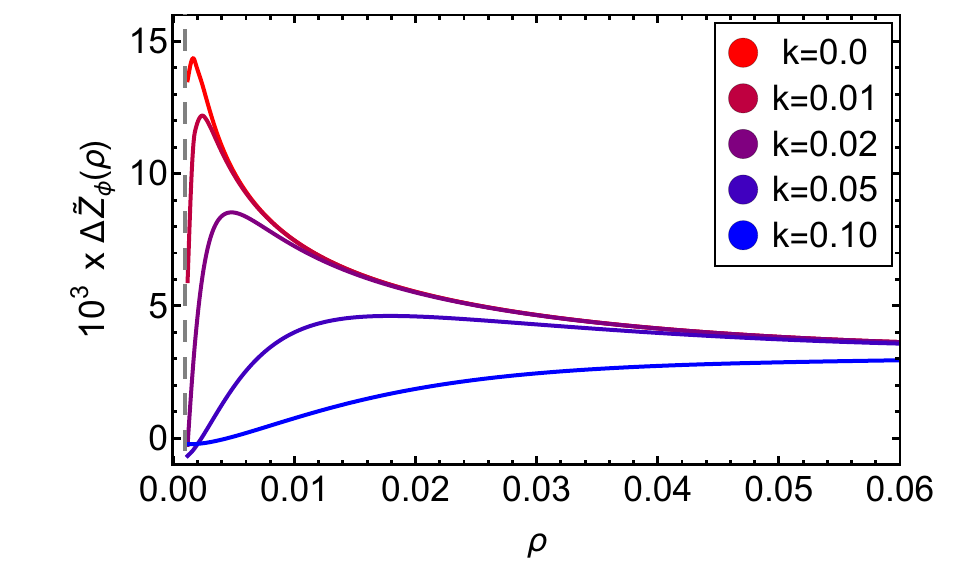}
		\subcaption{Near criticality: $\rho_{\mathrm{EoM}} = 0$. \hspace*{\fill}}
		\label{fig:criticalZ1}
	\end{subfigure}%
	\caption{Modification of the wave function $\Delta \tilde{Z}_\phi$ \labelcref{eq:Zphi}, computed from feed-down flows. To the right of the equations of motion $\rho_{\mathrm{EoM}}$ the wave function becomes very large and negative, which we do not show since it is in the unphysical part of the potential.\hspace*{\fill}} 
	\label{fig:feeddownZ}
\end{figure*}
In future works we also plan to employ smooth regulator functions, which are more adapted to momentum-dependent truncations of the effective action, see in particular \cite{Zorbach:2024zjx}. 

Finally we compute the modification of the wave function $\Delta \tilde Z_\phi$ using data from a finite difference implementation of the LPA flow \cite{Ihssen:2023qaq}. The initial action is given by the classical action \labelcref{eq:ClassicalAction1-ON} using
\begin{align}
	\Lambda = 1 \,, \hspace{2cm} \lambda_\varphi = 0.1 \,,
\end{align}
and varying initial masses $m_\varphi^2<0$ for $d=3$. The RG-time integration is performed up to $t=-20$, i.e.~$k < 10^{-8}$ with an implicit \textit{QBDF} time-stepping algorithm~\cite{doi:10.1137/S1064827594276424}.

Since the input data is generated using a finite difference scheme, it makes sense to evaluate $\tilde f_2$ in the same discretisation. The boundary condition, which implements locality (compare \Cref{sec:local} (a)), is implemented by setting $ f' (\phi \to \infty)=0$. In practice, the input data for $V_W$ is only available in some finite interval delimited by $\phi_{\mathrm{max}}$. We have tested that the results for the wave function are not sensitive to setting $f' (\phi_{\mathrm{max}})=0$ for large enough $\phi_{\mathrm{max}}$.

In \Cref{fig:feeddownZ}, we show the modification of the wave function in $d=3$ for initial masses deep in the broken phase (subscript $_b$) and tuned to the critical value (subscript $_c$)
\begin{align}
	m^2_{\varphi,b} &= -5.000 \times 10^{-2} \quad \Rightarrow \quad \phi_{\mathrm{EoM}} = 0.975 \,, \notag \\[1ex]
	m^2_{\varphi,c} &= -4.637 \times 10^{-3} \quad \Rightarrow \quad \phi_{\mathrm{EoM}} = 0\,.
\end{align}
Deep in the broken phase we find $\Delta \tilde{Z} = 0$ on the equations of motion, which is not surprising due to \labelcref{eq:Zphi}. However, as we can see from \labelcref{eq:f2tilde} this zero might be 'cancelled out' by an infinity in the definition of $\tilde{F}_2$. In particular in the symmetric phase we have $V_W^{(3)}=V_W^{(1)}=0$ on the equations of motion, necessitating some diverging behaviour in $\tilde{f}_2$ to fulfil \labelcref{eq:f2tilde} for $\tilde{f}_2' \neq 0$ in the broken phase. 
We find larger corrections to the wave function as we near criticality \Cref{fig:criticalZ1}.

\subsubsection{Interpretation of LPA feed-down flows} 
\label{sec:LPAFeeddownInterpretation}

We have implemented a trivial wave function $Z_\phi \equiv 1$ at all scales in the first order derivative expansion of the flowing field $\dot \phi_1$. 
The feed-back of $\mathcal{O}(\partial^2)$ fluctuations to the flow of the potential is removed, implementing a feed-down structure for the first order derivative expansion.

The composite field resulting from the feed-down flow can be interpreted as a reconstruction/error correction procedure answering the question:
Assuming the LPA (mean) field $\varphi$ is the \textit{physical} field we are interested in, what are the $\mathcal{O}(\partial^2)$ corrections to correlation functions?

This becomes apparent when considering the effects of the $\Delta \tilde Z$ correction \labelcref{eq:DelatZ}.
\begin{itemize}

\item[(1)] The solution to the equations of motion are not changed $\varphi_{\mathrm{EoM}} = \phi_{\mathrm{EoM}}$ due to $\dot \phi_0 = 0$.
\item[(2)] Correlation functions, for example the renormalised mass
\begin{align}\label{eq:mass2pt}
		\Gamma_\phi^{(2)}[\varphi]= \tilde{Z}_\phi \left( p^2 + \frac{V_{W}^{(2)}}{1 + \Delta \tilde Z_\phi}\right) + \mathcal{O}(p^4) \,,
\end{align}
are corrected by a factor $(1 + \Delta \tilde Z_\phi)$.
\end{itemize}
(1) is a reflection of the fact that the LPA mean field is not the full physical (mean) field, and that a reconstruction of momentum dependences using the derivative expansion of $\dot \phi$ does not change that.
Considering critical phenomena using the feed-down $\phi_{\mathrm{EoM}}$ as an order parameter can and should only yield the LPA solution by construction.
Consequently, the investigation of phase transitions (or physics) in a feed-down setup should always be performed in terms of correlation functions, for example in terms of the corrected mass \labelcref{eq:mass2pt}.
The size of correction factors, i.e.~$\Delta Z_\phi \gtrsim 1$, may signal the breakdown of the LPA approximation. A comprehensive discussion of convergence properties in the derivative expansion can be found in \cite{Balog:2019rrg}, see also the review \cite{Dupuis:2020fhh}. 

At this point it is also important to recall the difference between $\tilde \Gamma_\phi[\varphi]$ and the effective action of the fundamental field $\Gamma_\varphi[\phi]$, see \labelcref{eq:gammaNotGamma}: Whilst we have found a reconstruction of the wave function in terms of \labelcref{eq:DelatZ} the effective action in terms of the LPA mean field is not the full physical field in the first order derivative expansion. In fact, the reconstruction of correlation functions in terms of the cumulants of the fundamental field \labelcref{eq:varphi2Correlations} is needed to obtain full access to physical observables beyond LPA in this setup.
We defer the full reconstruction of momentum-dependent cumulants to future work. We also point out \Cref{app:ClassicalTargetd0}, where a reconstruction of the mass is performed in a zero-dimensional benchmark case. 

Finally, we also compare the implementation of a trivial wave function \labelcref{eq:Z1withphi1} using the momentum-dependent $\dot \phi_1$ within a feed-down structure to the ground state expansion using $\dot \phi_0$. The ground state expansion was recapitulated in \labelcref{eq:Zphi1} and preformed in \cite{Ihssen:2023nqd}.

The ground state expansion uses the flow of the wave function to determine the effective action $\Gamma_\phi[\phi]$ in terms of a composite operator $\dot \phi = \dot \phi_0$ with a classical dispersion relation
\begin{align}
	\Gamma_\phi^{(2)}[\phi]= p^2 + V^{(2)}(\phi)\,.
\end{align}
This is a field definition, which is closer to describing the physical particle than the original field $\varphi = \phi_\Lambda$. The composite operator is the field of interest for the computation of correlation functions, i.e.~there is no real necessity to compute the map $\phi[\varphi]$ other than verifying the global existence of the transformation. The price to pay are $\dot \phi$ and $\dot \phi'$ terms, modifying the flow of the potential.

By comparison, the absorption via $\dot \phi_1$ does not change the definition of the momentum-independent mean field $\varphi = \phi \vert_{p=0}$ and the LPA flow in the Wetterich equation remains unchanged. Here, the reconstruction is vital, because the information about $\mathcal{O}(\partial^2)$ fluctuations is stored in the map.

\section{Conclusion}
\label{sec:Outlook}

In this work we have suggested and worked out a novel functional renormalisation group approach with \textit{physics-informed renormalisation group flows} (PIRG flows), which is detailed in \Cref{sec:PIFlow}. The great potential of this novel approach has been illustrated at the example of key applications, and is based on a radical shift of perspective on RG flows: Instead of seeing RG flows as a mere tool for computing the effective action $\Gamma_\varphi[\varphi]$ of the fundamental field $\varphi$, they give access to pairs 
\begin{align}
\left( \Gamma_\phi[\phi]\,,\,\phi[\varphi]\right)\,, 
\label{eq:TargetPairConclusion}
\end{align}
where both, the composite field $\phi$ and its effective action, the target action, can be freely chosen and only their combination is governed by the generalised functional RG equation derived in \cite{Pawlowski:2005xe}. In short, the potential of PIRG flows lies in their generality: In principle, one can even define any \textit{target action} $\Gamma_T$ and derive the corresponding composite field. In practice this generality can be used to optimise the representation of a given theory. On the one hand this facilitates the access to physics phenomena and mechanisms in strongly correlated systems, on the other hand it minimises the computational costs. 

The generalised flow equation for the effective action \cite{Pawlowski:2005xe}, used in the present work, see \Cref{sec:flowingField}, as well as its counterpart for the Wilson effective action, the Wegner equation \cite{Wegner_1974}, allow for general reparametrisations of the theory under consideration. The full scope of these general reparametrisations has been addressed explicitly in \Cref{sec:FlowingFields} and \Cref{sec:Reconstruction}: The key property is the \textit{locality} of the infinitesimal transformation both in field and momentum space. Moreover, it has been shown in \Cref{sec:Reconstruction}, how the PIRG flows can be used to compute observables and even how to reconstruct the effective action of the fundamental field and hence the corresponding correlation functions. In combination this allows for fully controlled applications of the approach. 

A final conceptual and important development was discussed in \Cref{sec:FeedDown}, where \textit{feed-down} flows were constructed: In feed-down flows the higher orders of a given expansion scheme do not feed back into the lower order flow equations. Then, the next order of such an expansion scheme is simply computed with the input of the lower order results, effectively decoupling the infinite tower of flow equations. Apart from stabilising the hierarchy of flow equations, the feed-down structure can be used for successive order by order optimisation of expansion schemes as well as the order by order implementation of symmetry identities similar to that in perturbation theory. This intriguing possibility was worked out explicitly within the derivative expansion, one of the two main systematic expansion schemes used in the fRG approach, the other being the vertex expansion. 

The conceptual advances with PIRG flows were implemented and illustrated within selected applications in \Cref{sec:PI-Feed-down}. 
Specifically, we compared PIRG flows with the extreme case of a \textit{classical target action} in the local potential approximation (LPA) for an O(1) model with the standard LPA with the Wetterich equation in \Cref{sec:ClassicalTarget-ON}. The results are very promising, even though the reconstruction of the effective action of the fundamental field as well as its correlation functions requires more work. In \Cref{sec:CompleteFeedDown1stOrder} we computed the first order of the derivative expansion with feed-down PIRG flows, using the LPA effective action as a target action, obtained from the standard Wetterich flow. Both examples, PIRG flows with the \textit{classical target action} and feed-down flows, are conceptually different from previously employed transformations, such as emergent (fermionic) composites, dynamical hadronisation
or the ground state expansion.
From the perspective of general PIRG flows these applications use $\dot \phi$ to capture specific physics, either a resonant interaction or the ground state. This allows for an interpretation of the flowing field as the \textit{physical state}, i.e.~it needs no reconstruction in terms of the original field $\varphi$ to have an interpretation in terms of observables. On the other hand, \textit{classical target action} and feed-down flows trivialise the effective action in a way that efficiently and qualitatively reduces the computational costs of a given expansion scheme. The latter application is key to the quest for apparent convergence in strongly correlated systems, the resolution of which typically requires the use of advanced approximations. 

The functional renormalisation group approach with PIRG flows offers a novel RG approach, whose flexibility allows for the most direct access to physics phenomena and mechanisms as well as a significant minimisation of the involved computational costs. Unlocking the great potential of PIRG flows should allow us to significantly further our understanding and quantitative access to the physics of strongly correlated systems. There are many applications that are natural extensions of that presented here. Here we only name a few of them that are already under investigation: (1) The reparametrisation potential of PIRG flows can be used to construct fully gauge invariant flows, and this possibility originates in the very fact that the composite field effective action is \textit{not} identical to that in the original fields. This idea is discussed further in \cite{Ihssen:2025cff}. (2) Ground state expansions are highly interesting for an application to non-equilibrium evolutions of closed and open quantum systems, including systems with an initial far-from-equilibrium state. One may even base expansion schemes on time-dependent ground state expansions. (3) The flexibility of the choice of the target action allows for the systematic inclusion of results from other approaches such as (resummed) perturbation theory, hard thermal and dense loop results of low and high order, or density functional theory. (4) Finally, a combination of several of the present advances will be applied to the ongoing endeavour of resolving the physics of QCD at finite temperature and density. A specific first application is the phase structure of QCD, including the location of the critical end point with full apparent convergence and a small systematic error. We hope to report on these applications in the near future.

\begin{acknowledgements}
We thank K.~Falls, B.~Knorr, M.~Salmhofer, C.~Wetterich and N.~Wink for discussions. One of the first application of PIRG flows is its use in studies of the QCD phase structure undertaken in the fQCD-collaboration~\cite{fQCD}. We thank its members for discussions and collaboration on related work. This work is funded by the Deutsche Forschungsgemeinschaft (DFG, German Research Foundation) under Germany’s Excellence Strategy EXC 2181/1 - 390900948 (the Heidelberg STRUCTURES Excellence Cluster) and the Collaborative Research Centre SFB 1225 - 273811115 (ISOQUANT). 
\end{acknowledgements}

\appendix

\section{Symmetries and cumulants-preserving physics-informed RG flows}
\label{app:c1fixing}

In this Appendix we discuss the choice of the constant part $ \dot {\mathcal{C}}$ of the PIRG flow, see \labelcref{eq:GTC}. It carries the $k$-dependence of the normalisation $\mathcal{N}_\phi$ of the path integral \labelcref{eq:ZGenComposites}, as well as genuine deformations of the field $ \Delta \dot{\mathcal{C}}_k$. In \Cref{app:CumulantPreserve}, we show, how the rather formal cumulants-preserving property \labelcref{eq:nPointvarphi} turns into a practical cumulants-preserving property for $\dot{\cal C}$, based on the triviality property of the PIRG for asymptotically large composite fields. In \Cref{app:choice} the cumulants-preserving property of the flow is used to disentangle the physical part of ${\dot C}$ and the normalisation flow $\partial_t \ln {\cal N}_k$. In \Cref{app:Symmetry,app:TargetActionFreeTheory} we illustrate some of the specifics at the example of the free theory and discuss symmetry properties of the classical target action flows. The latter allows us to further simplify the cumulants-preserving flows. The conceptual findings of this Appendix are then used for the computation of observables in \Cref{app:ClassicalTargetd0}, which offers a comprehensive benchmark test.

\subsection{Cumulants-preserving PIRGs}
\label{app:CumulantPreserve}

In this Appendix we discuss, how cumulants-preserving PIRG flows are obtained. To that end we have to enforce the cumulants-preserving property \labelcref{eq:Nk-muIndep} on the flow. We shall use the property of the path integral, that the fundamental effective action reduces to the classical effective action for large field amplitudes, $\phi\to \infty$, 
\begin{align}
\Gamma_T[\phi\to \infty]= 	\Gamma_{T,\textrm{as}}[\phi] = S_\textrm{cl}[\varphi]+\ln {\cal N}_k(\boldsymbol{g}) \,, 
\label{eq:Gammaas}
\end{align}
where ${\cal N}_k$ is the possibly coupling-dependent normalisation of the path integral.
For the investigation of \labelcref{eq:Gammaas} we recall the flow for general target actions is given by \labelcref{eq:GenFlowT}. For the present purpose we re-arrange it as follows, 
\begin{align} 
	\frac{d  \Gamma_T[\phi]}{d t}  =\frac{1}{2} \Tr\Bigg[ G_T[\phi]\,\partial_t R_{\phi}\Bigg] + \Tr\Bigg[ G_T[\phi]\,  \dot{\phi}^{(1)}[\phi]\, R_{\phi}\Bigg] \,, 
	\label{eq:GenFlowRearranged}
\end{align}
with 
\begin{align}
	\frac{d  \Gamma_T[\phi]}{d t}   = \partial_t \Gamma_{T,k}[\phi] + \int_x \dot \phi \, \Gamma_{T,k}^{(1)}[\phi]\,.
	\label{eq:dtGamma}
\end{align}
For local transformations $\dot\phi[\phi]$ initiated with the fundamental field at $k\to \infty$, the diagrammatic right-hand side of the flow equation \labelcref{eq:GenFlowRearranged} vanishes identically for $\phi\to \infty$, and we arrive at 
\begin{align} 
	\frac{d  \Gamma_{T,\textrm{as}}[\phi]}{d t}  =0\,.
	\label{eq:GenFlowAsymptotic}
\end{align}
Inserting \labelcref{eq:Gammaas} in \labelcref{eq:GenFlowAsymptotic} leads to 
\begin{align}
	\frac{d}{d t}  \Gamma_T[\phi\to \infty]=\partial_t \ln {\cal N}_k(\boldsymbol{g}) \,,  
\label{eq:ASFlow}
\end{align}
and we can translate the cumulants-preserving property \labelcref{eq:Nk-muIndep} into a property of the constant part $\dot{\mathcal{ C}}$ cumulants-preserving flows, \labelcref{eq:GTC},
\begin{align}
\dot{\cal C}^{(c)}:\ 	\frac{d^n \left(\frac{d}{d t} \Gamma_\textrm{as}[\phi,\dot {\cal C}^{(c)}]\right)}{d^{m_1} g_1 \cdots d^{m_n} g_m }=0\,, 
	\label{eq:CumulantPreservingFlows} 
\end{align}
for all $n$ and $m_1,...,m_n \in \mathbbm{N}^+$. We shall see below, that \labelcref{eq:ASFlow} not only allows us to define cumulants-preserving flows, but also allows us to disentangle the non-trivial part of the flow from that of the normalisation. \Cref{eq:CumulantPreservingFlows} is the final, practical form of the cumulants-preserving property \labelcref{eq:Nk-muIndep}. 

We first illustrate the considerations above at the example of the Wetterich flow, before we discuss the general case of PIRG flows.

\subsubsection{Cumulants-preserving Wetterich flow}
\label{eq:CumulantsPreservingWet}

In the standard flow equation with $\dot \phi=0$, the normalisation drops out in the computation of correlation functions via their definition \labelcref{eq:nPointvarphi}. Accordingly, it is usually omitted or chosen for numerical convenience. Hence, general normalisations are functions of the classical coupling parameters of the theory, ${\cal N}={\cal N}(\boldsymbol{g})$. For example, the effective potential $V_W$, obtained from the Wetterich flow, is commonly normalised such that $V_{W,k}(\phi_{\textrm{EoM},k})=0$. This normalisation is achieved by subtracting the flow of the potential on the minimum, $\partial_t V_{W,k}(\phi_{\textrm{EoM},k})$, from the full flow. This is a $\boldsymbol{g}$-dependent shift of the effective potential, which reads in terms of the flow of the normalisation ${\cal N}_k$, 
\begin{align} 
\partial_t \ln {\cal N}_k \to \partial_t \ln {\cal N}'_k =\partial_t \ln {\cal N}_k + \partial_t V_{W,k}(\phi_{\textrm{EoM},k})\,.
\label{eq:NWshift}
\end{align}
Evidently, not both normalisations, ${\cal N}_k, {\cal N}'_k$ can be $\boldsymbol{g}$-independent and can satisfy the cumulants-preserving property \labelcref{eq:Nk-muIndep}. \Cref{eq:NWshift} indeed entails that $\log {\cal N}'_k$ is $\boldsymbol{g}$-dependent, as the implicit normalisation, induced by the Wetterich flow is not: For the respective proof we use that the effective action reduces to the classical one for large amplitudes, $\varphi \to \infty$, 
%
%
\begin{align}
		\Gamma_{\varphi,k} [\varphi \to \infty] = S_\textrm{\tiny{UV}}[\varphi ] +\ln  {\cal N}_{W,k}\,, 
\end{align}
where $S_\textrm{\tiny{UV}}[\varphi ]$ is the UV-effective action that only contains the UV-relevant terms, potentially including their UV running. Accordingly, the flow of the normalisation can be directly extracted from the flow, evaluated at asymptotically large fields. In the case of the Wetterich flow we find 
\begin{align}
		\partial_t \Gamma_{\varphi,k} [\varphi \to \infty] = \partial_t \ln \mathcal{N}_{W,k} =0\,. 
\label{eq:WetterichNorm}
\end{align}
Moreover, for $k\to \infty$, all theories tend towards the Gau\ss ian theory with the momentum-dependent mass $R_{\phi}$, and the normalisation there is simply ${\cal N}_{W,k\to \infty}=1$. In combination this entails 
\begin{align}
	{\cal N}_W=1\,.
	\label{eq:NW=1}
\end{align} 
It follows from \labelcref{eq:NW=1}, that ${\cal N}'_{W,k}$ in \labelcref{eq:NWshift} is $\boldsymbol{g}$-dependent, 
\begin{align} 
	{\cal N}'_{W,k}= \exp \left(V_{W,k}(\phi_{\textrm{EoM},k})\right)\,, 
\end{align}
and is not cumulants-preserving. This completes our discussion of the normalisation of Wetterich flows. 

\begin{figure*}
	\centering
	\begin{subfigure}{.48\linewidth}
		\centering
		\includegraphics[width=\textwidth]{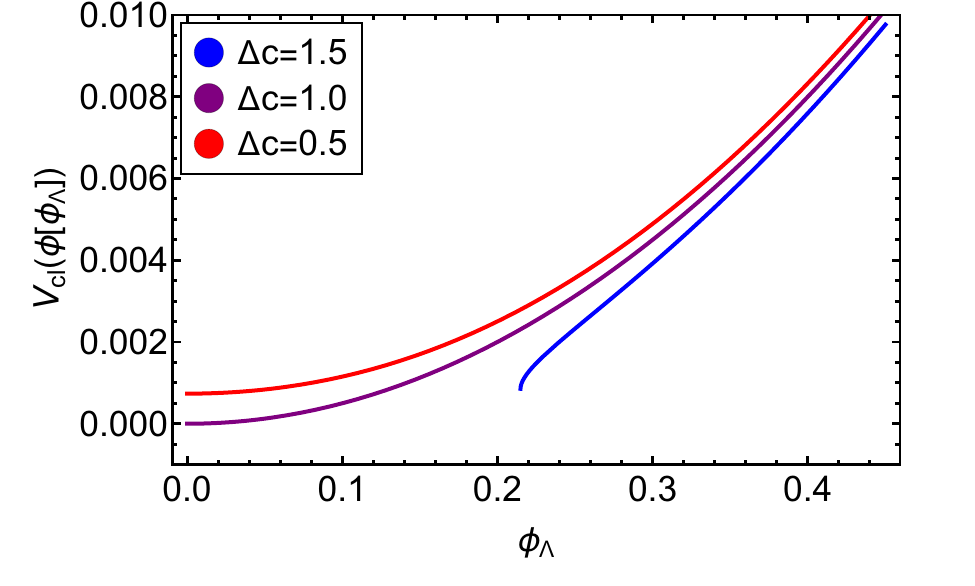}
		\subcaption{Effective potential. \hspace*{\fill}}
		\label{fig:freePot}
	\end{subfigure}%
	\hspace{0.02\linewidth}%
	\begin{subfigure}{.48\linewidth}
		\centering
		\includegraphics[width=\textwidth]{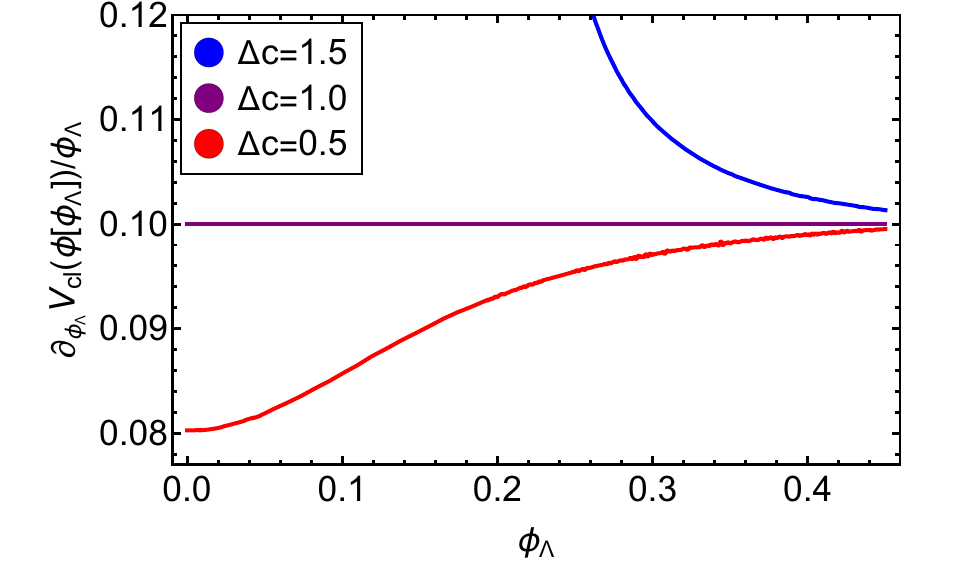}
		\subcaption{Mass. \hspace*{\fill}}
		\label{fig:freepotd}
	\end{subfigure}%
	\caption{The potential and its first derivative in a free theory for different values of the shift $\Delta c$, as introduced in \labelcref{eq:Cchoice}. One can clearly discern that there is one specific value for $\Delta c$ which leads to the trivial result of the 0th order derivative expansion of the Wetterich flow (\textit{purple}). The figure clearly shows how a naive choice of $\Delta c$ creates field dependences in the effective action $\Gamma_\phi[\phi(\varphi)]$ in the free theory.\hspace*{\fill}} 
	\label{fig:freeTheory}
\end{figure*}
%

\subsubsection{Cumulants-preserving PIRG flow}
\label{eq:CumulantsPreservingPIRG}

In the PIRG setup, effective actions
$\Gamma_\phi[\phi,\dot {\cal C}]$ are distinguished by their trajectories $\dot{\cal C}_k$. A trajectory $\dot{\cal C}_k$ does not only lead to a shift of the effective action but also to a reparametrisation of the theory. In particular, the composite field $\phi$ depends on it, 
\begin{align} 
	\phi= \phi[\varphi,\dot{\cal C}]\,, 
	\label{eq:phidotC}
\end{align}
where $\dot{\cal C}$ stands for the whole $k$-trajectory $\dot{\cal C}_k$. Given the locality of the transformation defined in \Cref{sec:local}, the flow of the composite for a given trajectory $\dot{\cal C}$, reads asymptotically
\begin{align}
	\lim_{\phi \to \infty}\dot \phi =  - \frac{2}{\lambda_\varphi }\frac{\Delta \dot{\mathcal{C}}_k (\boldsymbol{g})}{ \phi^3} \,,
\label{eq:shiftMap}
\end{align}
where $\Delta \dot{\mathcal{C}}_k$ is a scale-dependent constant in field space. Consequently the flowing field generates a constant flow 
\begin{align}
	\lim_{\phi \to \infty} \int_x \dot \phi \, \Gamma_{T,k}^{(1)}[\phi] = -\Delta \dot{\mathcal{C}}_k (\boldsymbol{g})\,.
\label{eq:ConstPhidot}
\end{align}
With \labelcref{eq:GTC} and \labelcref{eq:ConstPhidot} we are led to 
\begin{align}
	\frac{d}{dt}\Gamma_{T,k}[\phi \to  \infty] =    \dot {\mathcal{ C}}_k(\boldsymbol{g})  - \Delta \dot{\mathcal{C}}_k (\boldsymbol{g}) = \partial_t \ln \mathcal{N}_k\,,
\label{eq:deltaCs}
\end{align}
where we have used explicitly that the flow of the target action with $\dot{\mathcal{C}}=0$ is also local, i.e.~$\lim_{\phi \to \infty} \partial_t \Gamma_{T}[\phi,0]=0$.
The cumulants-preserving property requires that all coupling dependences of \labelcref{eq:deltaCs} have to cancel in the subtraction, i.e.~the flow of the normalisation $\mathcal{N}_k$ is independent of $\boldsymbol{g}$.

In conclusion, to verify the cumulants-preserving property for a specific calculation one has to compute the difference in \labelcref{eq:deltaCs} and check, that
\begin{align}
	\int_0^\Lambda \frac{dk}{k} \left(\dot {\mathcal{ C}}_k(\boldsymbol{g})  - \Delta \dot{\mathcal{C}}_k (\boldsymbol{g})\right) = \mathrm{const.}\,,
\label{eq:check}
\end{align}
with a $\boldsymbol{g}$-independent constant on the right-hand side of \labelcref{eq:check}. Finally we remark, that although the coupling dependence in \labelcref{eq:check} has to cancel, the specific value of the single terms $\dot {\mathcal{ C}}_k(\boldsymbol{g}) $ and $\Delta \dot{\mathcal{C}}_k (\boldsymbol{g})$ is not fixed for a specific set of coupling parameters $\boldsymbol{g}$. This freedom opens the gateway for optimisation procedures. The condition used in the present work is discussed in the following Section.

\subsection{Present implementation}
\label{app:choice}

As discussed previously, the cancellation in \labelcref{eq:deltaCs} manifests the freedom we have in choosing $\dot{\mathcal{C}}_k (\boldsymbol{g})$ throughout the flow.

The present work uses an implicit condition to determine $\dot{\mathcal{C}}_k (\boldsymbol{g})$, which allows a straightforward extraction of correlation functions. 
We proceed by illustrating this at the example of the $\int \varphi^2$ cumulants in the $\phi^4$-theory. They are obtained from the total $\boldsymbol{g}$-derivatives on the solution $\phi_\textrm{EoM}$, see \labelcref{eq:varphi2Correlations}.
The flow of \labelcref{eq:varphi2Correlations} is split into two terms, 
\begin{align}
	&\partial_t \left[	\left. \frac{d^n \Gamma_\phi [\phi(\varphi)]}{d (\mu_\varphi)^n}\right|_{\varphi = \varphi_{\mathrm{EoM}}} \right] \notag \\[1ex]
	&\hspace{4mm} = \left. \frac{d^n  }{d (\mu_\varphi)^n} \Big(\partial_t + \partial_t \phi_{\mathrm{EoM}} \,  \partial_\phi \Big)\Gamma_\phi [\phi(\varphi)]\right|_{\varphi = \varphi_{\mathrm{EoM}}}\,,
\label{eq:optC}
\end{align}
where the first term in the second row is the flow of the target action, evaluated on the solution of the composite field. 
The second term corresponds to a shift of the equations of motion. While it can be accommodated, we eliminate it for symmetric target actions for the sake of convenience, i.e.~$\phi_{\mathrm{EoM}}=0$ and thus $\varphi_{\mathrm{EoM}}=0$ at all times, see the calculation in \Cref{app:ClassicalTargetd0}. This fixes $\dot{\cal C}_k$ completely 
\begin{align}
\dot {\mathcal{C}}^{(c)}_k: 	\dot \phi [\phi=0] =0 \,.
\label{eq:NormaliseMap}
\end{align}
We have verified the cumulants-preserving property \labelcref{eq:check} of \labelcref{eq:NormaliseMap} for the zero-dimensional benchmark computation from \Cref{app:ClassicalTargetd0}.
We also note that \labelcref{eq:NormaliseMap} keeps the $\varphi \to - \varphi $ symmetry of the map intact. This is discussed further in \Cref{app:Symmetry}, also in regard to the symmetry broken phase of the O(N) model.

We close this discussion with repeating the remark, that \Cref{eq:NormaliseMap} is not a strict requirement on the field transformation. A detailed investigation of other choices is deferred to future work.

\subsection{Classical target action flow of a free theory}
\label{app:TargetActionFreeTheory}

In this Appendix, we discuss the implementation the classical target action flow \labelcref{eq:FlowGTClassical} in a free theory. In particular, we show the field dependences generated by sub-optimal choices of the constant $\dot{\mathcal{ C}}_k$ \labelcref{eq:NormaliseMap}.
	 
Using the expression \labelcref{eq:Vclassical} for the classical potential with $\mu_\varphi > 0$ and $\lambda = 0$ gives a general formulation for the flowing field transformation of the free theory
\begin{align}
		& 	\dot{\phi}\, \phi \mu_\varphi -\frac{2}{d+2} \dot{\phi}' \frac{A_d k^{d+2}}{k^2 + \mu_\varphi} 
		= \frac{A_d k^{d+2}}{k^2 + \mu_\varphi} 
		- \dot {\cal C}_k\,,
\label{eq:freeTheoryTargetAction}
\end{align}
where we have used the general expression for a target action flow in LPA for a scalar theory with one component \labelcref{eq:LPAtarget}. We parametrise the constant \labelcref{eq:FlowGTClassical} with
\begin{align}
	 \dot {\cal C}_k	= \frac{A_d k^{d+2}}{k^2 + \mu_\varphi}   \Delta c  \,.
\label{eq:Cchoice}
\end{align}
For $\Delta c=1$, \labelcref{eq:freeTheoryTargetAction} is a simple homogeneous ODE with the solution
	\begin{align}
		\dot \phi[\phi] = B_0  \exp \left[c_0 \frac{\phi^2}{2} \right]\,, \qquad c_0 =
		 \frac{(d+2) \mu_\varphi}{2 A_d}  \frac{k^2 + \mu_\varphi}{ k^{d+2} } \,,
\end{align}
where only the choice $B_0 = 0$ satisfies the conditions of locality in \Cref{sec:local}. Consequently, \labelcref{eq:Cchoice} ensures that in a free theory the Wetterich flow and the classical target action flow coincide. Hence, the PIRG flows are able to recover the trivial result in case of the free theory for this choice of the constant $\mathcal{C}_k$. The constraint \labelcref{eq:NormaliseMap} is also fulfilled.

This is, however, not the case for general choices of $\mathcal{C}_k$: In fact one can create a field dependence in a free theory by choosing $\Delta c \neq 1$. \Cref{eq:freeTheoryTargetAction} is again integrated using the locality argument as boundary condition $\dot \phi[\infty] = 0$, and we obtain the solution for general $\Delta c$
\begin{align}
	 	\dot \phi = -\frac{d+2 }{2} \sqrt{\frac{\pi}{2 c_0}} \, &(1- \Delta c) \exp(\frac{c_0  \phi^2}{2})\notag \\[1ex]
	 	&\times  \left(\erf\left[ \sqrt{\frac{c_0}{2}}  \phi \right]-1\right)\,,
\end{align}
which is evidently field-dependent  and also creates a field dependence in the reconstructed potential $V_{\mathrm{cl}}[\phi(\varphi)]$ \labelcref{eq:reconstruction}, within a 0th order derivative expansion.
We show the result for the potential in terms of the integrated composite operator in \Cref{fig:freeTheory} for different choices of the constant $\Delta c$.
Note, that although $V_{\mathrm{cl}}[\phi(\varphi)]$ has non-trivial field dependences, we should be able to reconstruct the free theory from the cumulants via \labelcref{eq:varphi2Correlations}.

\subsection{Symmetry of the field transformation in the classical target action flow}
\label{app:Symmetry}

\begin{figure}[t]
	\centering
	\includegraphics[width=\linewidth]{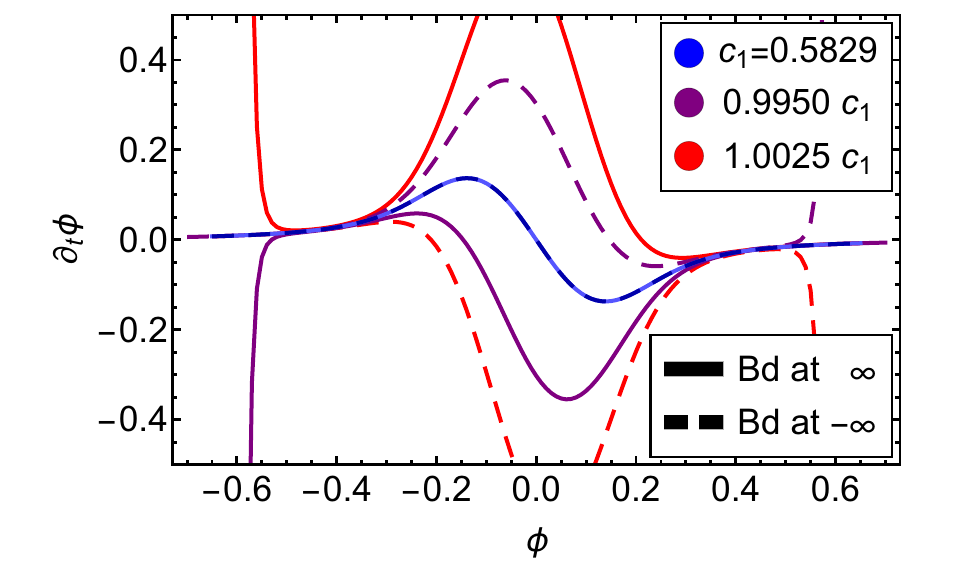}
	\caption{Flowing field transformation $\partial_t \phi =\dot \phi$ for different choices of boundary conditions. One corresponds to the upper integration bound in \labelcref{eq:numSol} which is set to $+ \infty$ (or $- \infty$ for the dashed lines), whereas $c_1$ parametrises the constant part of the flow \labelcref{eq:ChoiceCk}. It implements the choice of constant discussed in \Cref{app:c1fixing}. The blue curve indicates the value of $c_1$ for which $\dot \phi (\phi = \pm \infty) =0 $ and \labelcref{eq:NormaliseMap} are fulfilled. The constant is a scale-dependent quantity; the figure uses $k = 0.5$ as an example for the initial conditions specified in \Cref{sec:ClassicalTarget-ON}. \hspace*{\fill}}
	\label{fig:boundaries} 
\end{figure}

The implicit constraint \labelcref{eq:NormaliseMap} is implemented in all classical target action flows throughout this work. We also find that it is the only constraint which implements the $\phi \to - \phi$ symmetry of the classical action in the map and in the transformation $\dot \phi$.
The flowing field transformation is depicted in \Cref{fig:boundaries} for the choice of the constant which fulfils \labelcref{eq:NormaliseMap} and slightly modified values. The figure shows data from the example discussed in \Cref{sec:ClassicalTarget-ON}.

The optimisation criterion \labelcref{eq:NormaliseMap} is less straightforward in the symmetry broken phase with $\mu_\varphi<0$, where the change of the EoM with the RG-scale is physical and the field value $\phi = 0$ is unphysical. In this case, the flow restores convexity by mapping field values in the unphysical part of the potential to the minimum of $V_{\mathrm{cl}}$ and adjusting the set of allowed values $\mathcal{S}_\phi \neq \mathbbm{R}$.
Nevertheless $\phi = 0$ can still be implemented in the flow, since \labelcref{eq:LPAtarget} does not know about the previous deformation of fields. This has been done in \Cref{sec:ClassicalTarget-ON} and reproduces a solution to the equations of motion which is close to the result from the Wetterich equation. 

A more throughout investigation of optimisation criteria for the choice of normalisation constant goes beyond the scope of the present work and is deferred to the future.

\section{Cumulants from the classical target action in $d=0$}
\label{app:ClassicalTargetd0} 

The aim of this Appendix is to verify the suggested reconstruction mechanisms for correlation functions in \Cref{sec:Reconstruction}, with further insights provided in \Cref{app:c1fixing}. For this purpose we use the zero-dimensional case as a simple but fully conclusive benchmark.

\begin{figure*}
	\centering
	\begin{subfigure}{.48\linewidth}
		\centering
		\includegraphics[width=\textwidth]{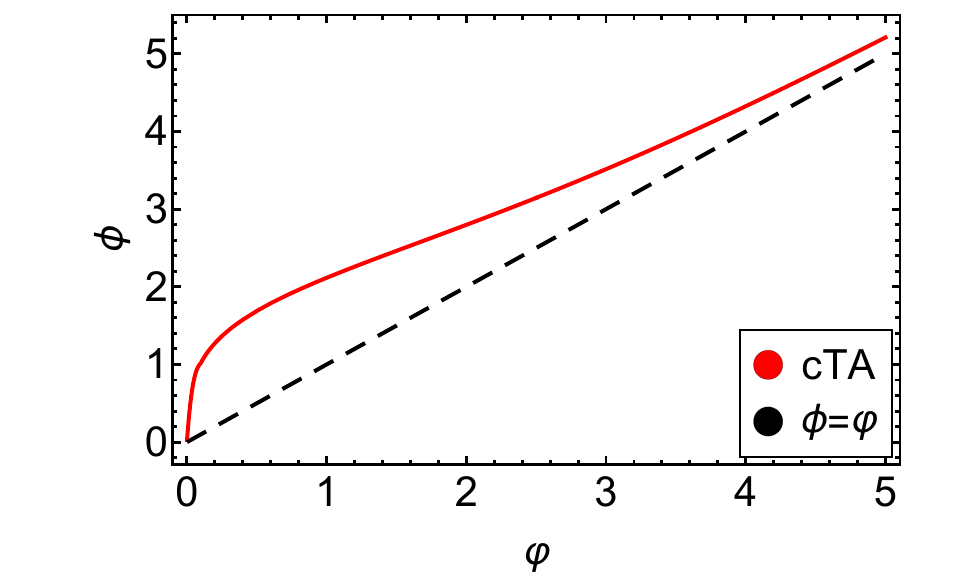}
		\subcaption{Reconstruction of the map for a classical target action. The mapping $\phi = \varphi$ is drawn in dashed black lines for better visibility.\hspace*{\fill}}
		\label{fig:d0map}
	\end{subfigure}%
	\hspace{0.02\linewidth}%
	\begin{subfigure}{.48\linewidth}
		\centering
		\includegraphics[width=\textwidth]{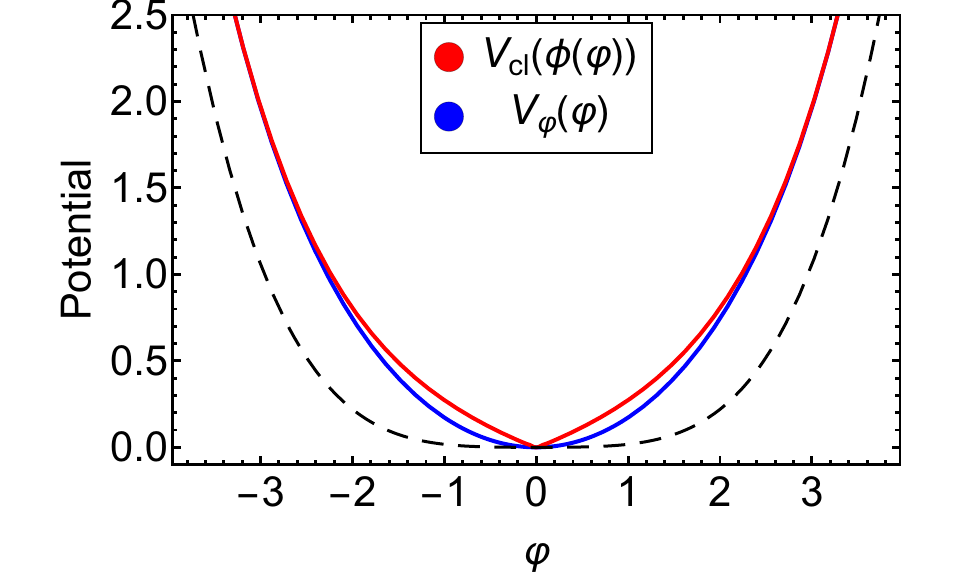}
		\subcaption{Full potential: The solution using a classical target action is compared with the standard solution $V_\varphi$. The classical potential is drawn in dashed black lines for reference. \hspace*{\fill}}
		\label{fig:d0pot}
	\end{subfigure}%
	\caption{Comparison of the \textit{classical target action} flow and the result obtained from a direct integration of the path integral \labelcref{eq:CumulantZd=0} in $d=0$. \hspace*{\fill}} 
	\label{fig:d0}
\end{figure*}
\subsection{Reference solution}
 
To begin with, the integral representation of the generating functions $Z_\phi(\phi)$ can be computed numerically with arbitrary precision. This also allows the computation of the effective action with arbitrary precision. Moreover, for the cumulants we only consider cumulants-preserving $Z^{(c)}_\phi(0)$, \labelcref{eq:carphi2Cumulants}, with a coupling-parameter independent normalisation ${\cal N}^{(c)}$, \labelcref{eq:Nk-muIndep}. For the same normalisation all $Z_\phi(0)$ agree, \labelcref{eq:ZEoM-ID}. Finally, all flows, the standard Wetterich flows and general PIRG flows can be solved without truncation artefacts as LPA is exact. 

All these properties originate in the fact that in $d=0$ dimensions the generating functional $Z_\phi[J_\phi]$ reduces to a generating function. For our example case of a $\phi^4$-theory with the classical action \labelcref{eq:ClassicalAction1-ON} the generating function is given by a one-dimensional integral, 
\begin{align}
	Z_\phi^{(c)}(J_\phi) = \int d \hat \varphi \,  e^{-S[ \hat \varphi] + J_\phi \hat \phi } \quad \textrm{with}\quad {\cal N}^{(c)} =1\,, 
	\label{eq:CumulantZd=0}
\end{align}
where general composites $\hat\phi[\hat\varphi]$ are coupled to the current $J_\phi$. The trivial normalisation $ {\cal N}^{(c)} =1$ implements the cumulants-preserving property \labelcref{eq:Nk-muIndep}, and \labelcref{eq:CumulantZd=0} encompasses the standard integral $Z_\varphi^{(c)}$. 

All normalised correlation functions of the fundamental field $\hat\varphi$ can be computed readily from $J_\varphi$-derivatives of the generating function with, 
\begin{align} 
\langle \hat\varphi^{2 n}\rangle 	 =  \frac{1}{Z_\varphi(0)} \left. \frac{d^{2n} Z_\varphi(J_\varphi)}{d J_\varphi^{2n}}\right|_{J_\varphi =0}\,, 
\label{eq:Correlations-varphi}
\end{align}
while the odd order correlation functions vanish for the action \labelcref{eq:ClassicalAction1-ON}. Evidently, all correlation functions simply require the computation of even moments of the $\hat\varphi$- integral \labelcref{eq:CumulantZd=0}, 
\begin{align}\nonumber 
	\int d \hat \varphi \, \hat\varphi^{2 n}  e^{-S[ \hat \varphi] } =&\, (-2)^{2 n}  \frac{d^{2 n} \int d \hat \varphi \, e^{-S[ \hat \varphi] } } {d\mu_\varphi^{2 n}} \\[1ex]
	=&\, (-2)^{2 n} \frac{d^{2 n} Z_\phi^{(c)}(0)} {d\mu_\varphi^{2 n}} \,.
	\label{eq:Correlations-Cumulants}
\end{align}
\Cref{eq:Correlations-Cumulants} relates the correlation functions to the cumulants, and the cumulants can be computed from $\boldsymbol{g}$-derivatives of $\log Z^{(c)}_\phi(0)$ with $\boldsymbol{g}=(\mu_\varphi,\lambda_\varphi)$. For the classical $\phi^4$-action, $Z^{(c)}_\phi(0)$ can be computed analytically and we find for $\mu_\phi>0$, 
\begin{align} 
	Z_\phi^{(c)}(0) = \sqrt{\frac{4 x}{\mu_\varphi}}\,
	K_{-\frac{1}{4}}\left(x\right)\,e^{x}\,, \quad\textrm{with} \quad x=\frac{\mu_\varphi ^2}{4 \lambda_\varphi }\,,
	\label{eq:Z0Bessel}
\end{align} 
valid for all $\phi$. A similar equation holds for $\mu_\phi<0$, but for the benchmark and illustration purposes of the present Appendix we restrict ourselves to convex classical target actions with $\mu_\varphi>0$. Additionally, we fix the normalisation constant such that the solution to the equation of motion remain unchanged in the coordinate map: We chose the constraint \labelcref{eq:NormaliseMap} and hence 
\begin{align}
	\phi_\textrm{EoM}=0\,.
\end{align}
The explicit results for the first and second cumulants are computed for the coupling parameter $\boldsymbol{g}$ with 
\begin{align}
	\lambda_\varphi=0.1 \,, \quad \quad \mu_\varphi \in(0, 0.2]\,, 
	\label{eq:gInd=0}
\end{align}
and we obtain from \labelcref{eq:Z0Bessel}
\begin{align}\nonumber 
\langle \hat\varphi^2\rangle^{(c)}= &\, -2 \,\frac{d \log Z_\phi^{(c)}(0)}{d\mu_\varphi} \\[1ex]
=&\, \frac{4}{\mu_\varphi} \left[x \left( \frac{ K_{\frac{5}{4}}\left(x\right)}{ K_{\frac{1}{4}}\left(x\right)}-1\right) -\frac12\right]\,, 
\label{eq:1ndCumulant}
\end{align}
for the first cumulant and
\begin{align}\nonumber 
\langle (\hat\varphi^2)^2\rangle^{(c)}=	&\,4\, \frac{d^2 \log \left[Z_\phi^{(c)}(0)\right]}{d\mu_\varphi^2}\\[1ex]
 &\hspace{-1.2cm}= -\frac{2}{\mu_\varphi } \langle \hat\varphi^2\rangle^{(c)}- \frac{8 x}{\mu_\varphi}\frac{ K_{\frac{5}{4}}\left(x\right)}{ K_{\frac{1}{4}}\left(x\right)} \partial_x \log \frac{K_{\frac{5}{4}}\left(x\right)}{K_{\frac{1}{4}}\left(x\right)}\,,
\label{eq:2ndCumulant}
\end{align}
for the second cumulant.

The analytic results \labelcref{eq:1ndCumulant,eq:2ndCumulant} are compared with those computed from the $\mu_\varphi$-derivatives of the classical target action $\Gamma_\phi^{(c)}(0)$, which is numerically computed with PIRG flows in the following.

\subsection{Benchmark with PIRG flows}

In the present case we consider cumulants-preserving variants of classical target action, recall the discussion in \Cref{app:c1fixing}. In particular we implement \labelcref{eq:NormaliseMap}, for which we have verified \labelcref{eq:check}. The computation is initiated at $\Lambda = \exp(5)\approx 150$ and the RG-time is integrated up to $t = -15$, corresponding to $k_{\mathrm{fin}} = 4 \times 10^{-5}$.
The map $\phi(\varphi)$ is reconstructed using \labelcref{eq:phi-varphi} and depicted in \Cref{fig:d0map}. Using the map we can also depict the potential in terms of the original field $V_{\mathrm{cl}}(\phi(\varphi))$. This is shown in \Cref{fig:d0pot}, where we have explicitly omitted the constant shift in the target action \labelcref{eq:GTClassical} for better visibility. One can clearly discern \labelcref{eq:gammaNotGamma} in the Figure: The solutions for the effective actions are genuinely different for different composites $\hat \phi$.

In the following, we use \labelcref{eq:carphi2Cumulants} to compute the mass at $k=0$ from the PIRG flow result and compare it with the analytic result derived from \labelcref{eq:Z0Bessel}. Moreover, apart from their relation to the moments of the path integral measure, the cumulants can be used to directly compute observables. For example, in $d=0$ the inverse mass squared is simply given by the two-point function, and hence the first order cumulant, 
\begin{align}
\frac{1}{m^2} = \langle \left(\hat\varphi^2\right)\rangle^{(c)} = 2   \frac{d V^{(c)}_\phi(\phi_\textrm{EoM}) }{d \mu_\varphi} \,. 
\label{eq:massd0}
\end{align}
Using \labelcref{eq:GTClassical}, the target action follows as
\begin{align}
	V^{(c)}_\phi(\phi) =  V_{\mathrm{cl}}(\phi)  + \mathcal{ C}^{(c)}_0(\boldsymbol{g}) \,. 
\end{align}
In particular we have
\begin{align}
V^{(c)}_{\phi}(\phi_{\mathrm{EoM}}) =  \mathcal{ C}^{(c)}_0(\boldsymbol{g}) \,,
\label{eq:DeltaV}
\end{align}
due to our choice of constant \labelcref{eq:NormaliseMap}, which implements $\phi_{\mathrm{EoM}}=0$ at all scales.
This finalises the setup for computations of cumulants. We emphasise that none of the steps was specific to $d=0$ and the setup can readily be used in general theories in general dimensions. We restrict ourselves to $d=0$ as it offers an analytic benchmark. 

The benchmark is performed in the range \labelcref{eq:gInd=0}. Here we compute $\mathcal{ C}_0^{(c)}(\boldsymbol{g})$ on a discrete $\mu_\varphi$ grid, with stencils spaced by $\Delta \mu_\varphi = 0.0005 $. Next, the data is interpolated using a higher order interpolation function in Mathematica. Since the function does not contain any non-analyticities, we can evaluate precise derivatives and compute the mass $m^2$ and the second cumulant $\langle (\hat\varphi^2)^2\rangle^{(c)}_\textrm{cTA}$ from \labelcref{eq:varphi2Correlations}. The result is depicted in the main text, \Cref{fig:d0correlations}, and matches the analytic result \labelcref{eq:1ndCumulant} and \labelcref{eq:2ndCumulant} within its numerical precision.

In the following, we give a pedagogic example where we compute the cumulants for a single mass parameter $\mu_\varphi$. Since we generate less data-points in $\mu_\varphi$, the $\mu_\varphi$-derivatives will be less precise than those depicted in \Cref{fig:d0correlations} and include a large numerical error from the finite difference scheme used: Using \labelcref{eq:DeltaV} and $\phi_\textrm{EoM}=0$ in \labelcref{eq:massd0} yields the following expressions for the cumulants
\begin{align}
\frac{d V^{(c)}_\phi(0) }{d \mu_\varphi} =  \frac{\mathcal{ C}^{(c)}_0(\mu_\varphi^+,\lambda_\varphi) -\mathcal{C}^{(c)}_0(\mu_\varphi^- \lambda_\varphi)}{2 \Delta \mu_\varphi}\,,
\label{eq:massTocorrelation}
\end{align}
with 
\begin{align} 
	\mu_\varphi^\pm =\mu_\varphi  \pm  \Delta \mu_\varphi\,.
\label{eq:mupm}
\end{align}
The second order cumulant can be computed by the second order derivative w.r.t.~$\mu_\varphi$ and we have used a symmetric discrete quadratic derivative 
\begin{align}
	\frac{d^2 V^{(c)}_\phi(0) }{d \mu_\varphi^2} =  \frac{\mathcal{ C}^{(c)}_0(\mu_\varphi^+,\lambda_\varphi) +\mathcal{C}^{(c)}_0(\mu_\varphi^-,\lambda_\varphi)-2\mathcal{ C}^{(c)}_0(\mu_\varphi ,\lambda_\varphi) }{ \Delta \mu_\varphi^2}\,.
	\label{eq:KurtosiscTA}
\end{align}
For this simple setup with discrete derivatives, using a classical target action with $\mu_\varphi=0.01$ and $\lambda_\varphi=0.1$, we obtain the mass and second cumulant
\begin{align}
	m^2_{\mathrm{cTA}} = 0.33674(4) \,, \qquad \langle (\hat\varphi^2)^2\rangle^{(c)}_\textrm{cTA} =10.4(2)\,,
	\label{eq:mcTA+KurtosiscTA}
\end{align}
where the error is obtained by considering the results for spacings $\Delta \mu_\varphi = 0.0005 \pm 0.0002$.

\Cref{eq:mcTA+KurtosiscTA} is corroborated with the analytic results obtained from \labelcref{eq:1ndCumulant,eq:2ndCumulant}, 
 \begin{align}
 	m^2= 0.336763\,, \qquad  \langle (\hat\varphi^2)^2\rangle^{(c)}=10.5885 \,.
 	\label{eq:1+2CumulantAnalytic}
 \end{align}
In conclusion we have shown that PIRG flows with even classical target actions can be used to compute relevant observables in the underlying quantum theory. We emphasise that we also could have used the standard flow equation to compute the full mass $m^2$ and $\langle (\hat\varphi^2)^2\rangle^{(c)}$ as a reference solution. This requires solving a partial differential equation numerically with qualitatively larger computational costs. In our opinion this illustrates very impressively the great potential of PIRG flows. 
The accuracy of the target action results may be improved even further by replacing the finite difference/interpolation function derivatives with an analytical flow comparable to the one used in \cite{Faigle-Cedzich:2023rxd}.

\section{Method of characteristics vs. classical target action flows in the large N limit}
\label{app:LargeN}

We give a brief outline of how the method of characteristics is used to obtain an analytic solution for LPA in the large N limit. This is followed by a discussion of the flowing fields solution in the same limit.

The Wetterich flow of the potential $V_{\infty}$ in the large N limit (i.e.~$N \to \infty$) reads
\begin{align}
	\partial_t V_\infty(\rho) = \frac{A_d k^{d+2}}{k^2 + V_\infty'(\rho)} + \mathcal{C}_k \,,
\end{align}
where we have suggestively added a constant $\mathcal{C}_k$ to show the similarity to the general flow \labelcref{eq:GenFlow}. 

By taking a derivative with respect to $\rho$, the constant is immediately removed in the derivation. The equation for $u_\infty = \partial_\rho V_\infty$ now reads
\begin{align}
	\partial_t u_{\infty} + \frac{A_d k^{d+2}}{(k^2 + u_{\infty})^2}\partial_\rho u_{\infty} = 0\,,
\label{eq:dtu}
\end{align}
which is a quasi-linear partial differential equation in $u_\infty$. Given \labelcref{eq:dtu}, a characteristic is a parametrised curve $(t, \rho(t))$ on which $u_\infty$ stays constant. The characteristic can be obtained from
\begin{align}
	\rho(k, \rho_\Lambda) = \rho_\Lambda - A_d  \int_{k}^{\Lambda} \frac{dk'}{k'} \frac{k'^{d+2}}{(k'^2 + u_\infty (\rho_\Lambda))^2} \,.
\label{eq:mofcharMap}
\end{align}
Using the initial condition at $t = 0$, the full solution is then given by
\begin{align}
	u_\infty(t, \rho(t,\rho_\Lambda)) = V_{\mathrm{cl}}^{(1)}(\rho_\Lambda) \,,
\end{align}
i.e.~the full solution is simply a deformation of the classical potential $V_{\mathrm{cl}}$. Importantly, one finds that there is no solution at $k\to 0$ for $\rho_\Lambda < \rho_{\Lambda,\mathrm{EoM}}$, where the latter indicates the classical equations of motion. This corresponds to the flattening of the potential.

The PIRG flows with a classical target action use a similar idea, but include a slight change of perspective. By using the classical action as target action $\Gamma_{T,k} \equiv S_\Lambda$ the fields are genuinely changed with the RG-time integration. 

We begin by considering the target action flow for general O(N) theories in a zeroth order derivative expansion for a flowing field without derivative terms, i.e.~\labelcref{eq:phidot0}. In the notation of \labelcref{eq:dotPhiDerExpCoeffs} the field transformation is given by
\begin{align}
	\dot \phi^a = \phi^a f_0(\rho)\,, \quad \frac{\delta \dot \phi^a}{\delta \phi_a} = 
	\left(f_0 + 2 \rho f_0', f_0, \dots , f_0 \right) \,.
	\label{eq:dotphilargeN}
\end{align}
Generally $\frac{\delta \dot \phi}{\delta \phi}$ also contains off-diagonal elements, which are not evaluated in the flow of the potential.
The general classical target action flow for O(N) theories reads
\begin{align}
	& 	2 \rho f_0 \, V_T' -\frac{2}{d+2}A_d k^{d+2} \left[\frac{ f_0 + 2 \rho f_0'}{k^2 + V_T' + 2 \rho V_T''} + \frac{(N-1) f_0}{k^2 + V_T'} \right] \notag\\[1ex]
	&\hspace{6mm}= A_d k^{d+2} \left[\frac{1}{k^2 + V_T' + 2 \rho V_T''}  +\frac{N-1}{k^2 + V_T'}  \right]
	- \partial_t V_T \,.
	\label{eq:LPAtargetON} 
\end{align}
The large N limit can be taken by using the rescaling
\begin{align}
	\rho \to N \rho\,, \quad \mathrm{and} \quad V_T \to N V_T \,, 
\end{align}
dividing by $N$ and subsequently taking the limit $N \to \infty$.
We obtain the large N target action flow
\begin{align}
		2 \rho f_0 \, V_T' -\frac{2 f_0}{d+2}  \frac{A_d k^{d+2}}{k^2 + V_T'}=  \frac{A_d k^{d+2}}{k^2 + V_T'}  
	- \partial_t V_T \,,
	\label{eq:LPAtargetLN} 
\end{align}
which is an algebraic equation in $f_0$.
We immediately find that for the classical target action flow, i.e.~constant $ \partial_t V_T $, $f_0$ has a non-trivial singularity where $ 2 \rho V_T' -\frac{2}{d+2}  \frac{A_d k^{d+2}}{k^2 + V_T'}=0$.
This singularity can be lifted by choosing $\partial_t V_T $ accordingly
\begin{align}
 f_0 = -\frac{d+2}{2}\frac{ 1 
		- \frac{k^2 + V_T'(\rho)}{k^2 + V_T'(\rho_0)}}{1-	\frac{d+2}{A_d k^{d+2}} \rho V_T'(\rho)(k^2 + V_T'(\rho)) }\,,
	\label{eq:LPAtargetLNfixed} 
\end{align}
where $\rho_0$ is the value of the field at which the denominator becomes singular.
\Cref{eq:LPAtargetLNfixed} is finite for $\rho \to 0$. Together with the parametrisation \labelcref{eq:dotphilargeN} this implements the criterion discussed in \Cref{app:c1fixing}. 
An analysis of the $\rho \to \infty$ behaviour in \labelcref{eq:LPAtargetLNfixed} shows a decay $f_0 \propto \rho^{-2}$, which manifests the locality of the transformation.

For a direct comparison to \labelcref{eq:mofcharMap} we use the definition of the invariant $\rho = \phi^2/2$ and of $\dot \phi$ \labelcref{eq:dotphilargeN} to define the flow of $\rho$
\begin{align}
	\dot \rho = \phi^a \dot \phi^a = 2 \rho f_0 \,.
\end{align}
This yields the map
\begin{align}
	\rho_k = \rho_\Lambda - \int_{k}^{\Lambda} \frac{dk'}{k'} 2 \rho_k' f_0(\rho_k') \,.
\end{align}
An interesting difference between both approaches is that \labelcref{eq:mofcharMap} uses the \textit{classical/original} field in the computation of the transformation, whereas the target action uses the flowing field.

This is due to the fact that the method of characteristics uses the classical field $\rho_\Lambda$ as a computational tool, whereas in the present target action setup the flowing field $\rho_k$ is the tool and the original field $\rho_\Lambda$ is the link to the fundamental field and thus the physical observables.

\section{Regulator}
\label{app:thrs+reg}

The present work uses a simple flat or Litim regulator, \cite{Litim:2001up}. It is diagonal in field space and its entries are given by
\begin{align}
	R_{k} = \left(k^2 - p^2 \right) \Theta(1-p^2/k^2) \,,
\label{eq:reg}
\end{align}
This regulator choice turns the evaluation of momentum loops analytical.

\section{Feed-down flows in the first order of the derivative expansion for O(N)-theories}
\label{app:Gen1stOrder}

Feed-down flows within the first order derivative expansion for general $N>1$ have to satisfy the constraint \labelcref{eq:f1-barf1}. The first order part, $\dot \phi_1$, is parametrised in \labelcref{eq:dotPhiDerExpCoeffs} and the coefficients $L\,,\,\bar L$ in the constraint \labelcref{eq:f1-barf1} are given by 
\begin{align}
	\Omega_d \int_0^{k} \,{d q} q^{d-1}  G_W^{ab}[\phi,q^2]\, R_{\phi}(q^2) = L(\rho) \,\delta^{ab}  + \bar L(\rho) \,\phi^a \phi^b\,.
\end{align}
We use the flat or Litim regulator \labelcref{eq:reg} with the propagator 
\begin{align}\nonumber 
	G_W^{ab} [\phi,q^2\leq k^2] =&\,\left( \delta^{ab}-\frac{ \phi^a \phi^b}{ 2 \rho }\right) \frac{1}{k^2 + V_W'(\rho) } \\[1ex] 
	&\hspace{-.4cm}+\frac{\phi^a \phi^b}{2 \rho}\, \frac{1}{k^2 + V'_W(\rho) + 2 \rho V''_W(\rho)}\,.
	\label{eq:LPAGphi}  
\end{align}
This leads to 
\begin{align} \nonumber 
	L(\rho) =&\,  \frac{\Omega_d}{d+2} \frac{k^{d+2}}{k^2 + \mu_\varphi + \lambda_\varphi\rho}\,,\\[1ex] 
	\bar L(\rho) =&\,\frac{1}{2 \rho} \left[  \frac{\Omega_d}{d+2} 
	\frac{k^{d+2}}{k^2 + \mu_\varphi + 3 \lambda_\varphi\rho}-  L(\rho)\right] \,.
\end{align}
and the constraint \labelcref{eq:f1-barf1} reads explicitly 
\begin{align}
	2 \rho\,\bar f_1 =- f_1 \, \left[1 -(N-1) \frac{ k^2 + \mu_\varphi + \lambda_\varphi \rho}{k^2 + \mu_\varphi + 3 \lambda_\varphi \rho}\right]\,. 
	\label{eq:NoFeedback}
\end{align}
For $N=1$ this reduces to $\tilde f_1 =0$, see \labelcref{eq:f1-barf1O1}. With \labelcref{eq:NoFeedback}, the generalised flow for the effective potential reads
\begin{align}
	\partial_t V_W = \frac12 \textrm{Tr}\, G_W^{ab} \partial_t R^{ab}_k\,,
\end{align}
which is the Wetterich equation in LPA. The full first order of the derivative expansion in the target action is obtained by the constraint equation for $\dot\phi_1$. Its resolution is straightforward but beyond the scope of the present work.

	\begingroup
	\allowdisplaybreaks

\newpage
	
	\bibliographystyle{apsrev4-2}
	\bibliography{ref_lib}

\begin{thebibliography}{65}%
\makeatletter
\providecommand \@ifxundefined [1]{%
 \@ifx{#1\undefined}
}%
\providecommand \@ifnum [1]{%
 \ifnum #1\expandafter \@firstoftwo
 \else \expandafter \@secondoftwo
 \fi
}%
\providecommand \@ifx [1]{%
 \ifx #1\expandafter \@firstoftwo
 \else \expandafter \@secondoftwo
 \fi
}%
\providecommand \natexlab [1]{#1}%
\providecommand \enquote  [1]{``#1''}%
\providecommand \bibnamefont  [1]{#1}%
\providecommand \bibfnamefont [1]{#1}%
\providecommand \citenamefont [1]{#1}%
\providecommand \href@noop [0]{\@secondoftwo}%
\providecommand \href [0]{\begingroup \@sanitize@url \@href}%
\providecommand \@href[1]{\@@startlink{#1}\@@href}%
\providecommand \@@href[1]{\endgroup#1\@@endlink}%
\providecommand \@sanitize@url [0]{\catcode `\\12\catcode `\$12\catcode
  `\&12\catcode `\#12\catcode `\^12\catcode `\_12\catcode `\%12\relax}%
\providecommand \@@startlink[1]{}%
\providecommand \@@endlink[0]{}%
\providecommand \url  [0]{\begingroup\@sanitize@url \@url }%
\providecommand \@url [1]{\endgroup\@href {#1}{\urlprefix }}%
\providecommand \urlprefix  [0]{URL }%
\providecommand \Eprint [0]{\href }%
\providecommand \doibase [0]{https://doi.org/}%
\providecommand \selectlanguage [0]{\@gobble}%
\providecommand \bibinfo  [0]{\@secondoftwo}%
\providecommand \bibfield  [0]{\@secondoftwo}%
\providecommand \translation [1]{[#1]}%
\providecommand \BibitemOpen [0]{}%
\providecommand \bibitemStop [0]{}%
\providecommand \bibitemNoStop [0]{.\EOS\space}%
\providecommand \EOS [0]{\spacefactor3000\relax}%
\providecommand \BibitemShut  [1]{\csname bibitem#1\endcsname}%
\let\auto@bib@innerbib\@empty
\bibitem [{\citenamefont {Ihssen}\ and\ \citenamefont
  {Pawlowski}(2023{\natexlab{a}})}]{Ihssen:2022xjv}%
  \BibitemOpen
  \bibfield  {author} {\bibinfo {author} {\bibfnamefont {F.}~\bibnamefont
  {Ihssen}}\ and\ \bibinfo {author} {\bibfnamefont {J.~M.}\ \bibnamefont
  {Pawlowski}},\ }\href {https://doi.org/10.21468/SciPostPhys.15.2.074}
  {\bibfield  {journal} {\bibinfo  {journal} {SciPost Phys.}\ }\textbf
  {\bibinfo {volume} {15}},\ \bibinfo {pages} {074} (\bibinfo {year}
  {2023}{\natexlab{a}})},\ \Eprint {https://arxiv.org/abs/2207.10057}
  {arXiv:2207.10057 [hep-th]} \BibitemShut {NoStop}%
\bibitem [{\citenamefont {Ihssen}\ and\ \citenamefont
  {Pawlowski}(2023{\natexlab{b}})}]{Ihssen:2023nqd}%
  \BibitemOpen
  \bibfield  {author} {\bibinfo {author} {\bibfnamefont {F.}~\bibnamefont
  {Ihssen}}\ and\ \bibinfo {author} {\bibfnamefont {J.~M.}\ \bibnamefont
  {Pawlowski}},\ }\href@noop {} {\  (\bibinfo {year} {2023}{\natexlab{b}})},\
  \Eprint {https://arxiv.org/abs/2305.00816} {arXiv:2305.00816 [hep-th]}
  \BibitemShut {NoStop}%
\bibitem [{\citenamefont {Dupuis}\ \emph {et~al.}(2021)\citenamefont {Dupuis},
  \citenamefont {Canet}, \citenamefont {Eichhorn}, \citenamefont {Metzner},
  \citenamefont {Pawlowski}, \citenamefont {Tissier},\ and\ \citenamefont
  {Wschebor}}]{Dupuis:2020fhh}%
  \BibitemOpen
  \bibfield  {author} {\bibinfo {author} {\bibfnamefont {N.}~\bibnamefont
  {Dupuis}}, \bibinfo {author} {\bibfnamefont {L.}~\bibnamefont {Canet}},
  \bibinfo {author} {\bibfnamefont {A.}~\bibnamefont {Eichhorn}}, \bibinfo
  {author} {\bibfnamefont {W.}~\bibnamefont {Metzner}}, \bibinfo {author}
  {\bibfnamefont {J.~M.}\ \bibnamefont {Pawlowski}}, \bibinfo {author}
  {\bibfnamefont {M.}~\bibnamefont {Tissier}},\ and\ \bibinfo {author}
  {\bibfnamefont {N.}~\bibnamefont {Wschebor}},\ }\href
  {https://doi.org/10.1016/j.physrep.2021.01.001} {\bibfield  {journal}
  {\bibinfo  {journal} {Phys. Rept.}\ }\textbf {\bibinfo {volume} {910}},\
  \bibinfo {pages} {1} (\bibinfo {year} {2021})},\ \Eprint
  {https://arxiv.org/abs/2006.04853} {arXiv:2006.04853 [cond-mat.stat-mech]}
  \BibitemShut {NoStop}%
\bibitem [{\citenamefont {Ihssen}\ \emph
  {et~al.}(2024{\natexlab{a}})\citenamefont {Ihssen}, \citenamefont
  {Pawlowski}, \citenamefont {Sattler},\ and\ \citenamefont
  {Wink}}]{Ihssen:2024miv}%
  \BibitemOpen
  \bibfield  {author} {\bibinfo {author} {\bibfnamefont {F.}~\bibnamefont
  {Ihssen}}, \bibinfo {author} {\bibfnamefont {J.~M.}\ \bibnamefont
  {Pawlowski}}, \bibinfo {author} {\bibfnamefont {F.~R.}\ \bibnamefont
  {Sattler}},\ and\ \bibinfo {author} {\bibfnamefont {N.}~\bibnamefont
  {Wink}},\ }\href@noop {} {\  (\bibinfo {year} {2024}{\natexlab{a}})},\
  \Eprint {https://arxiv.org/abs/2408.08413} {arXiv:2408.08413 [hep-ph]}
  \BibitemShut {NoStop}%
\bibitem [{\citenamefont {Wegner}(1974)}]{Wegner_1974}%
  \BibitemOpen
  \bibfield  {author} {\bibinfo {author} {\bibfnamefont {F.~J.}\ \bibnamefont
  {Wegner}},\ }\href {https://doi.org/10.1088/0022-3719/7/12/004} {\bibfield
  {journal} {\bibinfo  {journal} {Journal of Physics C: Solid State Physics}\
  }\textbf {\bibinfo {volume} {7}},\ \bibinfo {pages} {2098} (\bibinfo {year}
  {1974})}\BibitemShut {NoStop}%
\bibitem [{\citenamefont {Pawlowski}(2007)}]{Pawlowski:2005xe}%
  \BibitemOpen
  \bibfield  {author} {\bibinfo {author} {\bibfnamefont {J.~M.}\ \bibnamefont
  {Pawlowski}},\ }\href {https://doi.org/10.1016/j.aop.2007.01.007} {\bibfield
  {journal} {\bibinfo  {journal} {Annals Phys.}\ }\textbf {\bibinfo {volume}
  {322}},\ \bibinfo {pages} {2831} (\bibinfo {year} {2007})},\ \Eprint
  {https://arxiv.org/abs/hep-th/0512261} {arXiv:hep-th/0512261} \BibitemShut
  {NoStop}%
\bibitem [{\citenamefont {Gies}\ and\ \citenamefont
  {Wetterich}(2002)}]{Gies:2001nw}%
  \BibitemOpen
  \bibfield  {author} {\bibinfo {author} {\bibfnamefont {H.}~\bibnamefont
  {Gies}}\ and\ \bibinfo {author} {\bibfnamefont {C.}~\bibnamefont
  {Wetterich}},\ }\href {https://doi.org/10.1103/PhysRevD.65.065001} {\bibfield
   {journal} {\bibinfo  {journal} {Phys. Rev. D}\ }\textbf {\bibinfo {volume}
  {65}},\ \bibinfo {pages} {065001} (\bibinfo {year} {2002})},\ \Eprint
  {https://arxiv.org/abs/hep-th/0107221} {arXiv:hep-th/0107221} \BibitemShut
  {NoStop}%
\bibitem [{\citenamefont {Floerchinger}\ and\ \citenamefont
  {Wetterich}(2009)}]{Floerchinger:2009uf}%
  \BibitemOpen
  \bibfield  {author} {\bibinfo {author} {\bibfnamefont {S.}~\bibnamefont
  {Floerchinger}}\ and\ \bibinfo {author} {\bibfnamefont {C.}~\bibnamefont
  {Wetterich}},\ }\href {https://doi.org/10.1016/j.physletb.2009.09.014}
  {\bibfield  {journal} {\bibinfo  {journal} {Phys. Lett. B}\ }\textbf
  {\bibinfo {volume} {680}},\ \bibinfo {pages} {371} (\bibinfo {year}
  {2009})},\ \Eprint {https://arxiv.org/abs/0905.0915} {arXiv:0905.0915
  [hep-th]} \BibitemShut {NoStop}%
\bibitem [{\citenamefont {Gies}\ and\ \citenamefont
  {Wetterich}(2004)}]{Gies:2002hq}%
  \BibitemOpen
  \bibfield  {author} {\bibinfo {author} {\bibfnamefont {H.}~\bibnamefont
  {Gies}}\ and\ \bibinfo {author} {\bibfnamefont {C.}~\bibnamefont
  {Wetterich}},\ }\href {https://doi.org/10.1103/PhysRevD.69.025001} {\bibfield
   {journal} {\bibinfo  {journal} {Phys. Rev. D}\ }\textbf {\bibinfo {volume}
  {69}},\ \bibinfo {pages} {025001} (\bibinfo {year} {2004})},\ \Eprint
  {https://arxiv.org/abs/hep-th/0209183} {arXiv:hep-th/0209183} \BibitemShut
  {NoStop}%
\bibitem [{\citenamefont {Braun}(2009)}]{Braun:2009ewx}%
  \BibitemOpen
  \bibfield  {author} {\bibinfo {author} {\bibfnamefont {J.}~\bibnamefont
  {Braun}},\ }\href {https://doi.org/10.1140/epjc/s10052-009-1136-6} {\bibfield
   {journal} {\bibinfo  {journal} {Eur. Phys. J. C}\ }\textbf {\bibinfo
  {volume} {64}},\ \bibinfo {pages} {459} (\bibinfo {year} {2009})},\ \Eprint
  {https://arxiv.org/abs/0810.1727} {arXiv:0810.1727 [hep-ph]} \BibitemShut
  {NoStop}%
\bibitem [{\citenamefont {Mitter}\ \emph {et~al.}(2015)\citenamefont {Mitter},
  \citenamefont {Pawlowski},\ and\ \citenamefont
  {Strodthoff}}]{Mitter:2014wpa}%
  \BibitemOpen
  \bibfield  {author} {\bibinfo {author} {\bibfnamefont {M.}~\bibnamefont
  {Mitter}}, \bibinfo {author} {\bibfnamefont {J.~M.}\ \bibnamefont
  {Pawlowski}},\ and\ \bibinfo {author} {\bibfnamefont {N.}~\bibnamefont
  {Strodthoff}},\ }\href {https://doi.org/10.1103/PhysRevD.91.054035}
  {\bibfield  {journal} {\bibinfo  {journal} {Phys. Rev. D}\ }\textbf {\bibinfo
  {volume} {91}},\ \bibinfo {pages} {054035} (\bibinfo {year} {2015})},\
  \Eprint {https://arxiv.org/abs/1411.7978} {arXiv:1411.7978 [hep-ph]}
  \BibitemShut {NoStop}%
\bibitem [{\citenamefont {Braun}\ \emph {et~al.}(2016)\citenamefont {Braun},
  \citenamefont {Fister}, \citenamefont {Pawlowski},\ and\ \citenamefont
  {Rennecke}}]{Braun:2014ata}%
  \BibitemOpen
  \bibfield  {author} {\bibinfo {author} {\bibfnamefont {J.}~\bibnamefont
  {Braun}}, \bibinfo {author} {\bibfnamefont {L.}~\bibnamefont {Fister}},
  \bibinfo {author} {\bibfnamefont {J.~M.}\ \bibnamefont {Pawlowski}},\ and\
  \bibinfo {author} {\bibfnamefont {F.}~\bibnamefont {Rennecke}},\ }\href
  {https://doi.org/10.1103/PhysRevD.94.034016} {\bibfield  {journal} {\bibinfo
  {journal} {Phys. Rev. D}\ }\textbf {\bibinfo {volume} {94}},\ \bibinfo
  {pages} {034016} (\bibinfo {year} {2016})},\ \Eprint
  {https://arxiv.org/abs/1412.1045} {arXiv:1412.1045 [hep-ph]} \BibitemShut
  {NoStop}%
\bibitem [{\citenamefont {Rennecke}(2015)}]{Rennecke:2015eba}%
  \BibitemOpen
  \bibfield  {author} {\bibinfo {author} {\bibfnamefont {F.}~\bibnamefont
  {Rennecke}},\ }\href {https://doi.org/10.1103/PhysRevD.92.076012} {\bibfield
  {journal} {\bibinfo  {journal} {Phys. Rev. D}\ }\textbf {\bibinfo {volume}
  {92}},\ \bibinfo {pages} {076012} (\bibinfo {year} {2015})},\ \Eprint
  {https://arxiv.org/abs/1504.03585} {arXiv:1504.03585 [hep-ph]} \BibitemShut
  {NoStop}%
\bibitem [{\citenamefont {Cyrol}\ \emph {et~al.}(2018)\citenamefont {Cyrol},
  \citenamefont {Mitter}, \citenamefont {Pawlowski},\ and\ \citenamefont
  {Strodthoff}}]{Cyrol:2017ewj}%
  \BibitemOpen
  \bibfield  {author} {\bibinfo {author} {\bibfnamefont {A.~K.}\ \bibnamefont
  {Cyrol}}, \bibinfo {author} {\bibfnamefont {M.}~\bibnamefont {Mitter}},
  \bibinfo {author} {\bibfnamefont {J.~M.}\ \bibnamefont {Pawlowski}},\ and\
  \bibinfo {author} {\bibfnamefont {N.}~\bibnamefont {Strodthoff}},\ }\href
  {https://doi.org/10.1103/PhysRevD.97.054006} {\bibfield  {journal} {\bibinfo
  {journal} {Phys. Rev. D}\ }\textbf {\bibinfo {volume} {97}},\ \bibinfo
  {pages} {054006} (\bibinfo {year} {2018})},\ \Eprint
  {https://arxiv.org/abs/1706.06326} {arXiv:1706.06326 [hep-ph]} \BibitemShut
  {NoStop}%
\bibitem [{\citenamefont {Fu}\ \emph {et~al.}(2020)\citenamefont {Fu},
  \citenamefont {Pawlowski},\ and\ \citenamefont {Rennecke}}]{Fu:2019hdw}%
  \BibitemOpen
  \bibfield  {author} {\bibinfo {author} {\bibfnamefont {W.-j.}\ \bibnamefont
  {Fu}}, \bibinfo {author} {\bibfnamefont {J.~M.}\ \bibnamefont {Pawlowski}},\
  and\ \bibinfo {author} {\bibfnamefont {F.}~\bibnamefont {Rennecke}},\ }\href
  {https://doi.org/10.1103/PhysRevD.101.054032} {\bibfield  {journal} {\bibinfo
   {journal} {Phys. Rev. D}\ }\textbf {\bibinfo {volume} {101}},\ \bibinfo
  {pages} {054032} (\bibinfo {year} {2020})},\ \Eprint
  {https://arxiv.org/abs/1909.02991} {arXiv:1909.02991 [hep-ph]} \BibitemShut
  {NoStop}%
\bibitem [{\citenamefont {Fukushima}\ \emph {et~al.}(2022)\citenamefont
  {Fukushima}, \citenamefont {Pawlowski},\ and\ \citenamefont
  {Strodthoff}}]{Fukushima:2021ctq}%
  \BibitemOpen
  \bibfield  {author} {\bibinfo {author} {\bibfnamefont {K.}~\bibnamefont
  {Fukushima}}, \bibinfo {author} {\bibfnamefont {J.~M.}\ \bibnamefont
  {Pawlowski}},\ and\ \bibinfo {author} {\bibfnamefont {N.}~\bibnamefont
  {Strodthoff}},\ }\href {https://doi.org/10.1016/j.aop.2022.169106} {\bibfield
   {journal} {\bibinfo  {journal} {Annals Phys.}\ }\textbf {\bibinfo {volume}
  {446}},\ \bibinfo {pages} {169106} (\bibinfo {year} {2022})},\ \Eprint
  {https://arxiv.org/abs/2103.01129} {arXiv:2103.01129 [hep-ph]} \BibitemShut
  {NoStop}%
\bibitem [{\citenamefont {Floerchinger}\ \emph {et~al.}(2008)\citenamefont
  {Floerchinger}, \citenamefont {Scherer}, \citenamefont {Diehl},\ and\
  \citenamefont {Wetterich}}]{Floerchinger:2008qc}%
  \BibitemOpen
  \bibfield  {author} {\bibinfo {author} {\bibfnamefont {S.}~\bibnamefont
  {Floerchinger}}, \bibinfo {author} {\bibfnamefont {M.}~\bibnamefont
  {Scherer}}, \bibinfo {author} {\bibfnamefont {S.}~\bibnamefont {Diehl}},\
  and\ \bibinfo {author} {\bibfnamefont {C.}~\bibnamefont {Wetterich}},\ }\href
  {https://doi.org/10.1103/PhysRevB.78.174528} {\bibfield  {journal} {\bibinfo
  {journal} {Phys. Rev. B}\ }\textbf {\bibinfo {volume} {78}},\ \bibinfo
  {pages} {174528} (\bibinfo {year} {2008})},\ \Eprint
  {https://arxiv.org/abs/0808.0150} {arXiv:0808.0150 [cond-mat.supr-con]}
  \BibitemShut {NoStop}%
\bibitem [{\citenamefont {Floerchinger}\ \emph {et~al.}(2010)\citenamefont
  {Floerchinger}, \citenamefont {Scherer},\ and\ \citenamefont
  {Wetterich}}]{Floerchinger:2009pg}%
  \BibitemOpen
  \bibfield  {author} {\bibinfo {author} {\bibfnamefont {S.}~\bibnamefont
  {Floerchinger}}, \bibinfo {author} {\bibfnamefont {M.~M.}\ \bibnamefont
  {Scherer}},\ and\ \bibinfo {author} {\bibfnamefont {C.}~\bibnamefont
  {Wetterich}},\ }\href {https://doi.org/10.1103/PhysRevA.81.063619} {\bibfield
   {journal} {\bibinfo  {journal} {Phys. Rev. A}\ }\textbf {\bibinfo {volume}
  {81}},\ \bibinfo {pages} {063619} (\bibinfo {year} {2010})},\ \Eprint
  {https://arxiv.org/abs/0912.4050} {arXiv:0912.4050 [cond-mat.quant-gas]}
  \BibitemShut {NoStop}%
\bibitem [{\citenamefont {Scherer}\ \emph {et~al.}(2011)\citenamefont
  {Scherer}, \citenamefont {Floerchinger},\ and\ \citenamefont
  {Gies}}]{Scherer:2010sv}%
  \BibitemOpen
  \bibfield  {author} {\bibinfo {author} {\bibfnamefont {M.~M.}\ \bibnamefont
  {Scherer}}, \bibinfo {author} {\bibfnamefont {S.}~\bibnamefont
  {Floerchinger}},\ and\ \bibinfo {author} {\bibfnamefont {H.}~\bibnamefont
  {Gies}},\ }\href@noop {} {\bibfield  {journal} {\bibinfo  {journal} {Phil.
  Trans. Roy. Soc. Lond. A}\ }\textbf {\bibinfo {volume} {368}},\ \bibinfo
  {pages} {2779} (\bibinfo {year} {2011})},\ \Eprint
  {https://arxiv.org/abs/1010.2890} {arXiv:1010.2890 [cond-mat.quant-gas]}
  \BibitemShut {NoStop}%
\bibitem [{\citenamefont {Lamprecht}(2007)}]{Lamprecht2007}%
  \BibitemOpen
  \bibfield  {author} {\bibinfo {author} {\bibfnamefont {F.}~\bibnamefont
  {Lamprecht}},\ }\href@noop {} {\bibinfo {title} {{Diploma thesis Heidelberg
  University}}} (\bibinfo {year} {2007})\BibitemShut {NoStop}%
\bibitem [{\citenamefont {Isaule}\ \emph {et~al.}(2018)\citenamefont {Isaule},
  \citenamefont {Birse},\ and\ \citenamefont {Walet}}]{Isaule:2018mxt}%
  \BibitemOpen
  \bibfield  {author} {\bibinfo {author} {\bibfnamefont {F.}~\bibnamefont
  {Isaule}}, \bibinfo {author} {\bibfnamefont {M.~C.}\ \bibnamefont {Birse}},\
  and\ \bibinfo {author} {\bibfnamefont {N.~R.}\ \bibnamefont {Walet}},\ }\href
  {https://doi.org/10.1103/PhysRevB.98.144502} {\bibfield  {journal} {\bibinfo
  {journal} {Phys. Rev. B}\ }\textbf {\bibinfo {volume} {98}},\ \bibinfo
  {pages} {144502} (\bibinfo {year} {2018})},\ \Eprint
  {https://arxiv.org/abs/1806.10373} {arXiv:1806.10373 [cond-mat.quant-gas]}
  \BibitemShut {NoStop}%
\bibitem [{\citenamefont {Isaule}\ \emph {et~al.}(2020)\citenamefont {Isaule},
  \citenamefont {Birse},\ and\ \citenamefont {Walet}}]{Isaule:2019pcm}%
  \BibitemOpen
  \bibfield  {author} {\bibinfo {author} {\bibfnamefont {F.}~\bibnamefont
  {Isaule}}, \bibinfo {author} {\bibfnamefont {M.~C.}\ \bibnamefont {Birse}},\
  and\ \bibinfo {author} {\bibfnamefont {N.~R.}\ \bibnamefont {Walet}},\ }\href
  {https://doi.org/10.1016/j.aop.2019.168006} {\bibfield  {journal} {\bibinfo
  {journal} {Annals Phys.}\ }\textbf {\bibinfo {volume} {412}},\ \bibinfo
  {pages} {168006} (\bibinfo {year} {2020})},\ \Eprint
  {https://arxiv.org/abs/1902.07135} {arXiv:1902.07135 [cond-mat.quant-gas]}
  \BibitemShut {NoStop}%
\bibitem [{\citenamefont {Daviet}\ and\ \citenamefont
  {Dupuis}(2022)}]{Daviet:2021whj}%
  \BibitemOpen
  \bibfield  {author} {\bibinfo {author} {\bibfnamefont {R.}~\bibnamefont
  {Daviet}}\ and\ \bibinfo {author} {\bibfnamefont {N.}~\bibnamefont
  {Dupuis}},\ }\href {https://doi.org/10.21468/SciPostPhys.12.3.110} {\bibfield
   {journal} {\bibinfo  {journal} {SciPost Phys.}\ }\textbf {\bibinfo {volume}
  {12}},\ \bibinfo {pages} {110} (\bibinfo {year} {2022})},\ \Eprint
  {https://arxiv.org/abs/2111.11458} {arXiv:2111.11458 [cond-mat.quant-gas]}
  \BibitemShut {NoStop}%
\bibitem [{\citenamefont {Salmhofer}(2007)}]{Salmhofer:2006pn}%
  \BibitemOpen
  \bibfield  {author} {\bibinfo {author} {\bibfnamefont {M.}~\bibnamefont
  {Salmhofer}},\ }\href {https://doi.org/10.1002/andp.200610223} {\bibfield
  {journal} {\bibinfo  {journal} {Annalen Phys.}\ }\textbf {\bibinfo {volume}
  {16}},\ \bibinfo {pages} {171} (\bibinfo {year} {2007})},\ \Eprint
  {https://arxiv.org/abs/cond-mat/0607289} {arXiv:cond-mat/0607289}
  \BibitemShut {NoStop}%
\bibitem [{\citenamefont {Baldazzi}\ and\ \citenamefont
  {Falls}(2021)}]{Baldazzi:2021orb}%
  \BibitemOpen
  \bibfield  {author} {\bibinfo {author} {\bibfnamefont {A.}~\bibnamefont
  {Baldazzi}}\ and\ \bibinfo {author} {\bibfnamefont {K.}~\bibnamefont
  {Falls}},\ }\href {https://doi.org/10.3390/universe7080294} {\bibfield
  {journal} {\bibinfo  {journal} {Universe}\ }\textbf {\bibinfo {volume} {7}},\
  \bibinfo {pages} {294} (\bibinfo {year} {2021})},\ \Eprint
  {https://arxiv.org/abs/2107.00671} {arXiv:2107.00671 [hep-th]} \BibitemShut
  {NoStop}%
\bibitem [{\citenamefont {Baldazzi}\ \emph
  {et~al.}(2022{\natexlab{a}})\citenamefont {Baldazzi}, \citenamefont
  {Zinati},\ and\ \citenamefont {Falls}}]{Baldazzi:2021ydj}%
  \BibitemOpen
  \bibfield  {author} {\bibinfo {author} {\bibfnamefont {A.}~\bibnamefont
  {Baldazzi}}, \bibinfo {author} {\bibfnamefont {R.~B.~A.}\ \bibnamefont
  {Zinati}},\ and\ \bibinfo {author} {\bibfnamefont {K.}~\bibnamefont
  {Falls}},\ }\href {https://doi.org/10.21468/SciPostPhys.13.4.085} {\bibfield
  {journal} {\bibinfo  {journal} {SciPost Phys.}\ }\textbf {\bibinfo {volume}
  {13}},\ \bibinfo {pages} {085} (\bibinfo {year} {2022}{\natexlab{a}})},\
  \Eprint {https://arxiv.org/abs/2105.11482} {arXiv:2105.11482 [hep-th]}
  \BibitemShut {NoStop}%
\bibitem [{\citenamefont {Christiansen}\ \emph {et~al.}(2015)\citenamefont
  {Christiansen}, \citenamefont {Knorr}, \citenamefont {Meibohm}, \citenamefont
  {Pawlowski},\ and\ \citenamefont {Reichert}}]{Christiansen:2015rva}%
  \BibitemOpen
  \bibfield  {author} {\bibinfo {author} {\bibfnamefont {N.}~\bibnamefont
  {Christiansen}}, \bibinfo {author} {\bibfnamefont {B.}~\bibnamefont {Knorr}},
  \bibinfo {author} {\bibfnamefont {J.}~\bibnamefont {Meibohm}}, \bibinfo
  {author} {\bibfnamefont {J.~M.}\ \bibnamefont {Pawlowski}},\ and\ \bibinfo
  {author} {\bibfnamefont {M.}~\bibnamefont {Reichert}},\ }\href
  {https://doi.org/10.1103/PhysRevD.92.121501} {\bibfield  {journal} {\bibinfo
  {journal} {Phys. Rev. D}\ }\textbf {\bibinfo {volume} {92}},\ \bibinfo
  {pages} {121501} (\bibinfo {year} {2015})},\ \Eprint
  {https://arxiv.org/abs/1506.07016} {arXiv:1506.07016 [hep-th]} \BibitemShut
  {NoStop}%
\bibitem [{\citenamefont {Wetterich}(1993)}]{Wetterich:1992yh}%
  \BibitemOpen
  \bibfield  {author} {\bibinfo {author} {\bibfnamefont {C.}~\bibnamefont
  {Wetterich}},\ }\href {https://doi.org/10.1016/0370-2693(93)90726-X}
  {\bibfield  {journal} {\bibinfo  {journal} {Phys. Lett. B}\ }\textbf
  {\bibinfo {volume} {301}},\ \bibinfo {pages} {90} (\bibinfo {year} {1993})},\
  \Eprint {https://arxiv.org/abs/1710.05815} {arXiv:1710.05815 [hep-th]}
  \BibitemShut {NoStop}%
\bibitem [{\citenamefont {Ellwanger}(1994)}]{Ellwanger:1993mw}%
  \BibitemOpen
  \bibfield  {author} {\bibinfo {author} {\bibfnamefont {U.}~\bibnamefont
  {Ellwanger}},\ }\href {https://doi.org/10.1007/BF01555911} {\bibfield
  {journal} {\bibinfo  {journal} {Z. Phys. C}\ }\textbf {\bibinfo {volume}
  {62}},\ \bibinfo {pages} {503} (\bibinfo {year} {1994})},\ \Eprint
  {https://arxiv.org/abs/hep-ph/9308260} {arXiv:hep-ph/9308260} \BibitemShut
  {NoStop}%
\bibitem [{\citenamefont {Morris}(1994)}]{Morris:1993qb}%
  \BibitemOpen
  \bibfield  {author} {\bibinfo {author} {\bibfnamefont {T.~R.}\ \bibnamefont
  {Morris}},\ }\href {https://doi.org/10.1142/S0217751X94000972} {\bibfield
  {journal} {\bibinfo  {journal} {Int. J. Mod. Phys. A}\ }\textbf {\bibinfo
  {volume} {9}},\ \bibinfo {pages} {2411} (\bibinfo {year} {1994})},\ \Eprint
  {https://arxiv.org/abs/hep-ph/9308265} {arXiv:hep-ph/9308265} \BibitemShut
  {NoStop}%
\bibitem [{\citenamefont {Fu}\ \emph {et~al.}(2023)\citenamefont {Fu},
  \citenamefont {Huang}, \citenamefont {Pawlowski},\ and\ \citenamefont
  {Tan}}]{Fu:2022uow}%
  \BibitemOpen
  \bibfield  {author} {\bibinfo {author} {\bibfnamefont {W.-j.}\ \bibnamefont
  {Fu}}, \bibinfo {author} {\bibfnamefont {C.}~\bibnamefont {Huang}}, \bibinfo
  {author} {\bibfnamefont {J.~M.}\ \bibnamefont {Pawlowski}},\ and\ \bibinfo
  {author} {\bibfnamefont {Y.-y.}\ \bibnamefont {Tan}},\ }\href
  {https://doi.org/10.21468/SciPostPhys.14.4.069} {\bibfield  {journal}
  {\bibinfo  {journal} {SciPost Phys.}\ }\textbf {\bibinfo {volume} {14}},\
  \bibinfo {pages} {069} (\bibinfo {year} {2023})},\ \Eprint
  {https://arxiv.org/abs/2209.13120} {arXiv:2209.13120 [hep-ph]} \BibitemShut
  {NoStop}%
\bibitem [{\citenamefont {Fu}\ \emph {et~al.}(2024)\citenamefont {Fu},
  \citenamefont {Huang}, \citenamefont {Pawlowski},\ and\ \citenamefont
  {Tan}}]{Fu:2024ysj}%
  \BibitemOpen
  \bibfield  {author} {\bibinfo {author} {\bibfnamefont {W.-j.}\ \bibnamefont
  {Fu}}, \bibinfo {author} {\bibfnamefont {C.}~\bibnamefont {Huang}}, \bibinfo
  {author} {\bibfnamefont {J.~M.}\ \bibnamefont {Pawlowski}},\ and\ \bibinfo
  {author} {\bibfnamefont {Y.-y.}\ \bibnamefont {Tan}},\ }\href
  {https://doi.org/10.21468/SciPostPhys.17.5.148} {\bibfield  {journal}
  {\bibinfo  {journal} {SciPost Phys.}\ }\textbf {\bibinfo {volume} {17}},\
  \bibinfo {pages} {148} (\bibinfo {year} {2024})},\ \Eprint
  {https://arxiv.org/abs/2401.07638} {arXiv:2401.07638 [hep-ph]} \BibitemShut
  {NoStop}%
\bibitem [{\citenamefont {Bonanno}\ \emph {et~al.}(2025)\citenamefont
  {Bonanno}, \citenamefont {Ihssen},\ and\ \citenamefont
  {Pawlowski}}]{Bonanno:2025mon}%
  \BibitemOpen
  \bibfield  {author} {\bibinfo {author} {\bibfnamefont {A.}~\bibnamefont
  {Bonanno}}, \bibinfo {author} {\bibfnamefont {F.}~\bibnamefont {Ihssen}},\
  and\ \bibinfo {author} {\bibfnamefont {J.~M.}\ \bibnamefont {Pawlowski}},\
  }\href@noop {} {\  (\bibinfo {year} {2025})},\ \Eprint
  {https://arxiv.org/abs/2504.03437} {arXiv:2504.03437 [hep-th]} \BibitemShut
  {NoStop}%
\bibitem [{\citenamefont {Baldazzi}\ \emph {et~al.}(2023)\citenamefont
  {Baldazzi}, \citenamefont {Falls}, \citenamefont {Kluth},\ and\ \citenamefont
  {Knorr}}]{Baldazzi:2023pep}%
  \BibitemOpen
  \bibfield  {author} {\bibinfo {author} {\bibfnamefont {A.}~\bibnamefont
  {Baldazzi}}, \bibinfo {author} {\bibfnamefont {K.}~\bibnamefont {Falls}},
  \bibinfo {author} {\bibfnamefont {Y.}~\bibnamefont {Kluth}},\ and\ \bibinfo
  {author} {\bibfnamefont {B.}~\bibnamefont {Knorr}},\ }\href@noop {} {\
  (\bibinfo {year} {2023})},\ \Eprint {https://arxiv.org/abs/2312.03831}
  {arXiv:2312.03831 [hep-th]} \BibitemShut {NoStop}%
\bibitem [{\citenamefont {Baldazzi}\ \emph
  {et~al.}(2022{\natexlab{b}})\citenamefont {Baldazzi}, \citenamefont {Falls},\
  and\ \citenamefont {Ferrero}}]{Baldazzi:2021fye}%
  \BibitemOpen
  \bibfield  {author} {\bibinfo {author} {\bibfnamefont {A.}~\bibnamefont
  {Baldazzi}}, \bibinfo {author} {\bibfnamefont {K.}~\bibnamefont {Falls}},\
  and\ \bibinfo {author} {\bibfnamefont {R.}~\bibnamefont {Ferrero}},\ }\href
  {https://doi.org/10.1016/j.aop.2022.168822} {\bibfield  {journal} {\bibinfo
  {journal} {Annals Phys.}\ }\textbf {\bibinfo {volume} {440}},\ \bibinfo
  {pages} {168822} (\bibinfo {year} {2022}{\natexlab{b}})},\ \Eprint
  {https://arxiv.org/abs/2112.02118} {arXiv:2112.02118 [hep-th]} \BibitemShut
  {NoStop}%
\bibitem [{\citenamefont {Knorr}(2022)}]{Knorr:2022ilz}%
  \BibitemOpen
  \bibfield  {author} {\bibinfo {author} {\bibfnamefont {B.}~\bibnamefont
  {Knorr}},\ }\href@noop {} {\  (\bibinfo {year} {2022})},\ \Eprint
  {https://arxiv.org/abs/2204.08564} {arXiv:2204.08564 [hep-th]} \BibitemShut
  {NoStop}%
\bibitem [{\citenamefont {Ohta}\ and\ \citenamefont
  {Yamada}(2025)}]{Ohta:2025xxo}%
  \BibitemOpen
  \bibfield  {author} {\bibinfo {author} {\bibfnamefont {N.}~\bibnamefont
  {Ohta}}\ and\ \bibinfo {author} {\bibfnamefont {M.}~\bibnamefont {Yamada}},\
  }\href@noop {} {\  (\bibinfo {year} {2025})},\ \Eprint
  {https://arxiv.org/abs/2506.03601} {arXiv:2506.03601 [hep-th]} \BibitemShut
  {NoStop}%
\bibitem [{\citenamefont {Wetterich}(2024)}]{Wetterich:2024uub}%
  \BibitemOpen
  \bibfield  {author} {\bibinfo {author} {\bibfnamefont {C.}~\bibnamefont
  {Wetterich}},\ }\href {https://doi.org/10.1016/j.nuclphysb.2024.116707}
  {\bibfield  {journal} {\bibinfo  {journal} {Nucl. Phys. B}\ }\textbf
  {\bibinfo {volume} {1008}},\ \bibinfo {pages} {116707} (\bibinfo {year}
  {2024})},\ \Eprint {https://arxiv.org/abs/2402.04679} {arXiv:2402.04679
  [hep-th]} \BibitemShut {NoStop}%
\bibitem [{\citenamefont {Litim}\ \emph {et~al.}(2006)\citenamefont {Litim},
  \citenamefont {Pawlowski},\ and\ \citenamefont {Vergara}}]{Litim:2006nn}%
  \BibitemOpen
  \bibfield  {author} {\bibinfo {author} {\bibfnamefont {D.~F.}\ \bibnamefont
  {Litim}}, \bibinfo {author} {\bibfnamefont {J.~M.}\ \bibnamefont
  {Pawlowski}},\ and\ \bibinfo {author} {\bibfnamefont {L.}~\bibnamefont
  {Vergara}},\ }\href@noop {} {\  (\bibinfo {year} {2006})},\ \Eprint
  {https://arxiv.org/abs/hep-th/0602140} {arXiv:hep-th/0602140} \BibitemShut
  {NoStop}%
\bibitem [{\citenamefont {Ihssen}\ and\ \citenamefont
  {Pawlowski}(2025{\natexlab{a}})}]{Ihssen:2025hyl}%
  \BibitemOpen
  \bibfield  {author} {\bibinfo {author} {\bibfnamefont {F.}~\bibnamefont
  {Ihssen}}\ and\ \bibinfo {author} {\bibfnamefont {J.~M.}\ \bibnamefont
  {Pawlowski}},\ }\href@noop {} {\  (\bibinfo {year} {2025}{\natexlab{a}})},\
  \Eprint {https://arxiv.org/abs/2507.13011} {arXiv:2507.13011 [hep-th]}
  \BibitemShut {NoStop}%
\bibitem [{\citenamefont {Pawlowski}\ \emph {et~al.}(2017)\citenamefont
  {Pawlowski}, \citenamefont {Scherer}, \citenamefont {Schmidt},\ and\
  \citenamefont {Wetzel}}]{Pawlowski:2015mlf}%
  \BibitemOpen
  \bibfield  {author} {\bibinfo {author} {\bibfnamefont {J.~M.}\ \bibnamefont
  {Pawlowski}}, \bibinfo {author} {\bibfnamefont {M.~M.}\ \bibnamefont
  {Scherer}}, \bibinfo {author} {\bibfnamefont {R.}~\bibnamefont {Schmidt}},\
  and\ \bibinfo {author} {\bibfnamefont {S.~J.}\ \bibnamefont {Wetzel}},\
  }\href {https://doi.org/10.1016/j.aop.2017.06.017} {\bibfield  {journal}
  {\bibinfo  {journal} {Annals Phys.}\ }\textbf {\bibinfo {volume} {384}},\
  \bibinfo {pages} {165} (\bibinfo {year} {2017})},\ \Eprint
  {https://arxiv.org/abs/1512.03598} {arXiv:1512.03598 [hep-th]} \BibitemShut
  {NoStop}%
\bibitem [{\citenamefont {Fu}(2022)}]{Fu:2022gou}%
  \BibitemOpen
  \bibfield  {author} {\bibinfo {author} {\bibfnamefont {W.-j.}\ \bibnamefont
  {Fu}},\ }\href {https://doi.org/10.1088/1572-9494/ac86be} {\bibfield
  {journal} {\bibinfo  {journal} {Commun. Theor. Phys.}\ }\textbf {\bibinfo
  {volume} {74}},\ \bibinfo {pages} {097304} (\bibinfo {year} {2022})},\
  \Eprint {https://arxiv.org/abs/2205.00468} {arXiv:2205.00468 [hep-ph]}
  \BibitemShut {NoStop}%
\bibitem [{\citenamefont {Fu}\ \emph {et~al.}(2025)\citenamefont {Fu},
  \citenamefont {Luo}, \citenamefont {Pawlowski}, \citenamefont {Rennecke},\
  and\ \citenamefont {Yin}}]{Fu:2023lcm}%
  \BibitemOpen
  \bibfield  {author} {\bibinfo {author} {\bibfnamefont {W.-j.}\ \bibnamefont
  {Fu}}, \bibinfo {author} {\bibfnamefont {X.}~\bibnamefont {Luo}}, \bibinfo
  {author} {\bibfnamefont {J.~M.}\ \bibnamefont {Pawlowski}}, \bibinfo {author}
  {\bibfnamefont {F.}~\bibnamefont {Rennecke}},\ and\ \bibinfo {author}
  {\bibfnamefont {S.}~\bibnamefont {Yin}},\ }\href
  {https://doi.org/10.1103/PhysRevD.111.L031502} {\bibfield  {journal}
  {\bibinfo  {journal} {Phys. Rev. D}\ }\textbf {\bibinfo {volume} {111}},\
  \bibinfo {pages} {L031502} (\bibinfo {year} {2025})},\ \Eprint
  {https://arxiv.org/abs/2308.15508} {arXiv:2308.15508 [hep-ph]} \BibitemShut
  {NoStop}%
\bibitem [{\citenamefont {Diehl}\ \emph {et~al.}(2008)\citenamefont {Diehl},
  \citenamefont {Krahl},\ and\ \citenamefont {Scherer}}]{Diehl:2007xz}%
  \BibitemOpen
  \bibfield  {author} {\bibinfo {author} {\bibfnamefont {S.}~\bibnamefont
  {Diehl}}, \bibinfo {author} {\bibfnamefont {H.~C.}\ \bibnamefont {Krahl}},\
  and\ \bibinfo {author} {\bibfnamefont {M.}~\bibnamefont {Scherer}},\ }\href
  {https://doi.org/10.1103/PhysRevC.78.034001} {\bibfield  {journal} {\bibinfo
  {journal} {Phys. Rev. C}\ }\textbf {\bibinfo {volume} {78}},\ \bibinfo
  {pages} {034001} (\bibinfo {year} {2008})},\ \Eprint
  {https://arxiv.org/abs/0712.2846} {arXiv:0712.2846 [cond-mat.stat-mech]}
  \BibitemShut {NoStop}%
\bibitem [{\citenamefont {Floerchinger}(2014)}]{Floerchinger:2013bty}%
  \BibitemOpen
  \bibfield  {author} {\bibinfo {author} {\bibfnamefont {S.}~\bibnamefont
  {Floerchinger}},\ }\href {https://doi.org/10.1016/j.nuclphysa.2014.04.013}
  {\bibfield  {journal} {\bibinfo  {journal} {Nucl. Phys. A}\ }\textbf
  {\bibinfo {volume} {927}},\ \bibinfo {pages} {119} (\bibinfo {year}
  {2014})},\ \Eprint {https://arxiv.org/abs/1301.6542} {arXiv:1301.6542
  [nucl-th]} \BibitemShut {NoStop}%
\bibitem [{\citenamefont {Tanizaki}(2013)}]{Tanizaki:2013uma}%
  \BibitemOpen
  \bibfield  {author} {\bibinfo {author} {\bibfnamefont {Y.}~\bibnamefont
  {Tanizaki}},\ }\href {https://doi.org/10.1093/ptep/ptt092} {\bibfield
  {journal} {\bibinfo  {journal} {PTEP}\ }\textbf {\bibinfo {volume} {2013}},\
  \bibinfo {pages} {113A01} (\bibinfo {year} {2013})},\ \Eprint
  {https://arxiv.org/abs/1304.3286} {arXiv:1304.3286 [cond-mat.quant-gas]}
  \BibitemShut {NoStop}%
\bibitem [{\citenamefont {Pawlowski}\ and\ \citenamefont
  {Rennecke}(2014)}]{Pawlowski:2014zaa}%
  \BibitemOpen
  \bibfield  {author} {\bibinfo {author} {\bibfnamefont {J.~M.}\ \bibnamefont
  {Pawlowski}}\ and\ \bibinfo {author} {\bibfnamefont {F.}~\bibnamefont
  {Rennecke}},\ }\href {https://doi.org/10.1103/PhysRevD.90.076002} {\bibfield
  {journal} {\bibinfo  {journal} {Phys. Rev. D}\ }\textbf {\bibinfo {volume}
  {90}},\ \bibinfo {pages} {076002} (\bibinfo {year} {2014})},\ \Eprint
  {https://arxiv.org/abs/1403.1179} {arXiv:1403.1179 [hep-ph]} \BibitemShut
  {NoStop}%
\bibitem [{\citenamefont {Braun}\ \emph {et~al.}(2019)\citenamefont {Braun},
  \citenamefont {Leonhardt},\ and\ \citenamefont {Pawlowski}}]{Braun:2018svj}%
  \BibitemOpen
  \bibfield  {author} {\bibinfo {author} {\bibfnamefont {J.}~\bibnamefont
  {Braun}}, \bibinfo {author} {\bibfnamefont {M.}~\bibnamefont {Leonhardt}},\
  and\ \bibinfo {author} {\bibfnamefont {J.~M.}\ \bibnamefont {Pawlowski}},\
  }\href {https://doi.org/10.21468/SciPostPhys.6.5.056} {\bibfield  {journal}
  {\bibinfo  {journal} {SciPost Phys.}\ }\textbf {\bibinfo {volume} {6}},\
  \bibinfo {pages} {056} (\bibinfo {year} {2019})},\ \Eprint
  {https://arxiv.org/abs/1806.04432} {arXiv:1806.04432 [hep-ph]} \BibitemShut
  {NoStop}%
\bibitem [{\citenamefont {Helmboldt}\ \emph {et~al.}(2015)\citenamefont
  {Helmboldt}, \citenamefont {Pawlowski},\ and\ \citenamefont
  {Strodthoff}}]{Helmboldt:2014iya}%
  \BibitemOpen
  \bibfield  {author} {\bibinfo {author} {\bibfnamefont {A.~J.}\ \bibnamefont
  {Helmboldt}}, \bibinfo {author} {\bibfnamefont {J.~M.}\ \bibnamefont
  {Pawlowski}},\ and\ \bibinfo {author} {\bibfnamefont {N.}~\bibnamefont
  {Strodthoff}},\ }\href {https://doi.org/10.1103/PhysRevD.91.054010}
  {\bibfield  {journal} {\bibinfo  {journal} {Phys. Rev. D}\ }\textbf {\bibinfo
  {volume} {91}},\ \bibinfo {pages} {054010} (\bibinfo {year} {2015})},\
  \Eprint {https://arxiv.org/abs/1409.8414} {arXiv:1409.8414 [hep-ph]}
  \BibitemShut {NoStop}%
\bibitem [{\citenamefont {Grossi}\ and\ \citenamefont
  {Wink}(2023)}]{Grossi:2019urj}%
  \BibitemOpen
  \bibfield  {author} {\bibinfo {author} {\bibfnamefont {E.}~\bibnamefont
  {Grossi}}\ and\ \bibinfo {author} {\bibfnamefont {N.}~\bibnamefont {Wink}},\
  }\href {https://doi.org/10.21468/SciPostPhysCore.6.4.071} {\bibfield
  {journal} {\bibinfo  {journal} {SciPost Phys. Core}\ }\textbf {\bibinfo
  {volume} {6}},\ \bibinfo {pages} {071} (\bibinfo {year} {2023})},\ \Eprint
  {https://arxiv.org/abs/1903.09503} {arXiv:1903.09503 [hep-th]} \BibitemShut
  {NoStop}%
\bibitem [{\citenamefont {Koenigstein}\ \emph
  {et~al.}(2022{\natexlab{a}})\citenamefont {Koenigstein}, \citenamefont
  {Steil}, \citenamefont {Wink}, \citenamefont {Grossi},\ and\ \citenamefont
  {Braun}}]{Koenigstein:2021rxj}%
  \BibitemOpen
  \bibfield  {author} {\bibinfo {author} {\bibfnamefont {A.}~\bibnamefont
  {Koenigstein}}, \bibinfo {author} {\bibfnamefont {M.~J.}\ \bibnamefont
  {Steil}}, \bibinfo {author} {\bibfnamefont {N.}~\bibnamefont {Wink}},
  \bibinfo {author} {\bibfnamefont {E.}~\bibnamefont {Grossi}},\ and\ \bibinfo
  {author} {\bibfnamefont {J.}~\bibnamefont {Braun}},\ }\href
  {https://doi.org/10.1103/PhysRevD.106.065013} {\bibfield  {journal} {\bibinfo
   {journal} {Phys. Rev. D}\ }\textbf {\bibinfo {volume} {106}},\ \bibinfo
  {pages} {065013} (\bibinfo {year} {2022}{\natexlab{a}})},\ \Eprint
  {https://arxiv.org/abs/2108.10085} {arXiv:2108.10085 [cond-mat.stat-mech]}
  \BibitemShut {NoStop}%
\bibitem [{\citenamefont {Koenigstein}\ \emph
  {et~al.}(2022{\natexlab{b}})\citenamefont {Koenigstein}, \citenamefont
  {Steil}, \citenamefont {Wink}, \citenamefont {Grossi}, \citenamefont {Braun},
  \citenamefont {Buballa},\ and\ \citenamefont
  {Rischke}}]{Koenigstein:2021syz}%
  \BibitemOpen
  \bibfield  {author} {\bibinfo {author} {\bibfnamefont {A.}~\bibnamefont
  {Koenigstein}}, \bibinfo {author} {\bibfnamefont {M.~J.}\ \bibnamefont
  {Steil}}, \bibinfo {author} {\bibfnamefont {N.}~\bibnamefont {Wink}},
  \bibinfo {author} {\bibfnamefont {E.}~\bibnamefont {Grossi}}, \bibinfo
  {author} {\bibfnamefont {J.}~\bibnamefont {Braun}}, \bibinfo {author}
  {\bibfnamefont {M.}~\bibnamefont {Buballa}},\ and\ \bibinfo {author}
  {\bibfnamefont {D.~H.}\ \bibnamefont {Rischke}},\ }\href
  {https://doi.org/10.1103/PhysRevD.106.065012} {\bibfield  {journal} {\bibinfo
   {journal} {Phys. Rev. D}\ }\textbf {\bibinfo {volume} {106}},\ \bibinfo
  {pages} {065012} (\bibinfo {year} {2022}{\natexlab{b}})},\ \Eprint
  {https://arxiv.org/abs/2108.02504} {arXiv:2108.02504 [cond-mat.stat-mech]}
  \BibitemShut {NoStop}%
\bibitem [{\citenamefont {Steil}\ and\ \citenamefont
  {Koenigstein}(2022)}]{Steil:2021cbu}%
  \BibitemOpen
  \bibfield  {author} {\bibinfo {author} {\bibfnamefont {M.~J.}\ \bibnamefont
  {Steil}}\ and\ \bibinfo {author} {\bibfnamefont {A.}~\bibnamefont
  {Koenigstein}},\ }\href {https://doi.org/10.1103/PhysRevD.106.065014}
  {\bibfield  {journal} {\bibinfo  {journal} {Phys. Rev. D}\ }\textbf {\bibinfo
  {volume} {106}},\ \bibinfo {pages} {065014} (\bibinfo {year} {2022})},\
  \Eprint {https://arxiv.org/abs/2108.04037} {arXiv:2108.04037
  [cond-mat.stat-mech]} \BibitemShut {NoStop}%
\bibitem [{\citenamefont {Ihssen}\ \emph
  {et~al.}(2024{\natexlab{b}})\citenamefont {Ihssen}, \citenamefont
  {Pawlowski}, \citenamefont {Sattler},\ and\ \citenamefont
  {Wink}}]{Ihssen:2022xkr}%
  \BibitemOpen
  \bibfield  {author} {\bibinfo {author} {\bibfnamefont {F.}~\bibnamefont
  {Ihssen}}, \bibinfo {author} {\bibfnamefont {J.~M.}\ \bibnamefont
  {Pawlowski}}, \bibinfo {author} {\bibfnamefont {F.~R.}\ \bibnamefont
  {Sattler}},\ and\ \bibinfo {author} {\bibfnamefont {N.}~\bibnamefont
  {Wink}},\ }\href {https://doi.org/10.1016/j.cpc.2024.109182} {\bibfield
  {journal} {\bibinfo  {journal} {Comput. Phys. Commun.}\ }\textbf {\bibinfo
  {volume} {300}},\ \bibinfo {pages} {109182} (\bibinfo {year}
  {2024}{\natexlab{b}})},\ \Eprint {https://arxiv.org/abs/2207.12266}
  {arXiv:2207.12266 [hep-th]} \BibitemShut {NoStop}%
\bibitem [{\citenamefont {Ihssen}\ \emph {et~al.}(2023)\citenamefont {Ihssen},
  \citenamefont {Sattler},\ and\ \citenamefont {Wink}}]{Ihssen:2023qaq}%
  \BibitemOpen
  \bibfield  {author} {\bibinfo {author} {\bibfnamefont {F.}~\bibnamefont
  {Ihssen}}, \bibinfo {author} {\bibfnamefont {F.~R.}\ \bibnamefont
  {Sattler}},\ and\ \bibinfo {author} {\bibfnamefont {N.}~\bibnamefont
  {Wink}},\ }\href {https://doi.org/10.1103/PhysRevD.107.114009} {\bibfield
  {journal} {\bibinfo  {journal} {Phys. Rev. D}\ }\textbf {\bibinfo {volume}
  {107}},\ \bibinfo {pages} {114009} (\bibinfo {year} {2023})},\ \Eprint
  {https://arxiv.org/abs/2302.04736} {arXiv:2302.04736 [hep-th]} \BibitemShut
  {NoStop}%
\bibitem [{\citenamefont {Ihssen}\ \emph {et~al.}(2025)\citenamefont {Ihssen},
  \citenamefont {Pawlowski}, \citenamefont {Sattler},\ and\ \citenamefont
  {Wink}}]{Ihssen:2023xlp}%
  \BibitemOpen
  \bibfield  {author} {\bibinfo {author} {\bibfnamefont {F.}~\bibnamefont
  {Ihssen}}, \bibinfo {author} {\bibfnamefont {J.~M.}\ \bibnamefont
  {Pawlowski}}, \bibinfo {author} {\bibfnamefont {F.~R.}\ \bibnamefont
  {Sattler}},\ and\ \bibinfo {author} {\bibfnamefont {N.}~\bibnamefont
  {Wink}},\ }\href {https://doi.org/10.1103/PhysRevD.111.036030} {\bibfield
  {journal} {\bibinfo  {journal} {Phys. Rev. D}\ }\textbf {\bibinfo {volume}
  {111}},\ \bibinfo {pages} {036030} (\bibinfo {year} {2025})},\ \Eprint
  {https://arxiv.org/abs/2309.07335} {arXiv:2309.07335 [hep-th]} \BibitemShut
  {NoStop}%
\bibitem [{\citenamefont {Grossi}\ \emph {et~al.}(2021)\citenamefont {Grossi},
  \citenamefont {Ihssen}, \citenamefont {Pawlowski},\ and\ \citenamefont
  {Wink}}]{Grossi:2021ksl}%
  \BibitemOpen
  \bibfield  {author} {\bibinfo {author} {\bibfnamefont {E.}~\bibnamefont
  {Grossi}}, \bibinfo {author} {\bibfnamefont {F.~J.}\ \bibnamefont {Ihssen}},
  \bibinfo {author} {\bibfnamefont {J.~M.}\ \bibnamefont {Pawlowski}},\ and\
  \bibinfo {author} {\bibfnamefont {N.}~\bibnamefont {Wink}},\ }\href
  {https://doi.org/10.1103/PhysRevD.104.016028} {\bibfield  {journal} {\bibinfo
   {journal} {Phys. Rev. D}\ }\textbf {\bibinfo {volume} {104}},\ \bibinfo
  {pages} {016028} (\bibinfo {year} {2021})},\ \Eprint
  {https://arxiv.org/abs/2102.01602} {arXiv:2102.01602 [hep-ph]} \BibitemShut
  {NoStop}%
\bibitem [{\citenamefont {Pawlowski}\ \emph {et~al.}(2023)\citenamefont
  {Pawlowski}, \citenamefont {Schneider},\ and\ \citenamefont
  {Wink}}]{Pawlowski:2021tkk}%
  \BibitemOpen
  \bibfield  {author} {\bibinfo {author} {\bibfnamefont {J.~M.}\ \bibnamefont
  {Pawlowski}}, \bibinfo {author} {\bibfnamefont {C.~S.}\ \bibnamefont
  {Schneider}},\ and\ \bibinfo {author} {\bibfnamefont {N.}~\bibnamefont
  {Wink}},\ }\href {https://doi.org/10.1016/j.cpc.2023.108711} {\bibfield
  {journal} {\bibinfo  {journal} {Comput. Phys. Commun.}\ }\textbf {\bibinfo
  {volume} {287}},\ \bibinfo {pages} {108711} (\bibinfo {year} {2023})},\
  \Eprint {https://arxiv.org/abs/2102.01410} {arXiv:2102.01410 [hep-ph]}
  \BibitemShut {NoStop}%
\bibitem [{\citenamefont {Zorbach}\ \emph {et~al.}(2024)\citenamefont
  {Zorbach}, \citenamefont {Stoll},\ and\ \citenamefont
  {Braun}}]{Zorbach:2024zjx}%
  \BibitemOpen
  \bibfield  {author} {\bibinfo {author} {\bibfnamefont {N.}~\bibnamefont
  {Zorbach}}, \bibinfo {author} {\bibfnamefont {J.}~\bibnamefont {Stoll}},\
  and\ \bibinfo {author} {\bibfnamefont {J.}~\bibnamefont {Braun}},\
  }\href@noop {} {\  (\bibinfo {year} {2024})},\ \Eprint
  {https://arxiv.org/abs/2401.12854} {arXiv:2401.12854 [hep-ph]} \BibitemShut
  {NoStop}%
\bibitem [{\citenamefont {Shampine}\ and\ \citenamefont
  {Reichelt}(1997)}]{doi:10.1137/S1064827594276424}%
  \BibitemOpen
  \bibfield  {author} {\bibinfo {author} {\bibfnamefont {L.~F.}\ \bibnamefont
  {Shampine}}\ and\ \bibinfo {author} {\bibfnamefont {M.~W.}\ \bibnamefont
  {Reichelt}},\ }\href {https://doi.org/10.1137/S1064827594276424} {\bibfield
  {journal} {\bibinfo  {journal} {SIAM Journal on Scientific Computing}\
  }\textbf {\bibinfo {volume} {18}},\ \bibinfo {pages} {1} (\bibinfo {year}
  {1997})},\ \Eprint
  {https://arxiv.org/abs/https://doi.org/10.1137/S1064827594276424}
  {https://doi.org/10.1137/S1064827594276424} \BibitemShut {NoStop}%
\bibitem [{\citenamefont {Balog}\ \emph {et~al.}(2019)\citenamefont {Balog},
  \citenamefont {Chat\'e}, \citenamefont {Delamotte}, \citenamefont
  {Marohnic},\ and\ \citenamefont {Wschebor}}]{Balog:2019rrg}%
  \BibitemOpen
  \bibfield  {author} {\bibinfo {author} {\bibfnamefont {I.}~\bibnamefont
  {Balog}}, \bibinfo {author} {\bibfnamefont {H.}~\bibnamefont {Chat\'e}},
  \bibinfo {author} {\bibfnamefont {B.}~\bibnamefont {Delamotte}}, \bibinfo
  {author} {\bibfnamefont {M.}~\bibnamefont {Marohnic}},\ and\ \bibinfo
  {author} {\bibfnamefont {N.}~\bibnamefont {Wschebor}},\ }\href
  {https://doi.org/10.1103/PhysRevLett.123.240604} {\bibfield  {journal}
  {\bibinfo  {journal} {Phys. Rev. Lett.}\ }\textbf {\bibinfo {volume} {123}},\
  \bibinfo {pages} {240604} (\bibinfo {year} {2019})},\ \Eprint
  {https://arxiv.org/abs/1907.01829} {arXiv:1907.01829 [cond-mat.stat-mech]}
  \BibitemShut {NoStop}%
\bibitem [{\citenamefont {Ihssen}\ and\ \citenamefont
  {Pawlowski}(2025{\natexlab{b}})}]{Ihssen:2025cff}%
  \BibitemOpen
  \bibfield  {author} {\bibinfo {author} {\bibfnamefont {F.}~\bibnamefont
  {Ihssen}}\ and\ \bibinfo {author} {\bibfnamefont {J.~M.}\ \bibnamefont
  {Pawlowski}},\ }\href@noop {} {\  (\bibinfo {year} {2025}{\natexlab{b}})},\
  \Eprint {https://arxiv.org/abs/2503.22638} {arXiv:2503.22638 [hep-th]}
  \BibitemShut {NoStop}%
\bibitem [{\citenamefont {Braun}\ \emph {et~al.}(2024)\citenamefont {Braun},
  \citenamefont {Chen}, \citenamefont {Fu}, \citenamefont {Gao}, \citenamefont
  {Ihssen}, \citenamefont {Geissel}, \citenamefont {Huang}, \citenamefont
  {Pawlowski}, \citenamefont {Rennecke}, \citenamefont {Sattler}, \citenamefont
  {Schallmo}, \citenamefont {Stoll}, \citenamefont {Tan}, \citenamefont
  {T{\"o}pfel}, \citenamefont {Turnwald}, \citenamefont {Wen}, \citenamefont
  {Wessely}, \citenamefont {Wink}, \citenamefont {Yin},\ and\ \citenamefont
  {Zorbach}}]{fQCD}%
  \BibitemOpen
  \bibfield  {author} {\bibinfo {author} {\bibfnamefont {J.}~\bibnamefont
  {Braun}}, \bibinfo {author} {\bibfnamefont {Y.-r.}\ \bibnamefont {Chen}},
  \bibinfo {author} {\bibfnamefont {W.-j.}\ \bibnamefont {Fu}}, \bibinfo
  {author} {\bibfnamefont {F.}~\bibnamefont {Gao}}, \bibinfo {author}
  {\bibfnamefont {F.}~\bibnamefont {Ihssen}}, \bibinfo {author} {\bibfnamefont
  {A.}~\bibnamefont {Geissel}}, \bibinfo {author} {\bibfnamefont
  {C.}~\bibnamefont {Huang}}, \bibinfo {author} {\bibfnamefont {J.~M.}\
  \bibnamefont {Pawlowski}}, \bibinfo {author} {\bibfnamefont {F.}~\bibnamefont
  {Rennecke}}, \bibinfo {author} {\bibfnamefont {F.~R.}\ \bibnamefont
  {Sattler}}, \bibinfo {author} {\bibfnamefont {B.}~\bibnamefont {Schallmo}},
  \bibinfo {author} {\bibfnamefont {J.}~\bibnamefont {Stoll}}, \bibinfo
  {author} {\bibfnamefont {Y.-y.}\ \bibnamefont {Tan}}, \bibinfo {author}
  {\bibfnamefont {S.}~\bibnamefont {T{\"o}pfel}}, \bibinfo {author}
  {\bibfnamefont {J.}~\bibnamefont {Turnwald}}, \bibinfo {author}
  {\bibfnamefont {R.}~\bibnamefont {Wen}}, \bibinfo {author} {\bibfnamefont
  {J.}~\bibnamefont {Wessely}}, \bibinfo {author} {\bibfnamefont
  {N.}~\bibnamefont {Wink}}, \bibinfo {author} {\bibfnamefont {S.}~\bibnamefont
  {Yin}},\ and\ \bibinfo {author} {\bibfnamefont {N.}~\bibnamefont {Zorbach}},\
  }\href@noop {} {\  (\bibinfo {year} {2024})}\BibitemShut {NoStop}%
\bibitem [{\citenamefont {Faigle-Cedzich}\ \emph {et~al.}(2023)\citenamefont
  {Faigle-Cedzich}, \citenamefont {Pawlowski},\ and\ \citenamefont
  {Wetterich}}]{Faigle-Cedzich:2023rxd}%
  \BibitemOpen
  \bibfield  {author} {\bibinfo {author} {\bibfnamefont {B.~M.}\ \bibnamefont
  {Faigle-Cedzich}}, \bibinfo {author} {\bibfnamefont {J.~M.}\ \bibnamefont
  {Pawlowski}},\ and\ \bibinfo {author} {\bibfnamefont {C.}~\bibnamefont
  {Wetterich}},\ }\href@noop {} {\  (\bibinfo {year} {2023})},\ \Eprint
  {https://arxiv.org/abs/2307.14787} {arXiv:2307.14787 [cond-mat.quant-gas]}
  \BibitemShut {NoStop}%
\bibitem [{\citenamefont {Litim}(2001)}]{Litim:2001up}%
  \BibitemOpen
  \bibfield  {author} {\bibinfo {author} {\bibfnamefont {D.~F.}\ \bibnamefont
  {Litim}},\ }\href {https://doi.org/10.1103/PhysRevD.64.105007} {\bibfield
  {journal} {\bibinfo  {journal} {Phys. Rev. D}\ }\textbf {\bibinfo {volume}
  {64}},\ \bibinfo {pages} {105007} (\bibinfo {year} {2001})},\ \Eprint
  {https://arxiv.org/abs/hep-th/0103195} {arXiv:hep-th/0103195} \BibitemShut
  {NoStop}%
\end{thebibliography}%
	
\end{document}